%% file: text.tex
\documentclass[showpacs,prb]{revtex4}

\input{headers}
\input{commands}

\begin{document}

% \title{Title}
% \author{Author A.\ Author}
% \email[Corresponding author: ]{name@sever.com}
% \affiliation{Affiliation}

\title{Different dimensionality trends in the Landau damping of magnons in iron, cobalt and nickel: time dependent density functional study}
\author{Pawe\l{} Buczek}
\email[Corresponding author: ]{pbuczek@mpi-halle.mpg.de}
\affiliation{Max-Planck-Institut f\"ur Mikrostrukturphysik, Weinberg 2, 06120 Halle (Saale), Germany}
\author{Arthur Ernst}
\affiliation{Max-Planck-Institut f\"ur Mikrostrukturphysik, Weinberg 2, 06120 Halle (Saale), Germany}
\author{Leonid M.\ Sandratskii}
\affiliation{Max-Planck-Institut f\"ur Mikrostrukturphysik, Weinberg 2, 06120 Halle (Saale), Germany}

\date{\today}

\begin{abstract}
We study the Landau damping of ferromagnetic magnons in Fe, Co, and Ni as the dimensionality of the system is reduced from three to two. We resort to the \textit{ab initio} linear response time dependent density functional theory in the adiabatic local spin density approximation. The numerical scheme is based on the Korringa-Kohn-Rostoker Green's function method. The key points of the theoretical approach and the implementation are discussed. We investigate the transition metals in three different forms: bulk phases, free-standing thin films and thin films supported on a nonmagnetic substrate. We demonstrate that the dimensionality trends in Fe and Ni are opposite: in Fe the transition from bulk bcc crystal to Fe/Cu(100) film reduces the damping whereas in Ni/Cu(100) film the attenuation increases compared to bulk fcc Ni. In Co, the strength of the damping depends relatively weakly on the sample dimensionality. We explain the difference in the trends on the basis of the underlying electronic structure. The influence of the substrate on the spin-wave damping is analyzed by employing Landau maps representing wave-vector resolved spectral density of the Stoner excitations.
\end{abstract}

\pacs{75.78.-n,75.30.Ds,75.70.Ak,75.50.Bb,75.50.Cc,71.15.Qe}
% 75.78.-n Magnetization dynamics
% 75.30.Ds Spin waves
% 75.70.Ak Magnetic properties of monolayers and thin films
% 75.50.Bb Fe and its alloys
% 75.50.Cc Other ferromagnetic metals and alloys
% 71.15.Qe Excited states: methodology

\maketitle
\tableofcontents

\section{Introduction}
\label{sec:Introduction}

The properties of excited magnetic states are of great importance in the fundamental and applied magnetism. Their spectrum determines the thermodynamical properties of magnets, including the Curie temperature \cite{Moriya1985,Pajda2001,Rusz2005}. The excitations contribute to the electronic specific heat \cite{Doniach1966} and the electrical and thermal conductivity, couple to charge degrees of freedom \cite{Vignale1985,Hertz1972,Hertz1973,Edwards1973,Kleinman1978,Schafer2004,Hofmann2009,Schmidt2010a}, limiting the life times and the mean free paths of excited electrons \cite{Gokhale1992,Plihal1995,Hong1999,Hong2000,Zhukov2004,Zhukov2004a,Zhukov2005}, and can provide a coupling mechanism in high temperature superconductors alternative to phonons \cite{Scalapino1995,Eschrig2006,Mazin2008,Mazin2009}. The control of spin dynamics and its attenuation are the central problems in the rapidly growing field of spintronics \cite{Zutic2004,Kaka2005,Neusser2009,Kajiwara2010,Khitun2010}.

Magnetically ordered materials feature a class of collective spin-flip excitations called \textit{spin-waves} or \textit{magnons} \cite{Vleck1958,Shirane1965,Cooke1980,Vollmer2003}. We focus on them in this paper. The spin-waves are usually pictured as a coherent precession of atomic moments around the direction of ground state magnetization \cite{Kittel2005}. Every magnon carries certain energy $\omega_{0}$ and crystal momentum $\vec{q}$ and changes the magnetization of the system by $2\mu_{\mathrm{B}}$. Metallic magnets feature another type of spin-flip excitation termed \textit{Stoner excitations}. They are electron transitions between one-electron states with opposite spin projections. The precessing magnetization can flip spins of single electrons, creating Stoner pairs, which leads to the damping of the moments' precession. The attenuation of collective excitations due to the interaction with single-particle continuum was first considered by Landau \cite{Landau1946,Fetter1971} and is commonly referred to as \textit{Landau damping}. This mechanism of attenuation is dominant in metals that are of our primary interest here. Other decay processes, which could be captured in a language of magnon-magnon, magnon-phonon or magnon-electron interaction, to mention the few, are not considered in this paper.

The theoretical description of spin dynamics that accurately treats both spin-wave and Stoner excitations has proven to be a challenge. Among different formalisms, a particularly transparent approach is offered by the \textit{linear response time dependent density functional theory} \cite{Gross1985}, relying on the evaluation of wave-vector dependent dynamic spin susceptibility. Such calculations are, however, very demanding both from the point of view of algorithmic complexity and computer resources and for a long time they were restricted to simple bulk systems. \cite{Savrasov1998,Staunton1999} Recently, we have developed an efficient numerical scheme allowing to evaluate the spin susceptibility of complex magnets and applied it to study energies and life-times of magnons in complex bulk phases \cite{Buczek2009} and ultrathin films \cite{Buczek2010d,Buczek2011}. In Sec.\ \ref{subsec:HistoricalPerpective} we give a brief historical overview of the developments in this field and relate our calculational approach to works of other authors. 

The properties of three transition metal ferromagnets Fe, Co and Ni have been for many years a subject of intensive theoretical studies (see, e.g., \cite{Kuebler2000,Lichtenstein2001}). Much attention has been also devoted to understanding spin-waves in the ultrathin films absorbed on metallic non-magnetic substrates \cite{Tang1998,Vollmer2003}, as such systems are of enormous practical importance. In the pioneering theoretical works \cite{Muniz2002,Costa2003,Muniz2003,Costa2004a,Costa2004,Costa2006} it was argued that the Landau damping of magnons in the ultrathin films should be generally more severe than in the corresponding bulk. \cite{Tang1998} The broken out-of-film-plane translational symmetry should increase the cross section for scattering of magnons and Stoner excitations. Additionally, the non-magnetic substrates feature infinite continuum of gapless Stoner excitations. Muniz \el \cite{Muniz2002} pointed to the Stoner transitions involving the electronic states of the substrate as an important contribution to the Landau damping. We have shown \cite{Buczek2011} that such simple dimensionality arguments are insufficient and the actual magnitude of Landau damping is sensitive to the details of the hybridization between the electron states of the film and the substrate.

In this paper, our goal is two-fold. First, we provide a thorough expostion of our theoretical approach and outline key details of its numerical implementation. Second, we report detailed study of the influence of the dimensionality of the system on the spin-wave damping in iron, cobalt and nickel. We show that the trend in the damping variation upon the transition from bulk to film is opposite for Fe and Ni whereas in Co the attenuation is practically unchanged. We provide an explanation of these trends on the basis of the properties of the electronic structure.

The paper is organized as follows. In Sec.\ \ref{sec:Formalism} we delineate the formalism of the time dependent density functional theory for spin-flip excitations in the linear regime. Sec.\ \ref{sec:Implementation} covers the details of the computer implementation. The results of the numerical studies are gathered and discussed in Sec.\ \ref{sec:Results}.

\section{Formalism}
\label{sec:Formalism}

\subsection{Historical perspective}
\label{subsec:HistoricalPerpective}

Until recently, the main body of the theoretical studies of spin-waves was based on the \textit{adiabatic} treatment of magnetic degrees of freedom \cite{Halilov1998a,Antropov1999}, where one maps the spin system onto an effective Heisenberg Hamiltonian of atomic moments which is much easier to study. The approach has acquired many different forms in the literature, namely the magnetic force theorem (MFT) \cite{Liechtenstein1987}, frozen magnon technique \cite{Sandratskii1991,Sandratskii1998,Halilov1998a} and static transverse susceptibility method (STSM) \cite{Grotheer2001}. Bruno \cite{Bruno2003} has shown that the MFT machinery in the form originally suggested gives a systematic error in the estimated Heisenberg exchange parameters. STSM method based on a consistent account for the external field is free from this error.

These adiabatic methods utilize density functional theory and, therefore, do not involve adjustable parameters. They are exact in ferromagnets in the limit of small magnon momenta, i.e.\ they allow to determine correctly the spin-wave stiffness constant. In this approach, however, the presence of the Stoner excitations is neglected and no prediction regarding magnon life-times can be made. Furthermore, for higher energies one expects that the coupling to the Stoner continuum, apart from leading to the decay of spin-waves, can also cause a renormalization of the magnon energy. \cite{Muniz2002} In the latter case the adiabatic description must fail also in the prediction of the dispersion.

As mentioned in the Introduction, the damping is captured in the calculations of wave-vector and frequency-dependent transverse magnetic susceptibility $\chi(\vec{q},\omega)$ where spin-waves and Stoner excitations are treated on an equal footing. One associates the spin-wave excitations with the poles of $\chi(\vec{q},\omega)$ in the complex energy plane. Roughly, the real part of the pole position gives the energy of the spin-wave and the imaginary part can be identified with the inverse life-time of the excitation.

Despite the commonly accepted importance of the dynamic susceptibility approach, very few \textit{ab initio} calculations along this line exist due to the complexity of corresponding numerical algorithms and high numerical costs of such calculations. The calculations performed so far deal mostly with elementary bulk systems. In 1980 Cooke \el \cite{Cooke1980} computed the susceptibility of Ni and Fe using the random phase approximation to the susceptibility starting from an idealized band structure. The dynamic method became particularly powerful after the parameter free linear response time dependent density functional theory (LRTDDFT) was developed in 1985 by Gross and Kohn. \cite{Gross1985} (An earlier work of Callaway and Wang \cite{Callaway1975} contains already all key ingredients of the magnetic response in the local density approximation.) From this time on, several works addressing the paramagnetic susceptibility appeared \cite{Stenzel1985,Winter1988,Staunton1986,Staunton1999,Staunton2000} taking advantage of the formalism. In 1998 Savrasov \cite{Savrasov1998} presented the first calculation of the susceptibility using linear response density functional theory for magnetically ordered systems. During the past few years we developed a novel and efficient computer implementation based on the Korringa-Kohn-Rostoker (KKR) Green's function (GF) method \cite{Buczek2009,Buczek2009a,Buczek2010d,Buczek2011} to address the spin susceptibilities of complex bulk systems and thin magnetic films.

Recently, Lounis \el \cite{Lounis2010,Lounis2011} reported on a considerable development of the earlier used empirical tight-binding scheme for the calculations of the transverse dynamic magnetic susceptibility, see e.g.\ \cite{Muniz2002} that takes advantage of KKR Green's functions. Several simplifications were introduced with respect to the standard KKR method such as the neglect of the energy dependence of the electronic wave functions as well as the restriction to the $d$-electron states only.

In our studies we rely on the treatment of the Kohn-Sham Hamiltonian of the DFT theory without simplifications beyond those used in the standard DFT calculations. In order to represent the response functions we use complete and numerically efficient \textit{Y-Ch basis} \cite{Buczek2009a} consisting from the products of the spherical harmonics and Chebyshev polynomials. In particular, we do not need to assume the rigid rotation of atomic magnetic moments \cite{Kotani2008} in the description of the spin response. Our experience shows that the properties of the spin-wave damping considered here depend sensitively on the structure of the Stoner continuum and therefore on the detailed form of the wave functions.

It is important to mention a number of works on the calculation of the dynamic magnetic susceptibility performed within the framework of the many body perturbation theory (MBPT). \cite{Hedin1965,Aryasetiawan1999} This approach has a strong potential for the study of spin-wave excitations, especially in the systems with strong electron correlations where MBPT approach can be advantageous with respect to the DFT theory in the local density framework. However, up to now, the MBPT was applied only to elemental bulk ferromagnets \cite{Karlsson2000,Sasiouglu2010} and time will show if the computational complexity of the scheme will allow the efficient study of nanoscopic systems. It should be also noted that the practical implementations of the method necessarily simplify original Hedin's equations, e.g., by retaining only the ladder diagrams in the $T$-matrix \cite{Zhukov2004,Zhukov2004a,Zhukov2005,Sasiouglu2010} used to construct spin-spin correlation function. The energy dependence and non-locality of the screened Coulomb interaction is often neglected as well. At this level of approximations the MBPT yields essentially the same physical picture of magnons and their damping as the adiabatic local spin density approximation (ALSDA) commonly used in time dependent density functional theory (TDDFT).

Concerning earlier theoretical studies of the spin-wave damping in the two-dimensional magnetic systems, up to our knowledge, there is only one series of investigations based on an empirical tight binding scheme and random phase approximation to the dynamic susceptibility. \cite{Tang1998,Muniz2002,Muniz2003,Costa2003,Costa2004,Costa2004a,Costa2008} These pioneering works yielded rich qualitative information on the spin-flip dynamics as well as on the relation between the parameters of the model and physical observables. Unfortunately, the strong sensitivity of the results to the parametrization of the underlying electronic stricture limits the usefulness of this approach.

In the context of comparing different calculational techniques it is important to address the question of preserving the spin rotational invariance. In the absence of magnetic anisotropy the rigid rotation of spin system costs no energy. In magnetically ordered materials this implies that the spin-wave spectrum becomes gapless and in the limit of the vanishing momentum the magnon energy should tend to zero, featuring a Goldstone mode. \cite{Goldstone1961,Goldstone1962} Formally, both LRTDDFT and MBPT approaches satisfy the conditions of the Goldstone theorem, albeit the concrete numerical realizations lead to deviations of the magnon energy from zero in the limit of $q=0$. A method to compensate for numerical errors and to recover the Goldstone mode was recently discussed in Refs.\ \cite{Lounis2010,Lounis2011}. Taking into account the importance of the issue and the absence of full agreement with the conclusions Lounis \el arrived at, we address the problem in some details in Sec.\ \ref{subsec:CollectiveModes}.

\subsection{Linear response time dependent density functional theory}

The theoretical description of the time evolution of magnetization poses a complex problem in the solid state physics. Time dependent density functional theory \cite{Runge1984} offers a tool for a parameter free description of time evolving system, also under the influence of dynamic external fields. The method is capable of describing arbitrarily large excitations, including non-linear effects like higher harmonic generation, but at present it is computationally too expensive to be efficiently applied to solids. In 1985 Gross and Kohn laid foundations of linear response time dependent density functional theory framework \cite{Gross1985}, which is applicable when the time dependent perturbation is small. The latter formalism allows to determine directly the density-density response function starting from the knowledge of Kohn-Sham (KS) ground state eigensystem. The knowledge of the response function is sufficient to characterize the excited states of the spin system. It is well suited for the \textit{ab initio} description of magnetic excitations in periodic solids. Our efforts in this field concentrated on the development of an accurate and effective calculational scheme applicable to complex magnets with many inequivalent atoms. In this and the next sections we expose the formalism and our numerical approach in some details emphasizing the aspects we consider important for the accuracy and efficiency. We resort to the adiabatic local spin density approximation (ALSDA). \cite{vonBarth1972,Callaway1975,Gross1985,Vignale1997,Onida2002} Unless specified otherwise Rydberg atomic units are used throughout.

We make use of the field operators \cite{Fetter1971}
\begin{align}
\fop\fbr{\vec{x}\alpha}  =
  \sum_{j} \phi_{j}\fbr{\vec{x}\alpha} \dOp_{j}, \quad
\fopc\fbr{\vec{x}\alpha} =
  \sum_{j} \phi_{j}\fbr{\vec{x}\alpha}^{\ast} \cOp_{j},
\end{align}
where $\vec{x}\alpha$ denote spatial and spin degrees of freedom, $\phi_{j}$ stands for a complete orthonormal set of single particle orbitals, and fermionic operators $\dOp_{j}$ and $\cOp_{j}$, respectively, destroy and create a particle in the orbital $j$. The particle and magnetization density operators read
\begin{align}
  \hat{\sigma}^{i}\fbr{\vec{x}} \equiv \sigma^{i}_{\alpha\beta}\fopc\fbr{\vec{x}\alpha}\fop\fbr{\vec{x}\beta},
\end{align}
where $\sigma^{0}_{\alpha\beta} \equiv \delta_{\alpha\beta}$ corresponds to the density operator and $\sigma^{1,2,3}\equiv\sigma^{x,y,z}$ are standard Pauli matrices. Einstein summation convention is used for $\alpha,\beta,\ldots$ indices.

We consider a time dependent perturbation Hamiltonian including oscillating electric potential and magnetic field coupled exclusively to the spin degree of freedom
\begin{align}
  \hat{H}^{\mathrm{ex}}\fbr{t}
    = \sum_{i=0}^{3} \int d\vec{x} \hat{\sigma}^{i}\fbr{\vec{x}t} \Xi_{i}\fbr{\vec{x}t}.
\label{eq:general_density_response}
\end{align}
Relativistic effects and diamagnetic response are neglected. All many-body operators in this work which contain time variable are assumed to be in the Heisenberg picture. The driving potential is defined as a four-vector
\begin{align}
  \Xi\fbr{\vec{x}t} = \fbr{ - \abs{e}\mathcal{V}\fbr{\vec{x}t}, - \frac{g\Bm}{2} \vec{B}\fbr{\vec{x}t}},
\label{eq:DrivingPot}
\end{align}
where $\mathcal{V}$ stands for the external electric potential, $e$ is the charge of electron, $\vec{B}$ is the external magnetic field, $\Bm$ denotes Bohr magneton and $g$ is the electron g-factor. The approximation $g=2$ is used in this work. We define the charge and magnetization density induced by the 
external time-dependent potential as
\begin{align}
   \delta n^{i}\fbr{\vec{x}t} \equiv
   \av{\hat{\sigma}^{i}\fbr{\vec{x}}}_\ex\fbr{t}
    - n^{i}\fbr{\vec{x}},
\end{align}
where $n^{i}\fbr{\vec{x}} \equiv \av{\hat{\sigma}^{i}\fbr{\vec{x}}}$ is the unperturbed ground state charge and magnetization density and $\av{\hat{o}}_\ex(t)$ stands for the expectation value of operator $\hat{o}$ at time $t$ when the perturbation is active. $\delta n^{i}$ is related to $\Xi_{j}\fbr{\vec{x}t}$ through the retarded density-density response function
\begin{align}
  \chi^{ij} \fbr{\vec{x},\vec{x}',t-t'} &= - \ii \Heav(t-t')
    \av{\comm%
      {\hat{\sigma}^{i}\fbr{\vec{x}t}}
      {\hat{\sigma}^{j}\fbr{\vec{x}'t'}}
    {}},
\label{eq:GeneralDensityResponse}
\end{align}
where $\comm{A}{B}{}\equiv AB - BA$. In the frequency domain one obtains
\begin{align}
  \delta n^{i}\fbr{\vec{x},\omega} \equiv \int dt \ee{\ii \omega t} \delta n^{i}\fbr{\vec{x},t}
    = \sum_{j=0}^{3} \int d\vec{x}' 
      \chi^{ij} \fbr{\vec{x},\vec{x}',\omega} \Xi_{j}\fbr{\vec{x}',\omega}.
\label{eq:ResponseDefinition}
\end{align}

LRTDDFT allows to compute the generalized susceptibility in the following two step procedure. \cite{Gross1985,Qian2002} First, one considers the \textit{Kohn-Sham susceptibility}
\begin{align}
\chi^{ij}\KS\fbr{\vec{x},\vec{x}',\omega} = 
  \sum_{km} \sigma^{i}_{\alpha\beta} \sigma^{j}_{\gamma\delta} 
    \fbr{f_{k} - f_{m}}
    \frac{
      \phi_{k}\fbr{\vec{x}\alpha}^{\ast}
      \phi_{m}\fbr{\vec{x}\beta}
      \phi_{m}\fbr{\vec{x}'\gamma}^{\ast}
      \phi_{k}\fbr{\vec{x}'\delta}
  }
  {
    \omega + \fbr{\eps_{k} - \eps_{m}} + \ii 0^{+}
  },
\label{eq:KSSusc}
\end{align}
giving the retarded response of the formally non-interacting Kohn-Sham system. The $\zp$ notation is introduced to stress that we deal with retarded quantities. In the above equation $\phi_{j}\fbr{\vec{x}\alpha}$'s and $\eps_{j}$'s denote respectively KS eigenfunctions and corresponding eigenenergies. $f_{j} \equiv \fFD(\eps_{j})$, where $\fFD(\eps)$ is the Fermi-Dirac distribution function. The induced charge and magnetization densities described by the Kohn-Sham susceptibility modify the Hartree and exchange-correlation potential, giving rise to a self-consistent problem: the induced densities contribute to the effective fields and are, simultaneously, induced by it. The self-consistency is reflected in the second step of the formalism
\begin{align}
  \chi^{ij} \fbr{\vec{x},\vec{x}',\omega} = \chi^{ij}\KS \fbr{\vec{x},\vec{x}',\omega} +
    \sum_{k,l=0}^{3}
    \iint d\vec{x}_{1}d\vec{x}_{2} \chi^{ik}\KS \fbr{\vec{x},\vec{x}_{1},\omega}
      \fbr{
        K\xc^{kl}\fbr{\vec{x}_{1},\vec{x}_{2},\omega}
        + \frac{2 \delta_{k0}\delta_{l0}}{\abs{\vec{x}_{1} - \vec{x}_{2}}}
      }
      \chi^{lj}\fbr{\vec{x}_{2},\vec{x}',\omega}.
\label{eq:GeneralDyson}
\end{align}
The last equation is referred to as ``susceptibility Dyson equation'', because of its characteristic form. $\chi$ is the density-density response function of the system and describes charge neutral excitations. It is often termed \textit{enhanced susceptibility}. The exchange-correlation kernel, $K\xc$, is defined as a functional derivative of exchange-correlation potential with respect to the density
\begin{align}
  K\xc^{ij}\sqbr{\av{\hat{\Psigma}\fbr{\vec{x}}}}\fbr{\vec{x},\vec{x}',t-t'} =
    \frac{\delta v\xc^{i}\fbr{\vec{x},t}}{\delta n^{j}\fbr{\vec{x}'t'}}
\end{align}
evaluated at the ground state values of electronic and magnetic densities.

The structure of Eq.~\eqref{eq:GeneralDyson} resembles the expression for the susceptibility obtained within the random phase approximation (RPA) to the many-electron Hamiltonian. \cite{Fetter1971,Cooke1980} However, it is important to note that equation \eqref{eq:GeneralDyson} is exact, providing the exact  $K\xc$ were known, whereas RPA corresponds to a specific approximation of the proper polarization function. 

The problem of constructing exact exchange-correlation kernel is equivalent to the exact solution of the many-body problem and one faces the necessity of approximating it. The most common choice, adopted also in this paper, is the ALSDA, in which $K\xc$ is approximated by frequency independent functional derivative of the LSDA exchange-correlation potential:
\begin{align}
  K\xc^{ij}\sqbr{\av{\hat{\Psigma}\fbr{\vec{x}}}}\fbr{\vec{x},\vec{x}',t - t'} =
    \frac{\delta v_{\textrm{LSDA}}^{i}\sqbr{\av{\hat{\Psigma}\fbr{\vec{x}}},\vec{x}}}{\delta n^{j}\fbr{\vec{x}}}
    \delta\fbr{\vec{x} - \vec{x}'} \delta\fbr{t - t'}.
\end{align}
The adiabaticity in this context pertains to the response of the electron gas, which is assumed to be given by the instantaneous values of the densities. Furthermore, adopting LSDA implies that the kernel is determined by the local value of densities. Recently, there has been a progress in constructing non-local magnetic exchange-correlation functionals \cite{Kurth2009,Eich2010}, but their implementation in practical band structure calculations has not been achieved yet.

From the computational point of view, the main challenge is the evaluation of the unenhanced susceptibility. The direct use of Lehmann representation given by eq.\ \eqref{eq:KSSusc} requires the summation over high energy unoccupied states and it is practically never used in concrete implementations \cite{Savrasov1998}, especially for metals. The use of KKR Green's functions in the construction of KS susceptibilities avoids this problem. There are several further advantages: it is not necessary to consider finite basis corrections and the description of systems without full three-dimensional discrete translational symmetry can be rather easily achieved. These issues are discussed deeper in Sec.\ \ref{sec:Implementation}.

Not only the calculation but also the analysis of the dynamic susceptibility of a complex system might become cumbersome. Because of large number of degrees of freedom the response is governed by a complex overlap of various excitation modes. \cite{Buczek2010d} In order to distinguish them, it is convenient to consider the \textit{loss matrix}, defined as the anti-Hermitian part of the susceptibility
\begin{align}
 \Loss\sqbr{\chi^{ij}(\vec{x},\vec{x}',\omega + \ii\zp)} \equiv \frac{1}{2\ii}
    \fbr{\chi^{ij}(\vec{x},\vec{x}',\omega + \ii\zp) - \chi^{ji}(\vec{x}',\vec{x},\omega + \ii\zp)^{\ast}}.
\end{align}
$\Loss\chi$ has a clear physical interpretation, as the power absorbed from the harmonic driving potential \cite{Callen1951}
\begin{align}
   \Xi_{i}\fbr{\vec{x}t} = \Xi_{i}\fbr{\vec{x}}\cos\omega t
\end{align}
reads
\begin{align}
  P = - \omega \sum_{ij} \iint d \vec{x} d \vec{x}'
    \Xi_{i}\fbr{\vec{x}}\Loss\sqbr{\chi^{ij}(\vec{x},\vec{x}',\omega)}\Xi_{j}\fbr{\vec{x}'}.
\end{align}
Linear response theory and the fluctuation-dissipation theorem \cite{Nyquist1928,Kubo1966} tell us that the energy absorption signifies the presence of excited states of the unperturbed system with the energy $\omega$. $\Loss\chi$ is a Hermitian matrix and features real eigenvalues and a set of orthogonal eigenvectors. The eigenvectors $\xi_{\lambda}(\vec{x})$ of $\Loss\sqbr{\chi^{ij}(\vec{x},\vec{x}',\omega)}$ represent the \emph{shapes} of \emph{natural modes} of the system. The associated eigenvalues, $\Loss\sqbr{\chi^{ij}(\vec{x},\vec{x}',\omega)}_{\lambda}$, give the rate of energy absorption from the external field with the shape $\xi_{\lambda}(\vec{x})$ oscillating with frequency $\omega$. Formally, the number of eigenvalues is infinite, but in reality only limited number of them are large, corresponding to physically relevant excitations. The analysis of $\Loss\chi$ allows their unambiguous identification. The contact with experiment is usually made by evaluating the Fourier transformation of the susceptibility as the $\Im\chi^{ij}_{\vec{K}\vec{K}}(\vec{q},\omega)$ is probed in the scattering experiments  \cite{VanHove1954,Vollmer2003}, cf.\ App.\ \ref{app:Basis}.

\subsection{Transverse magnetic susceptibility}

Within the non-relativistic LSDA, the Kohn-Sham states of the collinear magnets and paramagnetic systems can be characterized by a certain value of the spin-projection. We adopt the convention that the ground-state magnetization $\vec{m}\fbr{\vec{x}}$ points everywhere along the $z$ direction that is selected as the axis of spin-quantization.

In this case the susceptibility $\chi^{ij}\KS$ has only four independent elements and the following structure
\begin{align}
\chi\KS =
  \begin{pmatrix}
     \phantom{-}\chi\KS^{xx} & \chi\KS^{xy} &             0  &            0 \\
              - \chi\KS^{xy} & \chi\KS^{xx} &             0  &            0 \\
                           0 &            0 &  \chi\KS^{00}  & \chi\KS^{0z} \\
                           0 &            0 &  \chi\KS^{0z}  & \chi\KS^{00}
  \end{pmatrix}.
\end{align}
The response to the transverse (ie.\ with the direction lying in the $xy$-plane) magnetic field is transverse and does not involve charge density response as opposed to longitudinal magnetic field (along $z$ direction).

In ALSDA, the exchange-correlation kernel is given by the functional derivative of the LSDA exchange-correlation magnetic field
\begin{align}
 \vec{B}\xc(\vec{m})=B\xc(m)\frac{\vec{m}}{m}
\end{align}
with respect to magnetization \cite{Qian2002,Katsnelson2004}
\begin{align}
  \frac{\delta B^{i}\xc}{\delta m_{j}} &= B\xc \pder{}{m_{j}}\frac{m_{i}}{m}
    + \frac{\delta B\xc}{\delta m_{j}} \frac{m_{i}}{m}
   =  \frac{B\xc}{m}\fbr{\delta_{ij} - \frac{m_{i}m_{j}}{m^{2}}}
    + \frac{\delta B\xc}{\delta m} \frac{m_{i}m_{j}}{m^{2}}.
\end{align}
The first term gives the response in the direction perpendicular to $\hat{\vec{m}}\equiv\hat{\vec{z}}$ (the transverse response), while the second along the direction of ground state magnetization. We see that the induced transverse magnetization gives rise to an additional effective exchange-correlation magnetic field which is also transverse. Thus, in collinear magnets ALSDA decouples magnons and non-spin-flip excitations.

This is a useful property since it allows us to separate the Dyson equation for the transverse magnetic susceptibility from the one for the longitudinal and the charge response. In order to study spin-flip excitations in collinear systems it is sufficient to consider transverse components ($xx$, and $xy$) of the magnetic susceptibility. It is convenient to introduce circular coordinates for magnetization response, magnetic field, and the transverse susceptibility
\begin{align}
  m^{\pm} = m_{x} \pm \ii m_{y}, \quad
  B^{\pm} = B_{x} \pm \ii B_{y}, \quad
  \chi^{\pm} = \chi^{xx} \mp \ii \chi^{xy}.
\end{align}
They are connected through the following relation
\begin{align}
  m^{\pm}\fbr{\vec{x},\omega} = \int d\vec{x'}
    \chi^{\pm}\fbr{\vec{x},\vec{x}',\omega} B^{\pm}\fbr{\vec{x}',\omega}.
\end{align}
The explicit form of the KS $\chi^{\pm}$ susceptibility in the collinear system reads
\begin{align}
\chi\KS^{\pm}\fbr{\vec{x},\vec{x}',\omega} = 2
    \sum_{km} \fbr{f_{k}^{\uparrow\downarrow} - f_{m}^{\downarrow\uparrow}}
        \frac{
          \phi_{k}^{\uparrow\downarrow}\fbr{\vec{x}}^{\ast}
          \phi_{m}^{\downarrow\uparrow}\fbr{\vec{x}}
          \phi_{m}^{\downarrow\uparrow}\fbr{\vec{x}'}^{\ast}
          \phi_{k}^{\uparrow\downarrow}\fbr{\vec{x}'}
        }{
          \omega + \eps_{k}^{\uparrow\downarrow} - \eps_{m}^{\downarrow\uparrow} + \ii\zp
        }.
\label{eq:ChiKSpm}
\end{align}
Left arrows corresponds to $\chi\KS^{+}$ while the right arrows to $\chi\KS^{-}$.

The Dyson equation for enhanced susceptibility takes the form
\begin{align}
  \chi^{\pm} \fbr{\vec{x},\vec{x}',\omega} =
    \chi^{\pm}\KS \fbr{\vec{x},\vec{x}',\omega} +
    \int d\vec{x}_{1} \chi^{\pm}\KS\fbr{\vec{x},\vec{x}_{1},\omega} K\xc\fbr{\vec{x}_{1}}
      \chi^{\pm}\fbr{\vec{x}_{1},\vec{x}',\omega}.
\label{eq:SusceptibilityDysonEquation}
\end{align}
The use of circular coordinates allows us to work with a complex scalar equation instead of a matrix equation in Cartesian spin projections. The real exchange-correlation kernel amounts to
\begin{align}
  K\xc\fbr{\vec{x}} = - {\Bm} \frac{B\xc\fbr{\vec{x}}}{m\fbr{\vec{x}}},
\end{align}
where $B\xc\fbr{\vec{x}}$ and $m\fbr{\vec{x}}$ are local values of the exchange-correlation magnetic fields and magnetization density respectively. They can easily be found once a LSDA calculation of the ground state is performed.

If one casts the spatial dependence of $\chi^{\pm}$, $\chi^{\pm}\KS$ and $K\xc$ in an orthonormal basis, the Dyson equation takes a matrix form with the formal solution
\begin{align}
  \chi^{\pm} \fbr{\omega} = \fbr{\onemat - \chi^{\pm}\KS\fbr{\omega} K\xc\fbr{\omega}}^{-1}
    \chi^{\pm}\KS \fbr{\omega}.
\label{eq:SusceptibilityDysonEquationMatrix}
\end{align}
According to this equation the singularities of the enhanced (physical) susceptibility stem from two sources. The first are the singularities of $\chi^{\pm}\KS \fbr{\omega}$ corresponding to electronic transitions between occupied and empty KS states with opposite spin projections, so called \textit{Stoner transitions}; they form the \textit{Stoner continuum}. The second correspond to zeros of the $\onemat - \chi^{\pm}\KS K\xc$. The step from unenhanced susceptibility to the enhanced susceptibility results in a remarkable property that the energy absorption can now take place for frequencies outside the Stoner continuum. This signifies the new type of the excitations different from the Stoner transitions described by the unenhanced susceptibility. These are the spin-waves. Their formation in complex solids and interactions with Stoner excitations are the subject of the next section.

\subsection{Collective modes}
\label{subsec:CollectiveModes}

For the sake of subsequent analysis it is convenient to rewrite equation \eqref{eq:SusceptibilityDysonEquationMatrix} as
\begin{align}
  \chi^{-1}\fbr{\omega} = \chi^{-1}\KS\fbr{\omega} - K\xc\fbr{\omega}.
\label{eq:DysonInverse}
\end{align}
We assume that the inverses of KS and enhanced susceptibility matrices exist for the frequencies of interest. The following analysis does not require $K\xc$ to be adiabatic or local. If the Hamiltonian under consideration admits certain symmetries, the Hilbert space in which we represent the susceptibility can be decomposed into subspaces where the analysis can be performed independent of each other. For example, for periodic solids the quasi-momentum $\vec{q}\in\BZ$, where $\BZ$ stands for the first Brillouin zone, is a good quantum number and the susceptibility becomes block diagonal when represented in the Bloch basis. To simplify notation, $\vec{q}$ is omitted in this section.

\subsubsection{Frequency $\omega_{0}$ outside the Stoner continuum}

Away from the Stoner continuum and the singularities of the kernel the matrix $\chi^{-1}\fbr{\omega_{0}}$, $\omega_{0}\in\Reals$, is Hermitian. Let $\left\lbrace \nu_{\lambda},\ket{m_{\lambda}}\right\rbrace $ be an eigensystem of $\chi^{-1}\fbr{\omega_{0}}$. $\lambda$ labels eigenvalues. If all $\nu_{\lambda}\ne0$, the inverse $\chi\fbr{\omega_{0}}$ is non-singular Hermitian matrix and there are no magnetically excited states at $\omega_{0}$. However, if there can form a resonance between an external magnetic field $\ket{B_{\lambda}}$ and the induced effective exchange-correlation field, such that the conditions
\begin{subequations}
\label{eq:ResCond}
\begin{align}
  \ket{B_{\lambda}} &= \chi^{-1}\KS\fbr{\omega_{0}} \ket{m_{\lambda}}, \\
  \ket{B_{\lambda}} &= K\xc\fbr{\omega_{0}} \ket{m_{\lambda}}
\end{align}
\end{subequations}
are fulfilled, the eigenvalue $\nu_{\lambda}$ vanishes. The enhanced susceptibility becomes singular which signifies the formation of a collective magnetic excited state (spin-wave) with energy $\omega_{0}$. It is an exact eigenstate of the many-body Hamiltonian and its life-time is infinite.

Let us construct the loss matrix associated with $\chi\fbr{\omega + \ii\zp}$ close to $\omega_{0}$ in this case. It will allow us to find spatial shape of the resonant field and the fluctuating magnetization density associated with the magnon. We expand $\chi^{-1}\fbr{z}$ around $\omega_{0}$
\begin{align}
  \chi^{-1}\fbr{z} = \chi^{-1}\fbr{\omega_{0}} + \fbr{z - \omega_{0}} \delta \chi^{-1}\fbr{\omega_{0}} 
    + O(\fbr{z - \omega_{0}}^{2}).
\end{align}
We assume that $\omega_{0}$ is an isolated singular point, i.e.\ close to $\omega_{0}$ the matrix $\chi^{-1}\fbr{z}$ is invertible, providing $z\ne\omega_{0}$. Let $\mathcal{N}$ be the set of indices $\lambda$ corresponding to $\nu_{\lambda} = 0$. If there exists more than one vanishing eigenvalue (the case of degenerated magnon states), we shall work with linear combinations of corresponding eigenvectors fulfilling the condition
\begin{align}
  \bra{m_{\lambda}} \delta \chi^{-1}\fbr{\omega_{0}} \ket{m_{\lambda'}} = s_{\lambda} \delta_{\lambda\lambda'}, \quad
    \lambda,\lambda'\in\mathcal{N}.
\end{align}
If $s_{\lambda} \ne 0$ we can construct the $-1$st term of the Laurent series of $\chi\fbr{z}$ around $\omega_{0}$
\begin{align}
  \chi\fbr{z} = \sum_{\lambda\in\mathcal{N}} \frac{1}{s_{\lambda}} \frac{\ket{m_{\lambda}}\bra{m_{\lambda}}}{z - \omega_{0}}
    + O(1),
\end{align}
which is sufficient to determine the associated loss matrix
\begin{align}
  \Loss\chi\fbr{\omega} = - \pi \delta\fbr{\omega - \omega_{0}}
    \sum_{\lambda\in\mathcal{N}} \frac{1}{s_{\lambda}} \ket{m_{\lambda}}\bra{m_{\lambda}}.
\end{align}
The singularity of $\chi\fbr{\omega_{0}}$ of the form discussed above signifies that there exists a spin-wave of energy $\omega_{0}$ which is an exact excited eigenstate of the system. Because of the linear character of the response an external field $\ket{B}$ oscillating with the frequency $\omega_{0}$ will excite the spin-wave and result in a strong energy absorption providing that the spatial form of the field and the spatial form of the loss-matrix eigenstate are not orthogonal
\begin{align}
  \bra{B} m_{\lambda}\rangle \ne 0, \quad \lambda\in\mathcal{N}.
\label{eq:divergentField}
\end{align}
The magnetization response to such a field is infinite in the linear response approximation. Physically, condition \eqref{eq:divergentField} determine the fields able to excite mode $\lambda$.

A similar argument can be used to prove that the LRDFT features Goldstone mode, i.e.\ an excitation of vanishing energy \cite{Goldstone1961,Goldstone1962}, corresponding to a singularity of static enhanced susceptibility. We want to prove that there exists a small static transverse magnetic field $B_{0}\fbr{\vec{x}}$, pointing everywhere in the same direction, say along $x$-axis, for which the magnetization response is infinite. This corresponds to a divergence of susceptibility and the presence of Goldstone mode. It is easy to prove that \cite{Buczek2009a}
\begin{subequations}
\label{eq:GoldstoneResonance}
\begin{align}
  \ket{B\xc} &= \chi^{-1}\KS\fbr{0} \ket{m\GS}, \\
  \ket{B\xc} &= K\xc\fbr{0} \ket{m\GS},
\end{align}
\end{subequations}
where the $\ket{B\xc}$ is a field of the \textit{shape} of the ground state exchange correlation field, but pointing in the transverse direction and, correspondingly, $\ket{m\GS}$ has the shape of ground-state value of the magnetization. All the response functions above are static and the expression is valid for general exchange-correlation kernel; it is also fulfilled in ALSDA. This is turn means that the response to the transverse static field of the shape of ground-state exchange-correlation field is infinite. This proves that LRTDDFT formalism features Goldstone mode. When dealing with periodic (uniform) systems, Goldstone mode is necessarily associated with $\vec{q}=\vec{0}$, because $B\xc\fbr{\vec{x}}$ is a periodic (constant) function of $\vec{x}$.

\subsubsection{Frequency $\omega_{0}$ inside the Stoner continuum}

Now we turn to the case that there are Stoner excitations with energy $\omega_{0}$. In this case, the resonance condition (\ref{eq:ResCond}) cannot be fulfilled exactly anymore, as $\chi\KS\fbr{\omega_{0}}$ ceases to be Hermitian and a non-zero phase is introduced between the driving field and the Kohn-Sham response. Some eigenvalues of $\onemat - \chi^{\pm}\KS\fbr{\omega} K\xc\fbr{\omega}$ can nevertheless become small and strongly enhance certain eigenvalues of $\Loss\chi(\omega)$. The latter might feature a clear peak with maximum at $\omega_{0}$. If the density of Stoner excitations around $\omega_{0}$ is small and weakly depends on $\omega$, the peak will take the form of Lorentzian-like resonance. Its non-zero width signifies that the corresponding spin-wave is not an exact eigenstate and features a finite life time. Physically, the attenuation is interpreted as a consequence of hybridization of the spin-waves with Stoner excitations. It is called Landau damping. In the region of dense Stoner continuum, i.e.\ where the Hermitian part of $\chi\KS\fbr{\omega}$ becomes comparable with its anti-Hermitian part, no well defined spin-waves form, leading to the phenomenon of \textit{spin-wave disappearance}. We will visualize these different regimes on concrete examples in Sec.\ \ref{sec:Results}.

It is important to note that for a particular frequency $\omega$ both the real part, which determines the magnon energy in the first line, and the imaginary part of the KS susceptibility (responsible for damping) are determined by the Stoner transitions in the system. The imaginary part is given by ``actual transtions'' in the sense that only Stoner pairs with energy differences equal to $\omega$ contribute to it. On the other hand the real part is determined by both the ``actual transtions'' and ``virtual transtions'', as it ivolves energy integral over all Stoner excitations.

We remark that the Landau mechanism is the only one leading to the finite life-time of collective excitations when adiabatic approximation of the exchange-correlation kernel is invoked. \cite{Vignale1997} The neglected complex singularities of $K\xc\fbr{\omega}$ matrix could lead to the broadening of the spin-wave peaks also outside the Stoner continuum and account for the effects like magnon-magnon or magnon-electron scattering.

\subsection{Sum rules}

The following sum rule \cite{Edwards1978} provides further insight in the relation between Stoner continuum and formation of spin-waves. One shows that the integration of the the loss matrix associated with transverse Kohn-Sham susceptibility over all frequencies is related to the ground state magnetization
\begin{align}
  \int d \vec{x}' \Loss\sqbr{\int_{-\infty}^{\infty}d\omega\chi^{+}\KS\fbr{\vec{x},\vec{x}',\omega + \ii\zp}} =
    - 2 \pi m\GS\fbr{\vec{x}}.
\end{align}
The sum rule for $\chi^{-}$ differs only by sign in the right-hand side of the above equation. Applying Cauchy integral one proves that integrated loss matrix of the \textit{enhanced} transverse susceptibility is given by exactly the same expression providing the exchange-correlation kernel is taken to be frequency independent
\begin{align}
  \int d \vec{x}' \Loss\sqbr{\int_{-\infty}^{\infty}d\omega\chi^{\pm}\fbr{\vec{x},\vec{x}',\omega + \ii\zp}} =
    \mp 2 \pi m\GS\fbr{\vec{x}}.
\end{align}
The same relation holds true for the lattice Fourier transformed susceptibilities defined by Eq.\ \eqref{eq:latticeFTchi}
\begin{align}
  \int d \vec{r}' \Loss\sqbr{\int_{-\infty}^{\infty}d\omega\chi^{\pm}\KS\fbr{\vec{r},\vec{r}',\vec{q},\omega + \ii\zp}} =
  \int d \vec{r}' \Loss\sqbr{\int_{-\infty}^{\infty}d\omega\chi^{\pm}   \fbr{\vec{r},\vec{r}',\vec{q},\omega + \ii\zp}} =
    \mp 2 \pi m\GS\fbr{\vec{r}},
\end{align}
where $\vec{r}, \vec{r}'\in\WS$.

The above relation has a clear physical interpretation. The self-consistency condition expressed by the Dyson equation leads to the redistribution of the spectral density of spin flip excitations. As long as the frequency dependence of the exchange-correlation kernel can be neglected, the integrated spectral weight does not change. However, the character of the excitations is now very different including spin-waves and hybrids between spin-waves and Stoner transitions. In many cases, in particular for small spin-wave momenta, most of the spectral power is concentrated in the spin-wave peaks. The surprising result that the Stoner excitations present in the underlying system cannot be excited by the field of the corresponding frequency may be understood as follows. In the energy region of Stoner continuum there exists a significant phase shift between the external driving field and the magnetization response given by $\chi\KS$ due to the large anti-Hermitian part of the latter matrix. This results in the induced exchange-correlation field that is out-of-phase with the external driving field. In the next step, the induced exchange-correlation field adds up to the external field and produces again a contribution that is out of phase to both external field and initial exchange-correlation field. This process to be performed up to the self-consistency leads to the compensation of the external field by the contributions of the induced exchange correlation fields and the suppression of the Stoner transitions.

\section{Computer implementation} 
\label{sec:Implementation}

In this Section we shall discuss major issues concerning the actual numerical determination of the response functions. The problem can be split into two parts, reflecting the structure of LRTDDFT. First, one finds $\chi\KS$ on the basis of the knowledge of the KS eigensystem. Second, the Dyson equation is solved in order to determine the enhanced (physical) susceptibility.

The determination of the KS susceptibility is the most computationally demanding part of the calculations. We begin with formally non-interacting KS Green's function
\begin{align}
  G_{\alpha\beta}\fbr{\vec{x},\vec{x}',z} = \sum_{j}
    \frac{\phi_{j}\fbr{\vec{x}\alpha}\phi_{j}\fbr{\vec{x}'\beta}^{\ast}}
      {z - \eps_{j}}, \, z\in\Complex,
\label{eq:BareGreensFunction}
\end{align}
where $\phi_{j}\fbr{\vec{x}\alpha}$'s and $\eps_{j}$'s denote KS eigenfunctions and corresponding eigenenergies. The numerical evaluation of $G$ is based on Korringa-Kohn-Rostoker method. The actual representation of the GF in the multiple scattering formalism is given in App.\ \ref{app:KKRGFProduct}.

By applying Cauchy theorem one obtains the following expression for the KS susceptibility \cite{Abrikosov1975,Schmalian1996} ($\gamma > \gamma' = \zp$):
\begin{align}
  \chi^{ij}\KS&\fbr{\vec{x},\vec{x}',\omega + \ii\gamma}
  = - \frac{1}{2\pi\ii} \int_{-\infty}^{\infty} d\eps \times \nno\\
    (
       &\fFD\fbr{\eps + \ii\gamma'}
        S_{ij}\fbr{\vec{x},\vec{x}',\eps + \omega + \ii\gamma + \ii\gamma',\eps + \ii\gamma'}\nno\\
      -&\fFD\fbr{\eps - \ii\gamma'}
        S_{ij}\fbr{\vec{x},\vec{x}',\eps + \omega + \ii\gamma - \ii\gamma',\eps - \ii\gamma'}\nno\\
      +&\fFD\fbr{\eps + \ii\gamma'}
        S_{ij}\fbr{\vec{x},\vec{x}',\eps + \ii\gamma',\eps - \omega - \ii\gamma + \ii\gamma'}\nno\\
      -&\fFD\fbr{\eps - \ii\gamma'}
        S_{ij}\fbr{\vec{x},\vec{x}',\eps - \ii\gamma',\eps - \omega - \ii\gamma - \ii\gamma'}
    )
\label{eq:Schmalian}
\end{align}
where the $S$ is the product of two KKRGFs
\begin{align}
  S_{ij}\fbr{\vec{x},\vec{x}',z_{1},z_{2}} \equiv
  \sigma^{i}_{\alpha\beta} \sigma^{j}_{\gamma\delta}
    G_{\beta\gamma} \fbr{\vec{x} ,\vec{x}',z_{1}}
    G_{\delta\alpha}\fbr{\vec{x}',\vec{x} ,z_{2}}.
\label{eq:Sij}
\end{align}
When working in $\vec{k}$-space, the product above is transformed into a convolution of two Green's functions over the Brillouin zone, cf.\ App.\ \ref{app:KKRGFProduct}. The convolution converges badly for energies $z_{1,2}$ close to the real KS energies. This in turn precludes the direct determination of the retarded KS susceptibility. \cite{Staunton1999} For computational expediency one calculates the susceptibility for a set of the points in the upper complex semi-plane, away from the real axis. As shown in App.\ \ref{app:ComplexIntegration}, in this case, the necessary energy integrations in eq.~\eqref{eq:Schmalian} can be reduced to integrations along finite complex contours, away from the poles of the GFs. At the end of the calculations the resultant susceptibility can be analytically continued to the real axis to recover the real time dynamics.

Note the presence of the Fermi-Dirac weight in eq.\ \eqref{eq:Schmalian}. The formulation presented avoids integration over energies above Fermi level. It is so, because KKRGF contains information about \textit{all} KS states, cf.\ eq.\ \eqref{eq:BareGreensFunction}. Thus, the serious technical problem of including explicitly unoccupied KS orbitals, as it is the case when evaluating $\chi\KS$ directly from eq.~\eqref{eq:KSSusc}, is avoided. \cite{Savrasov1998} No finite basis set corrections are necessary, either, when working with KKRGF \cite{Savrasov1990,Savrasov1992,Savrasov1996,Savrasov1998}, as the multiple scattering problem is solved separately for every complex energy. 
In this context, it is not clear how large is the error introduced by the minimal energy independent $d$-symmetry basis used by Lounis \el \cite{Lounis2010,Lounis2011}, an approximation avoided in this work. In our current study the major advantage of using KKRGF is the possibility of an efficient description of systems featuring surfaces and interfaces \cite{Wildberger1997,Zabloudil2005}, in particular films absorbed on a substrate. The construction of $\chi\KS$ taking into account specific representation of KKRGF is given in App.\ \ref{app:KKRGFProduct}.

It is convenient to cast the $\fbr{\vec{x},\vec{x}'}$ dependence into a separable basis when solving the Dyson equation. This aspect is discussed in Appendix \ref{app:Basis}. Subsequently, the equation \eqref{eq:SusceptibilityDysonEquation} can be solved by matrix inversion or, for frequencies away from the spin-wave poles, iteratively. The CPU time necessary to solve the susceptibility Dyson equation is negligible comparing to the time of computing the KS susceptibility.

Below we turn to the important question of the consistent description of the Goldstone mode. As discussed in Sec.\ \ref{subsec:CollectiveModes}, $\chi\fbr{\vec{q} = \vec{0},\omega=0}$ is a singular matrix with a diverging eigenvalue signifying the formation of the Goldstone mode. Seen alternatively, cf.\ eqs.\ \eqref{eq:GoldstoneResonance}, the matrix
\begin{align}
  \mathcal{D} \equiv \onemat - \chi^{\pm}\KS\fbr{0} K\xc
\label{eq:DysonMatrix}
\end{align}
features a vanishing eigenvalue corresponding to the eigenvector representing ground state magnetization $\ket{m\GS}$. This is the consequence of the spin rotational invariance of the problem.

Because of the numerical inaccuracies the condition of zero eigenvalue in the limit of $q\to0$ is not satisfied exactly since the calculated Kohn-Sham susceptibility does not correspond exactly to the ground-state exchange-correlation magnetic field and ground state magnetization. Below we discuss the method to compensate for this error. \cite{Buczek2009a} Numerical diagonalization of matrix $\mathcal{D}$ gives one eigenvalue that is very close to zero and much smaller than other eigenvalues. In our calculations for systems discussed in this paper this eigenvalue is typically of the order of $10^{-3}$ whereas other eigenvalues are of the order of unity. It is easy to verify that this small eigenvalue corresponds to the eigenvector very close to $\ket{m\GS}$. To obtain the matrix with zero eigenvalue we proceed as follows. Upon the diagonalization of the calculated $\mathcal{D}$, the small eigenvalue is set to zero and the remaining eigenvalues are left unchanged. Using the original eigenvectors of $\mathcal{D}$ the corrected diagonal matrix is transformed back into non-diagonal form. Because of the high precision of the diagonalization routines the corrected matrix  $\mathcal{D}_{\mathrm{corr}}$ will have a zero eigenvalue with a very high accuracy. $\mathcal{D}_{\mathrm{corr}}$ is now used to find the corrected exchange-correlation kernel
\begin{align}
  K\xc^{\mathrm{corr}} = \fbr{\chi^{\pm}\KS\fbr{0}}^{-1} \fbr{\onemat - \mathcal{D}}
\end{align}
This corrected kernel corresponds to the calculated Kohn-Sham susceptibility $\chi^{\pm}\KS\fbr{0}$ in the sense of the fulfillment of the Goldstone theorem and is used for the calculation of the enhanced susceptibility for all $\vec{q}$ vectors and frequencies. For the example of bcc Fe the uncorrected small eigenvalue of $\mathcal{D}$ matrix in our calculation was $\epsilon\approx3.6\cdot10^{-3}$ which corresponded to the shift of the Goldstone mode from zero by about \unit{2.3}{\milli\electronvolt} (compare to the energy width of magnon band of order of \unit{0.5}{\electronvolt}).

The methods to correct for the numerical deviations from the requirement of Goldstone theorem suggested by Lounis \el \cite{Lounis2011}, Kotani \el \cite{Kotani2008} and \c{S}a\c{s}\i{}o\u{g}lu \el \cite{Sasiouglu2010} are all based on the same general idea of bringing in necessary correspondence of the unenhanced susceptibility and underlying potential expressed by equation \eqref{eq:GoldstoneResonance}. That is also the essence of our approach. Our correction of one eigenvalue of order of $10^{-3}$ should be compared with $2-12\%$ correction for the exchange-correlation kernel reported by Lounis \el \cite{Lounis2011} and 50\% correction for the screend Coulomb interaction of \c{S}a\c{s}\i{}o\u{g}lu \el \cite{Sasiouglu2010} that demonstrates the robustness and accuracy of our	 method.

Finally, we mention that the computations of KS susceptibility can be massively parallelized and we resorted to the Message Passing Interface (MPI) \cite{Graham2005,Graham2006,Hursey2007}. The calculations of non-enhanced susceptibility for every frequency and wave vector are essentially independent from each other and can be performed on separate processors. No inter-processor communication is involved in this mode and the effectiveness of parallelization increases linearly with the number of processors even for their number as big as several thousands. The most time consuming calculations presented in this paper required several days of computational time on a modern 64 processor machine.

\section{Numerical results and discussion}
\label{sec:Results}

The standard picture of magnons in itinerant ferromagnets is based on the random phase approximation treatment of the uniform electron gas.  \cite{Moriya1985} In this model the only spin-wave branch evolves from the zero energy Goldstone mode. At $q = 0$, the whole spectral power of the Stoner continuum is transferred to the magnon pole, cf.\ Sec.\ \ref{subsec:CollectiveModes}. As the momentum increases the energy of the magnon rises ($\sim q^2$ for small momenta) and its life-time remains infinite until the magnon band makes a contact with the Stoner continuum, where the spin-wave abruptly looses its resonant character and cannot be considered as a well defined excitation. The strength of the latter effect, called ``spin-wave disappearance'' \cite{Mook1973}, follows from the properties of the uniform electron gas. For the frequencies corresponding to Stoner transitions the imaginary part of the KS susceptibility is practically everywhere comparable in magnitude to its real part.

Real materials feature much richer spin dynamics. There are at least two important reasons for this qualitative difference with the uniform electron gas. The first is the formation of atomic magnetic moments resulting from highly non-uniform spatial distribution of charge and spin densities. Although the atomic magnetic moments are formed by the itinerant electrons, the strong intraatomic exchange interaction keeps them well defined even at elevated temperatures. This feature is the reason of the usefulness of the Heisenberg model of interacting atomic spins in the discussion of the physics of itinerant-electron magnets. The strongly non-uniform distribution of magnetization on the atomic length-scale leads to two further effects not present in the uniform electron gas. First, multiple spin-wave branches form for momenta from the first Brillouin zone of the system. \cite{Buczek2009,Buczek2010d} Second, not the entire spectral weight of the Stoner continuum is shifted to the Goldstone mode for $q = 0$; part of it resides in the optical spin-wave modes and the residual Stoner continuum.

The second reason is the complexity of Stoner continuum reflecting the rich electronic structure of real materials. The latter is always of a multiple-band type; the bands differ strongly from each other in the character of the hybridization of the atomic $s$, $p$ and $d$ states of the same and different atoms. It is common to characterize the electronic structure of a ferromagnetic system by the exchange splitting $\Delta=Im$ where $I$ is the so-called Stoner parameter of the material and $m$ is the magnetic moment per atom. It is, however, very important that $\Delta$ characterizes the energy splitting only between the $3d$ states of similar spatial form. On the other hand, the Stoner continuum contains the spin-flip transitions between all available pairs of electron states with opposite spin projections. For a given energy and wave vector the value of the spectral density of the Stoner transitions is determined by the number of the states available for the transitions weighted with the respective matrix element. It depends crucially on the overlap of the wave functions of the initial and final electronic states. For the states separated by the exchange splitting $\Delta$, the overlap of the wave functions is large which leads to high spectral density of the Stoner transitions around this energy. As follows from the discussion in Sec.\ \ref{subsec:CollectiveModes}, a strong destructive influence of the Stoner transitions on the spin-wave states is expected for spin-waves forming in this frequency range. In many systems, however, the value of $\Delta$ exceeds substantially the spin-wave energies and the corresponding Stoner transitions cannot contribute to the Landau damping. The latter is governed by the low-energy part of the Stoner spectrum involving the electron transitions between states with possibly very different orbital character and therefore with a low transition probability. The low intensity of the Stoner spectral density at given $\vec{q}$ and $\omega$ leads to the weak Landau damping. As a result, the spin-waves corresponding to these $\vec{q}$ and $\omega$ remain well-defined.

Obviously, the form of the Stoner continuum, and in particular its low-energy part, is strongly system dependent reflecting details of the electronic hybridizations in the system. Not only the chemical composition but also the dimensionality of the material exercises strong influence on these properties. In ultrathin films, the presence of the nonmagnetic substrate brings the states of the magnetic film in contact with the states of the substrate, which complicates further electron spectrum and correspondingly Stoner continuum. In Ref.\ \cite{Buczek2011} we showed that arguments based on the simplified analysis of the dimensionality aspects are not sufficient to predict the properties of the spin-wave damping in the magnetic films. The concrete details of the electronic structure are essential. Depending on these details, the interpretation of the features of the spectral density of spin-flip excitations varies from the infinitely living Heisenberg-type spin-wave modes, through moderately damped well defined magnons up to the spin-wave disappearance effect.

Below we report a systematic comparison of the spin-wave damping in the bulk and film forms of Fe, Co and Ni. We show that the trends in the dimensionality dependence of the spin-wave properties are different for different elements and interpret them on the basis of the analysis of the underlying electronic structure and spin-flip spectrum. We primarily analyze spin-wave spectra based on the eigensystem of the loss matrices. The energy $\omega_{0}$ of the magnon is identified with the energy position of the spin-wave peak maximum. The full width at the half maximum (FWHM) of the peak gives the inverse life-time of the excitation. We present also the cases where the energy dependence of the eigenvalues cannot be described by a well defined peak. We briefly assess the applicability of the adiabatic methods to estimate frequencies of short wave-length magnons.

\subsection{Fe}
\label{subsec:Fe}

\subsubsection{bulk bcc Fe}

\begin{figure}
  \centering
  \includegraphics[width=0.45\textwidth]{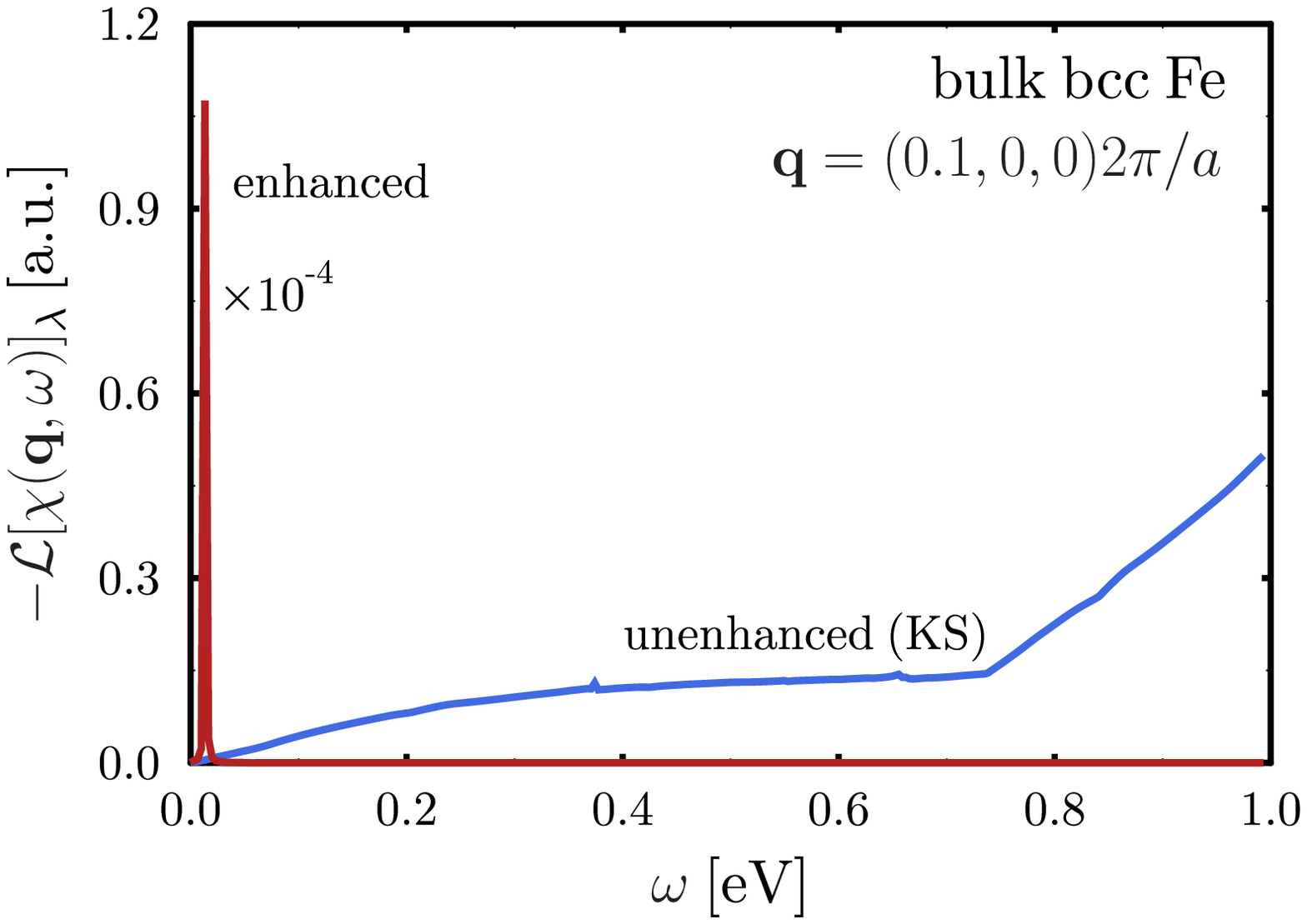} % 1a
  \includegraphics[width=0.45\textwidth]{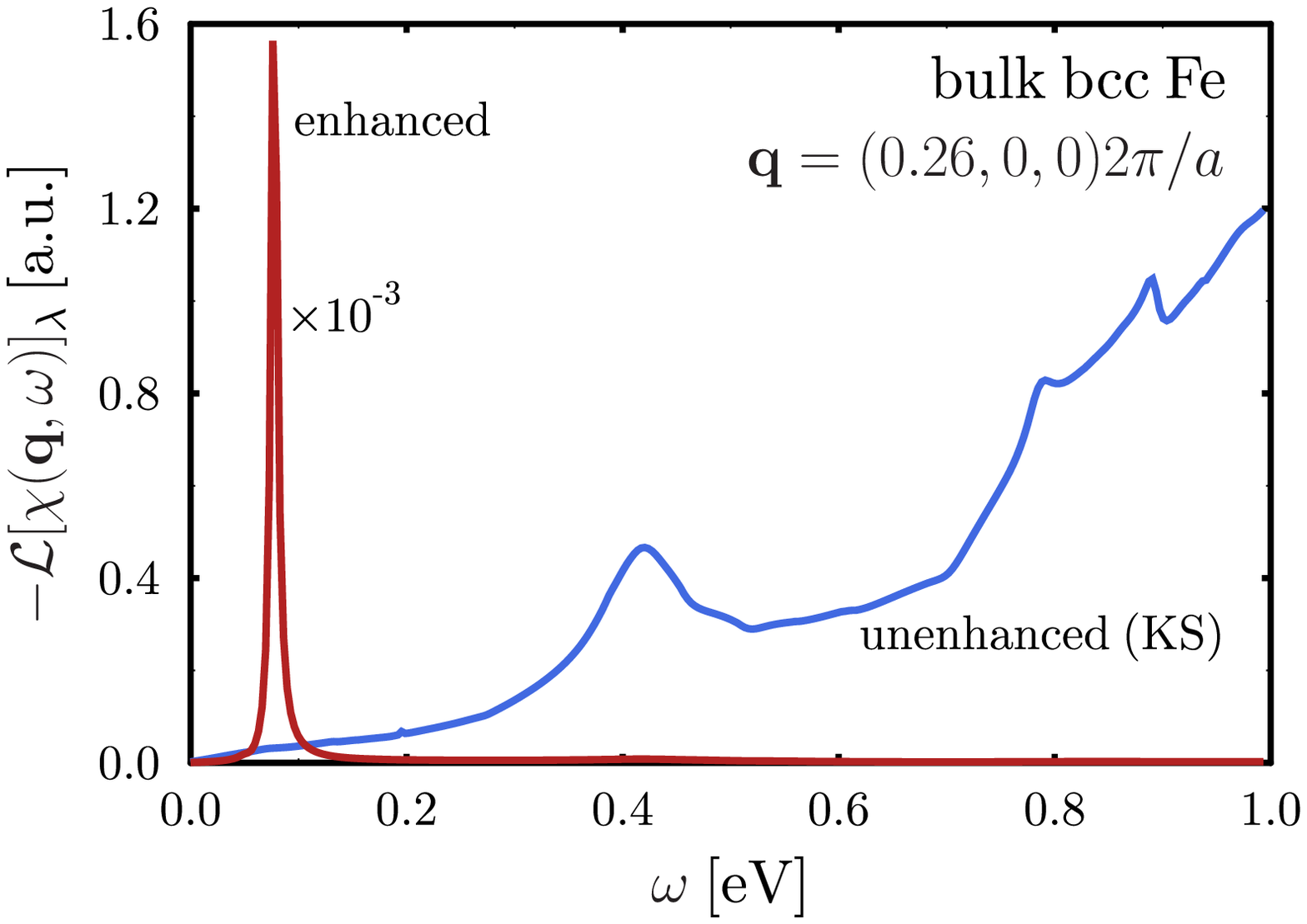} % 1b
  \includegraphics[width=0.45\textwidth]{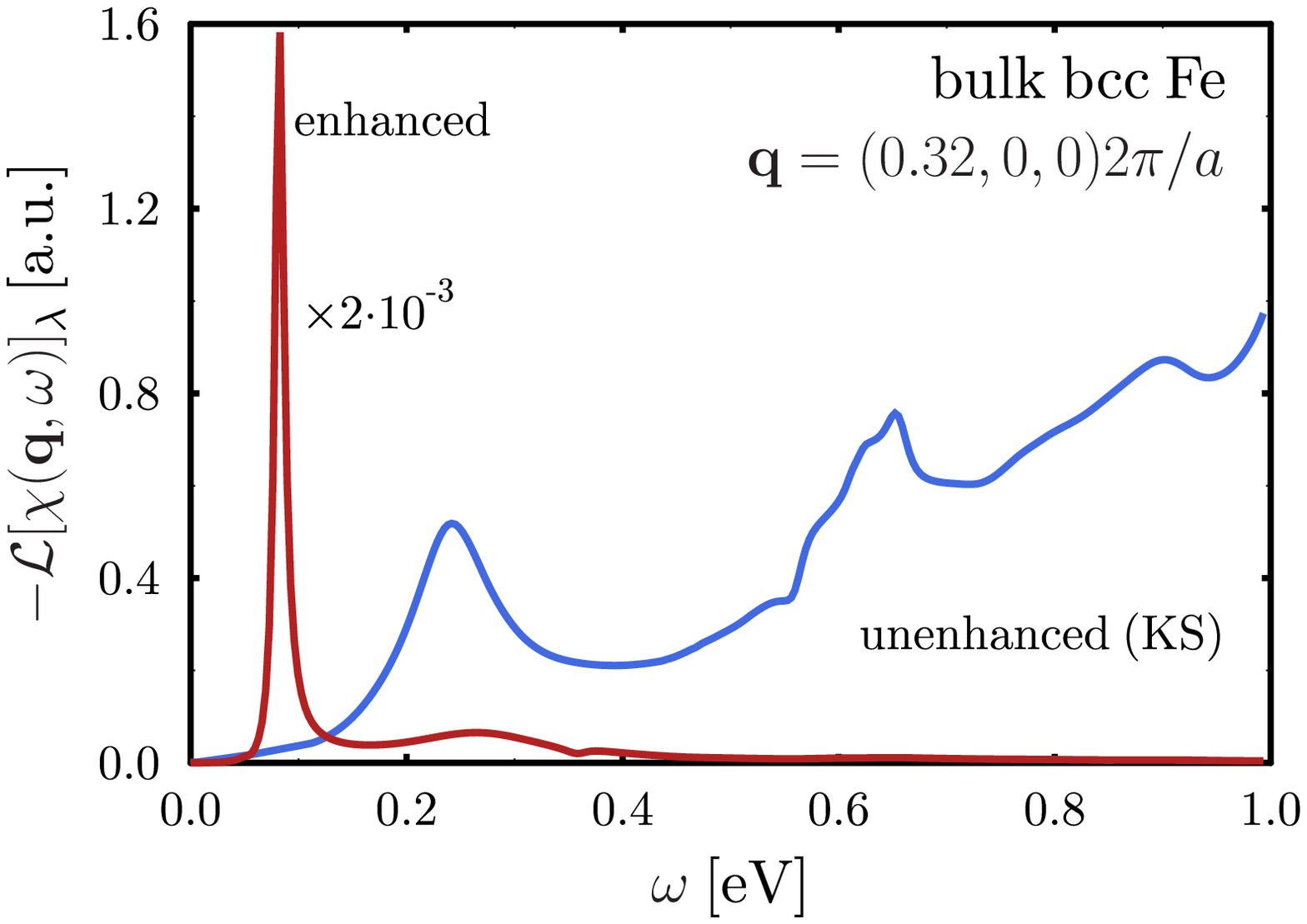} % 1c
  \includegraphics[width=0.45\textwidth]{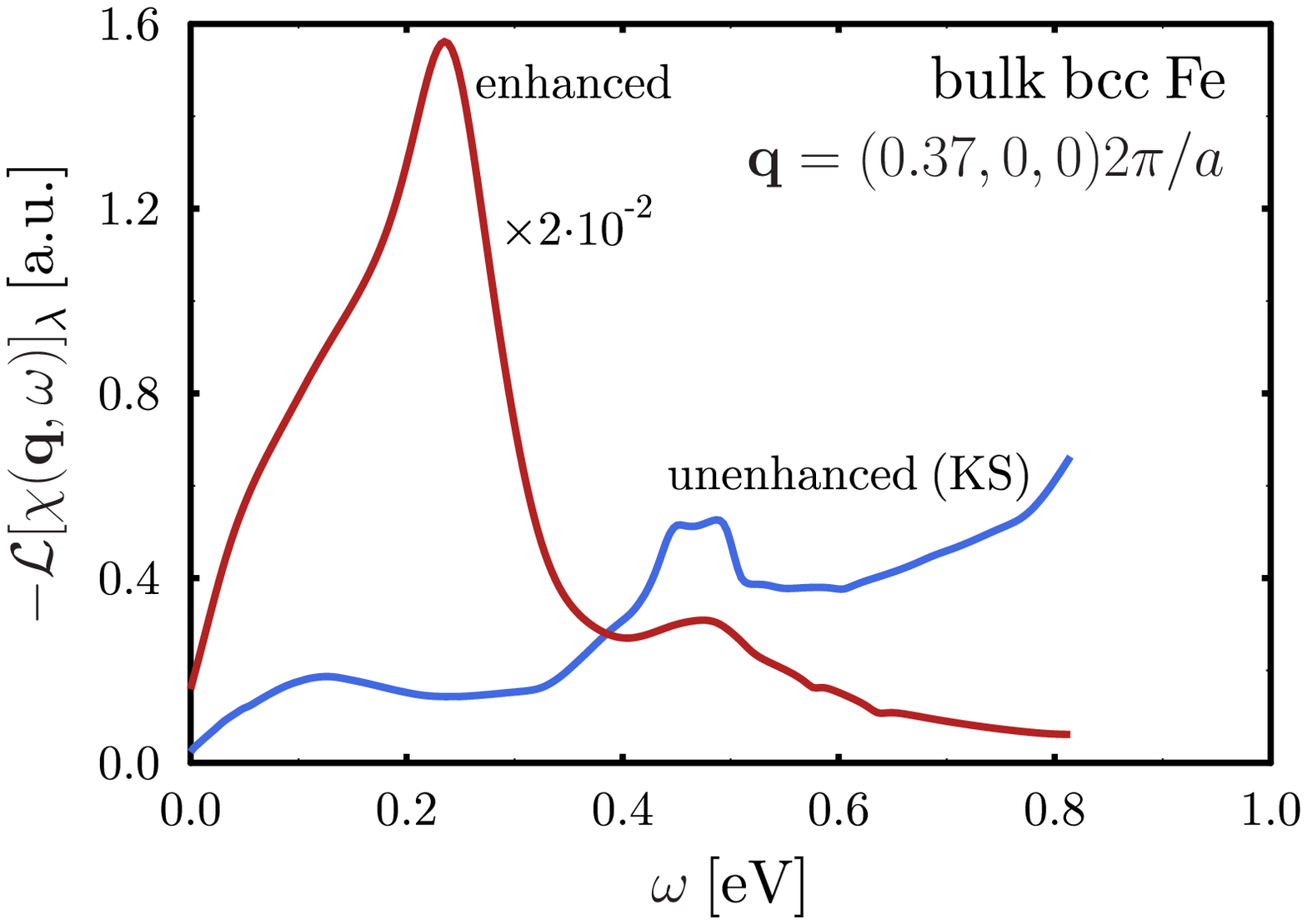} % 1d
  \includegraphics[width=0.45\textwidth]{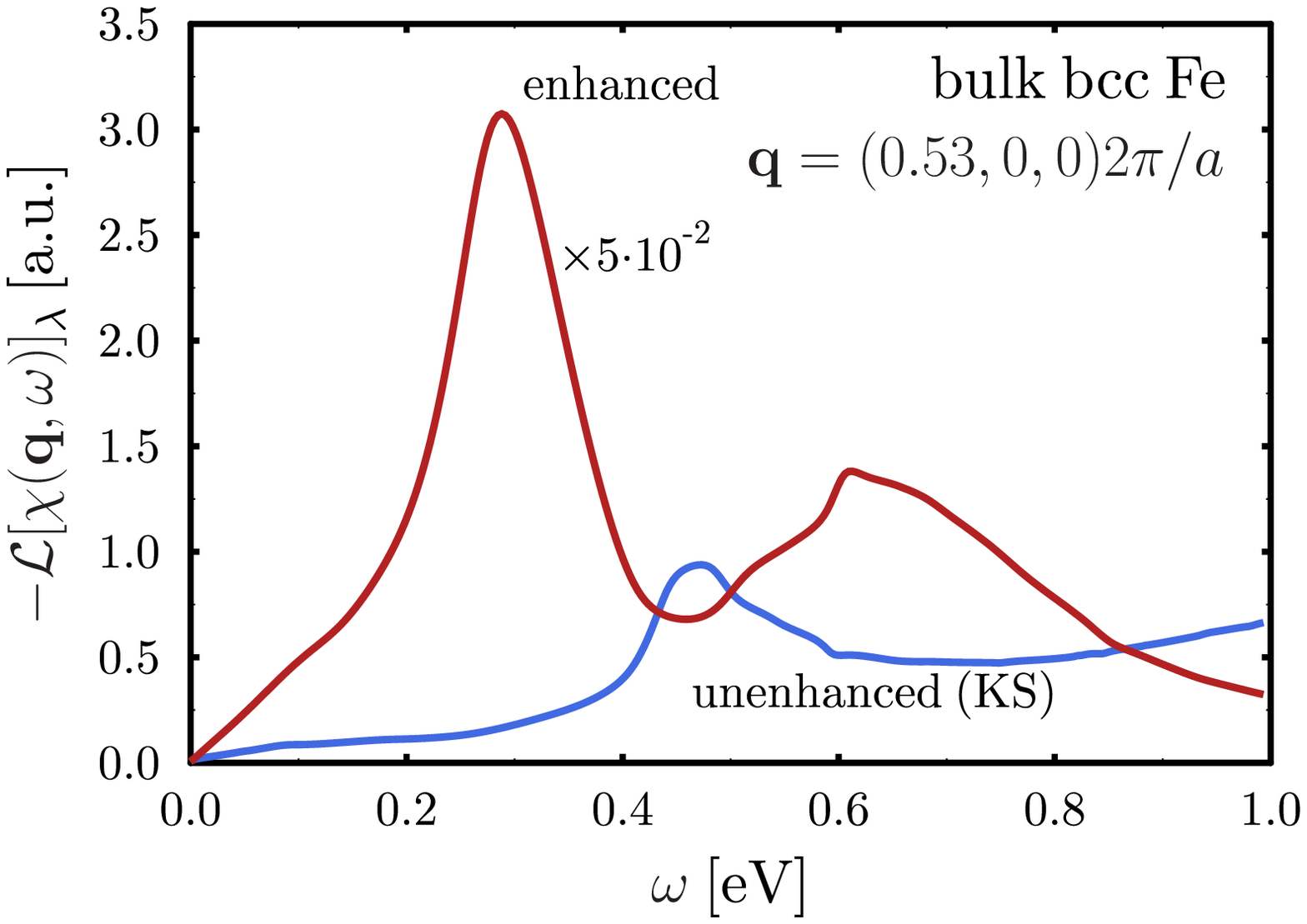} % 1e
  \caption{Examples of spin-flip spectra in bcc iron, atomic units, for different momenta along (100) direction. The largest eigenvalues of respectively enhanced and KS loss matrix are shown. Low energy spin-wave peaks have simple Lorentzian shapes small line-widths and carry substantial spectral power. Above the critical energy of \unit{82}{\milli\electronvolt} corresponding to $q_{c}=0.35\frac{2\pi}{a}$ wave-vector the spin-wave band enters abruptly a region of dense Stoner continuum. The spin-wave peaks become broad, acquire irregular non-Lorentzian shapes reflecting the energy dependence of the density of Stoner continuum, and carry much smaller spectral weight.}
  \label{fig:FeExample}
\end{figure}

\begin{figure}
  \centering
  \includegraphics[width=0.45\textwidth]{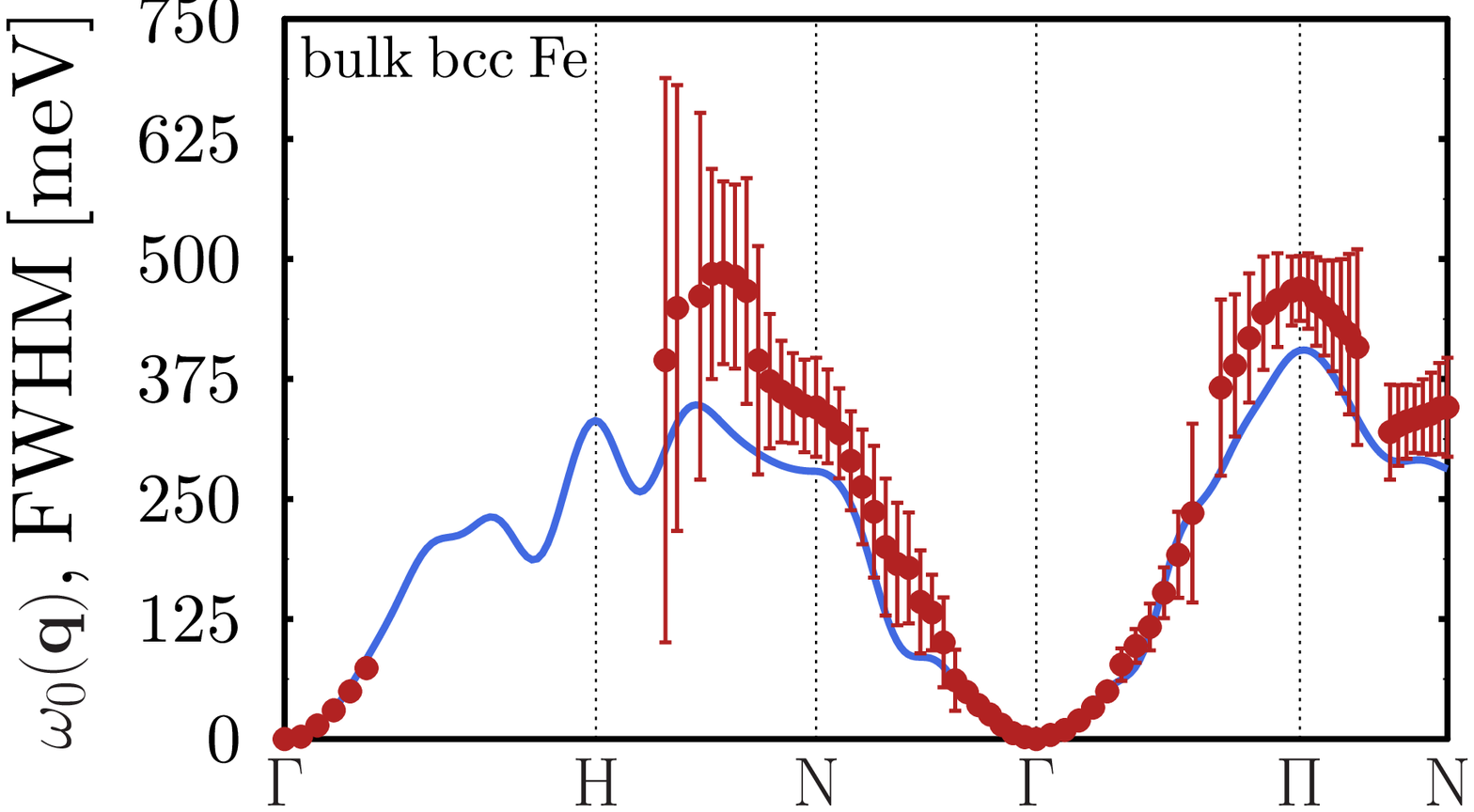} % 2
  \caption{Spin-waves of bcc iron obtained using LRTDDFT implementation described in this paper. Solid circles ({\Large\textbullet}) correspond to the maximum of spin-wave peak, while the error bars denote full width at half maximum (FWHM) of the peak. Solid line denotes spin-wave energies obtained from magnetic force theorem (MFT). Strong increase of Landau damping in all directions in the Brillouin zone (spin-wave disappearance effect) is seen for spin-wave energies above \unit{82}{\milli\electronvolt}. High energy spectrum along $\Gamma\mathrm{HN}$ directions is dominated by Stoner excitations and cannot be represented by a simple Lorentzian. Representative examples of the spectral power functions are given in Figure \ref{fig:FeExample}.}
  \label{fig:FeDS}
\end{figure}

According to our calculations, the spin-wave disappearance effect is particularly pronounced in the bcc Fe, cf.\ Figs.\ \ref{fig:FeExample} and \ref{fig:FeDS}. Above the critical energy of \unit{82}{\milli\electronvolt} corresponding to $q_{c}=0.35\frac{2\pi}{a}$ wave-vector the spin-wave band enters abruptly a region of dense Stoner continuum and the intensity contained in the spin-wave peak drops by an order of magnitude. The strong loss of spin-wave intensity above around \unit{80}{\milli\electronvolt} was clearly observed in the neutron scattering experiments, cf.\ e.g.\ Ref.\ \cite{Mook1973}. The effect is rather anisotropic and sets in mainly along $\Gamma$-$\mathrm{H}$ direction, which nicely matches the experiment as well. \cite{Mook1973} Fig.\ \ref{fig:FeExample} shows clearly that the strong damping is directly related to the appearance of low energy Stoner excitations at $q_{c}$. The origin the these low energy Stoner transitions can be traced back to relatively large density of majority $d$ spin electrons at the Fermi level in this material.

Also the energies of spin-waves in the adiabatic local density approximation correlate nicely with experimental values. \cite{Pajda2001,Grotheer2001} According to our data for $q<q_{c}$ the dispersion can be well represented by the biquadratic fit
\begin{align}
  \omega_{0}\fbr{q} = Dq^{2}\fbr{1-\gamma q^2}.
  \label{eq:biq_fit}
\end{align}
with parameters $D=\unit{252}{\milli\electronvolt\angstrom^{2}}$ and $\gamma=\unit{0.28}{\angstrom^{2}}$. Experimentally reported values for $D$ vary between \unit{266}{\milli\electronvolt\angstrom^{2}} and \unit{307}{\milli\electronvolt\angstrom^{2}} \cite{Shirane1965,Shirane1968,Collins1969,Mook1973,Loong1984}

\begin{figure}
  \centering
  \includegraphics[width=0.45\textwidth]{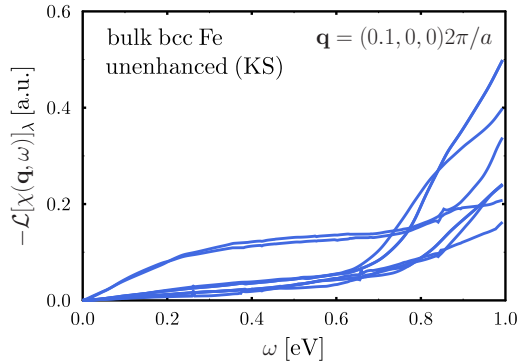} % 3
  \caption{The largest eigenvalues of loss matrix associated with unenhanced (KS) susceptibility for a selected momentum in bulk bcc Fe. There are several eigenvalues of comparable magnitude corresponding to transitions between different KS orbitals. On the contrary, the loss matrix associated with the enhanced susceptibility features only one eigenvalue, identified as a single spin-wave mode of the monatomic material.}
  \label{fig:chiq3evals}
\end{figure}

For the well defined, low energy spin-waves, the loss matrix of enhanced susceptibility features only one large eigenvalue. The associated eigenvector corresponds to the practically rigid rotation of magnetic moments. This results justifies the applicability of Heisenberg model to describe low energy magnons. For high energy strongly damped magnons, the spectrum contains only one dominating eigenvalue as well, but the corresponding eigenvectors involves small, but clear deformation of the atomic moment. On the other hand, the unenhanced susceptibility features multiple eigenvalues of comparable magnitude, cf.\ Fig.\ \ref{fig:chiq3evals}, corresponding to transitions between different KS orbitals.

\subsubsection{Fe films}

\begin{figure}
  \centering
  \includegraphics[width=0.45\textwidth]{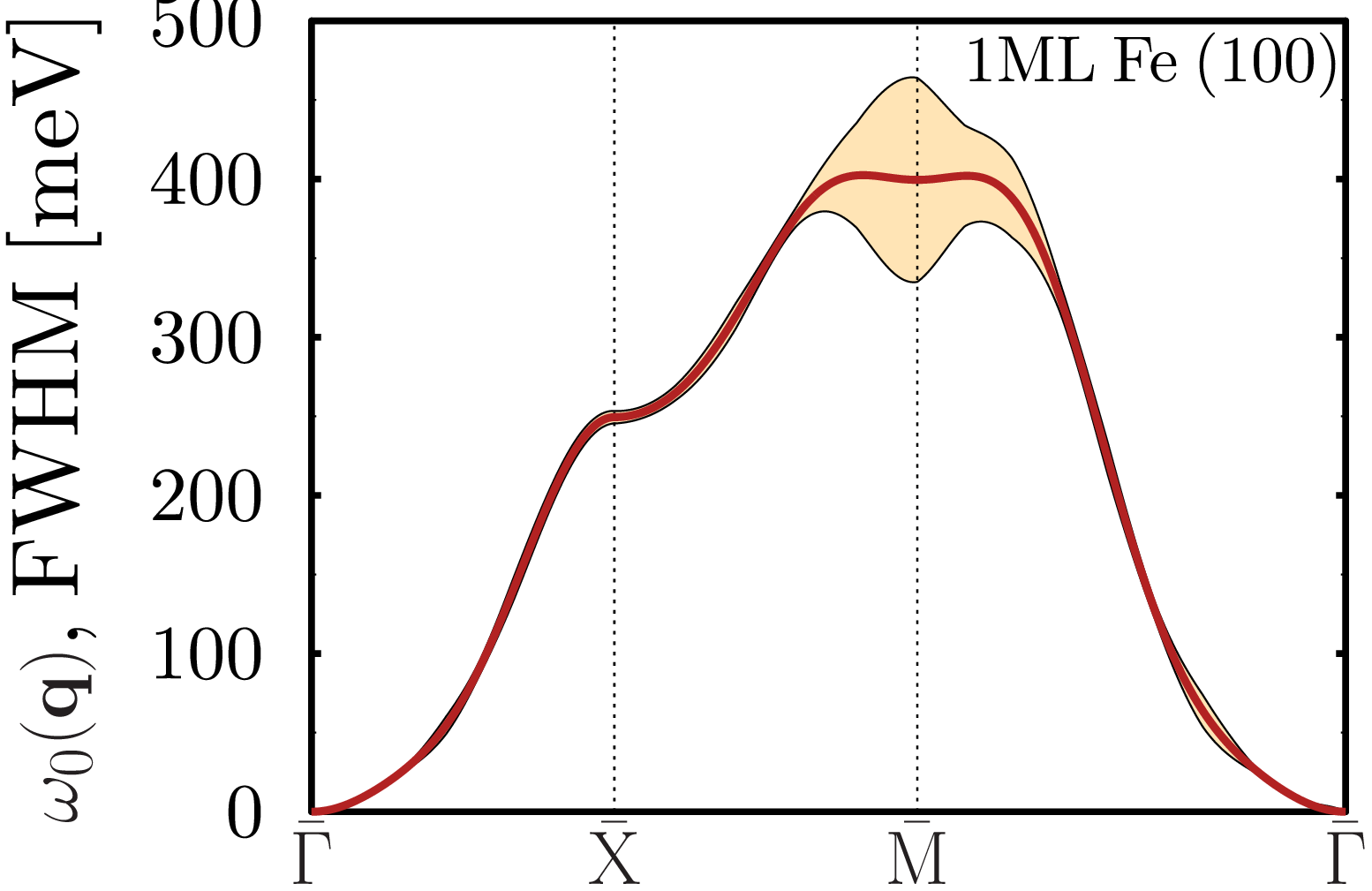}    % 4a
  \includegraphics[width=0.45\textwidth]{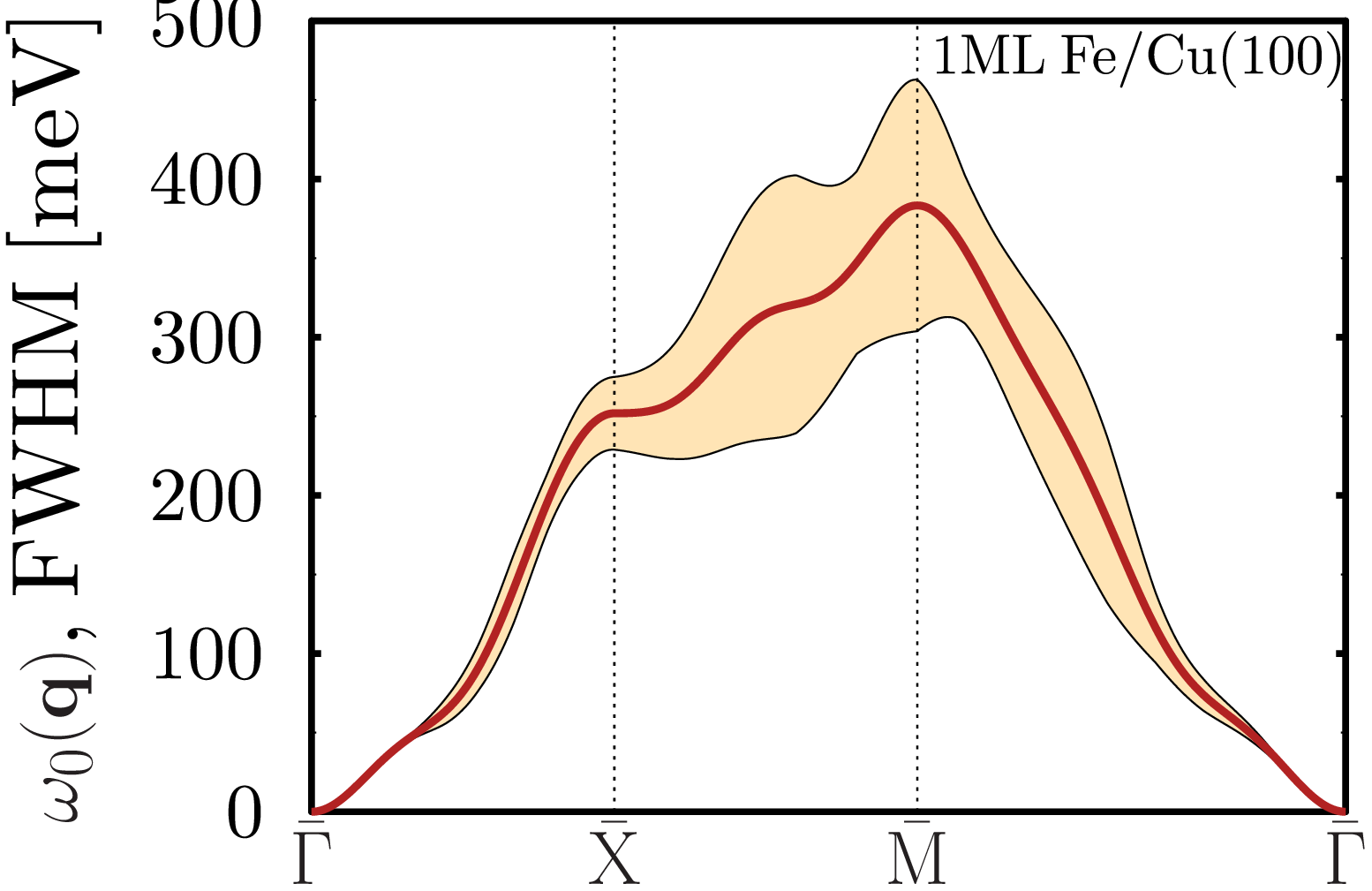} % 4b
  \caption{Spin-waves of Fe(100) monolayer free-standing and supported on Cu(100) surface. Solid lines correspond to $\omega_{0}\fbr{\vec{q}}$, while the width of the shaded region to FWHM. The spin dynamics of the free standing and supported monolayers differ weakly. From \cite{Buczek2011}.}
  \label{fig:FeMonolayer}
\end{figure}

\begin{figure}
  \centering
  \includegraphics[width=0.45\textwidth]{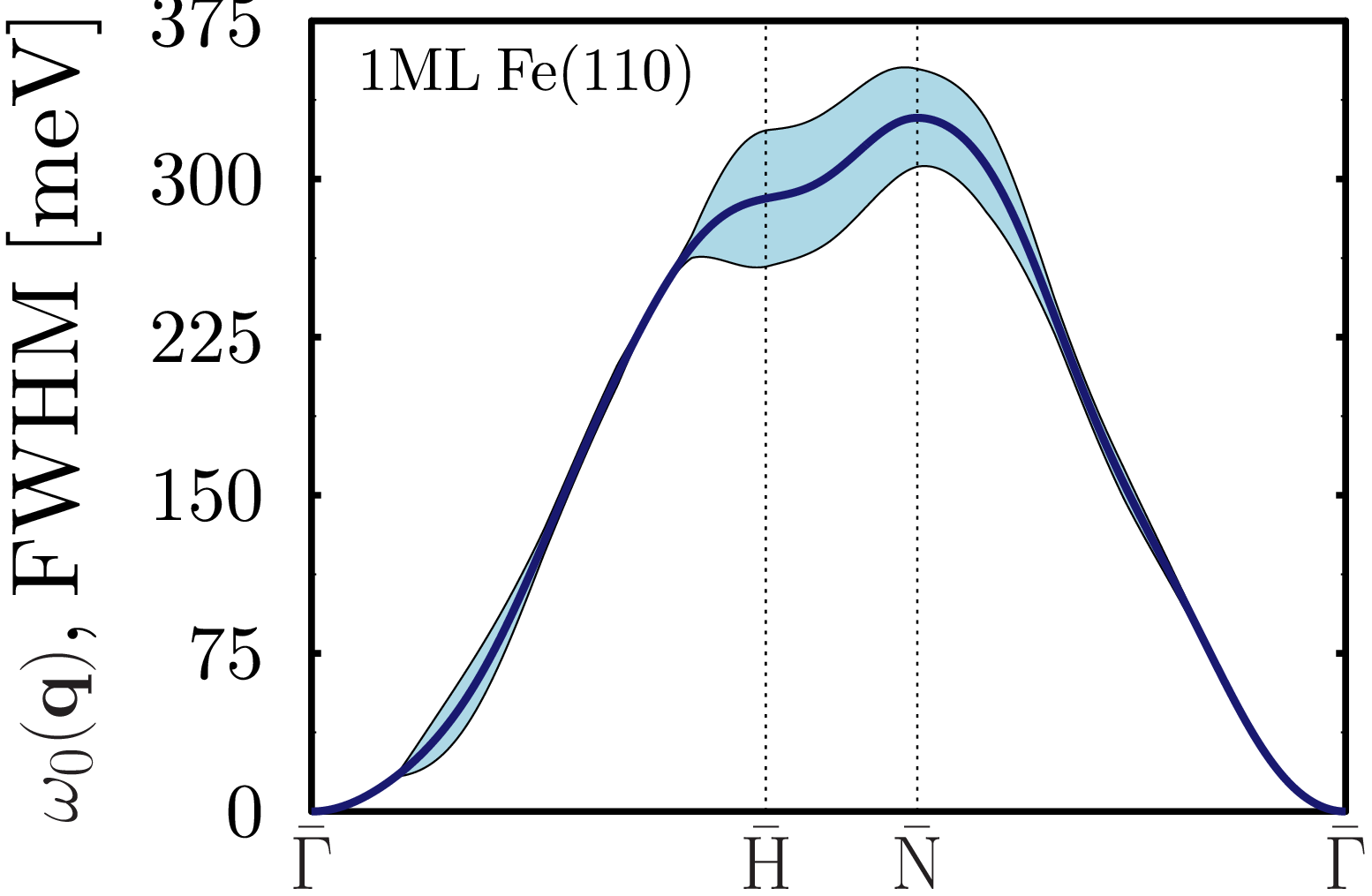}   % 5a
  \includegraphics[width=0.45\textwidth]{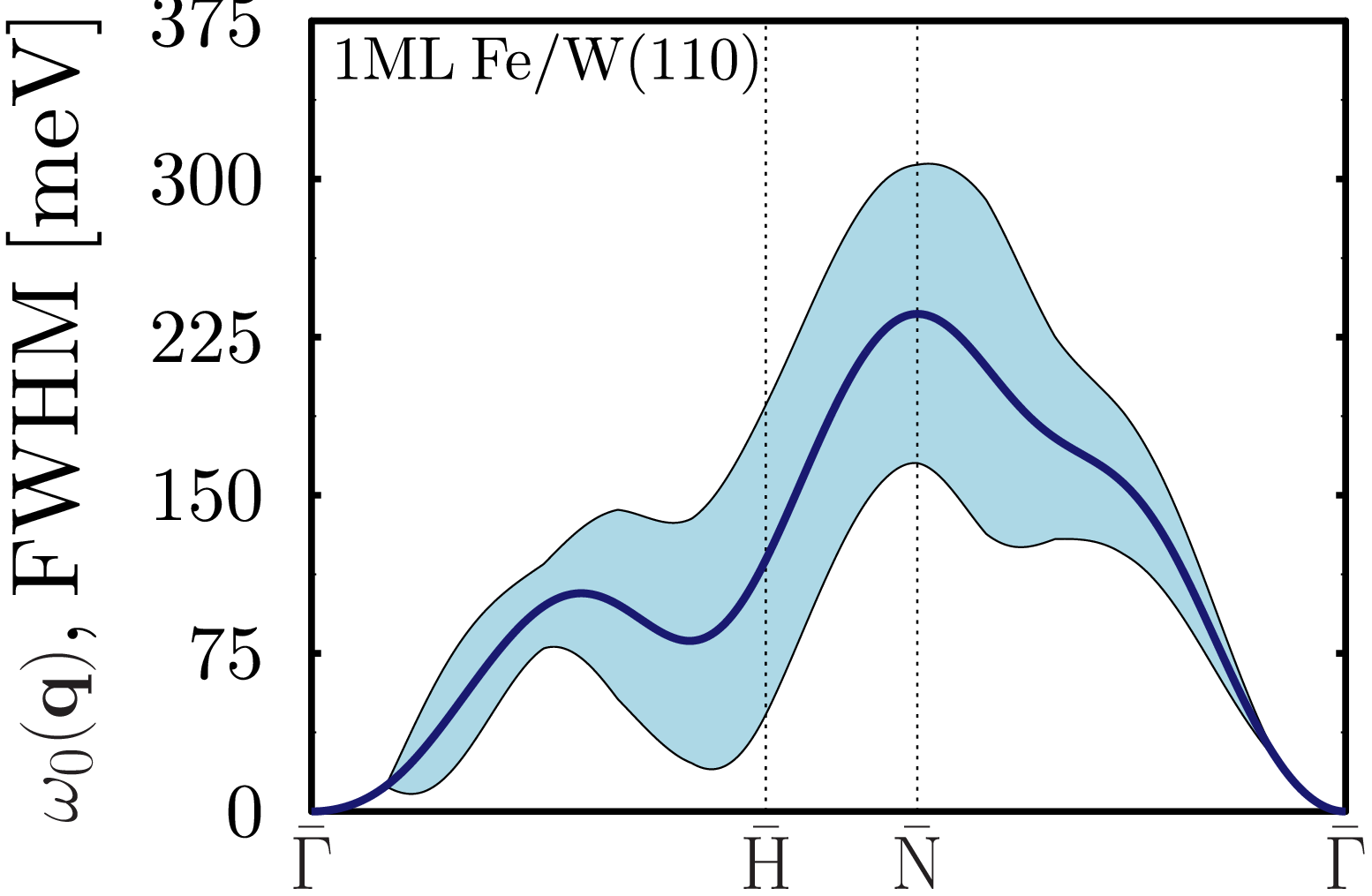} % 5b
  \caption{Spin-waves of Fe(110) monolayer free-standing and supported on W(110) surface. Solid lines correspond to $\omega_{0}\fbr{\vec{q}}$, while the width of the shaded region to FWHM. The surface state of W(110) leads to a qualitative change in the spin dynamics of the absorbed film. From \cite{Buczek2011}.}
  \label{fig:FeMonolayer110}
\end{figure}

Contrary to the bcc bulk case, the free standing monolayers of Fe, both of (100) and (110) crystallographic orientations, feature well defined spin-waves for all momenta \cite{Tang1998,Muniz2002,Costa2003,Muniz2003,Buczek2010,Buczek2011}, see also Figs.\ \ref{fig:FeMonolayer} and \ref{fig:FeMonolayer110}. This property can be traced back to the enhanced exchange splitting of $d$-symmetry orbitals and the band narrowing in the free film. It effectively removes the majority-spin $d$ states from the Fermi level \cite{Baldacchini2003,Achilli2007}, resulting in the small density of low-energy Stoner excitations. The same behavior is observed in unsupported Ni and Co (100) monolayers, which will be discussed later.

When the monolayers of Fe are deposited on a substrates the spin-dynamics is modified. We demonstrated recently \cite{Buczek2011} that the impact of the substrate varies strongly depending on the substrate and its orientation. Explicit LRTDDFT calculations for Fe/Cu(100) system have been performed by us for one \cite{Buczek2011} and three monolayers \cite{Schmidt2010a} coverages. The damping increases somewhat compared to the free standing case, but the spin-waves are well defined in the whole two-dimensional Brillouin zone, cf.\ Fig.\ \ref{fig:FeMonolayer}. The substrate-induced renormalization of magnon energies turns out to be small. In the case of Fe/Cu(100) the states of the substrate hybridize rather weakly with the states of the magnetic overlayer. The films feature small density of majority-spin states at the Fermi level and the spin dynamics of the films resembles qualitatively the one of free standing layers.

\begin{figure}[htbp]
  \centering
  \includegraphics[width=0.49\textwidth]{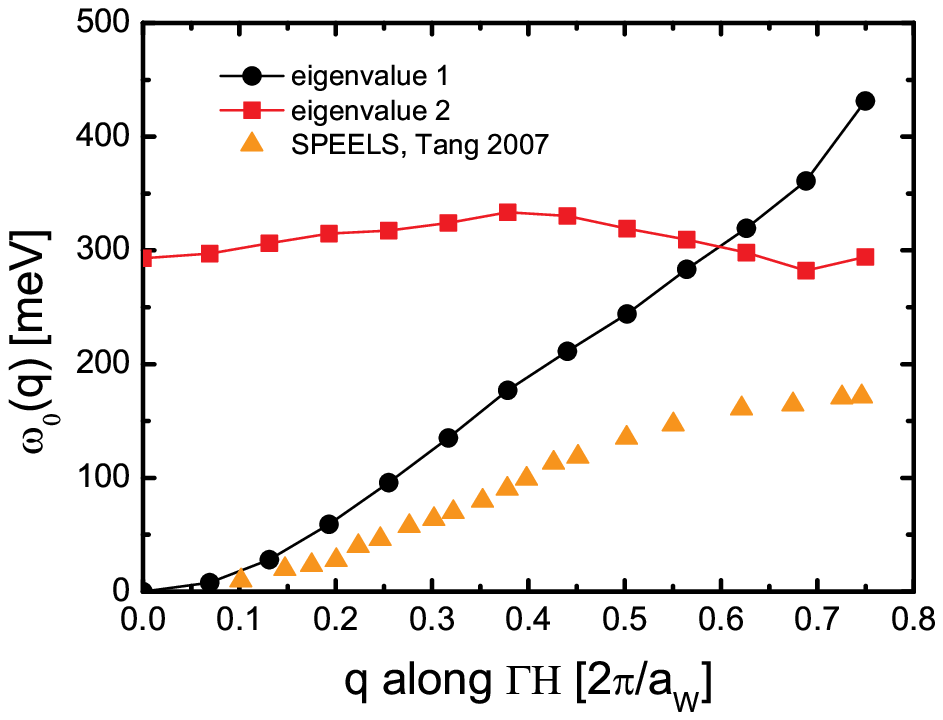} % 6a
  \includegraphics[width=0.49\textwidth]{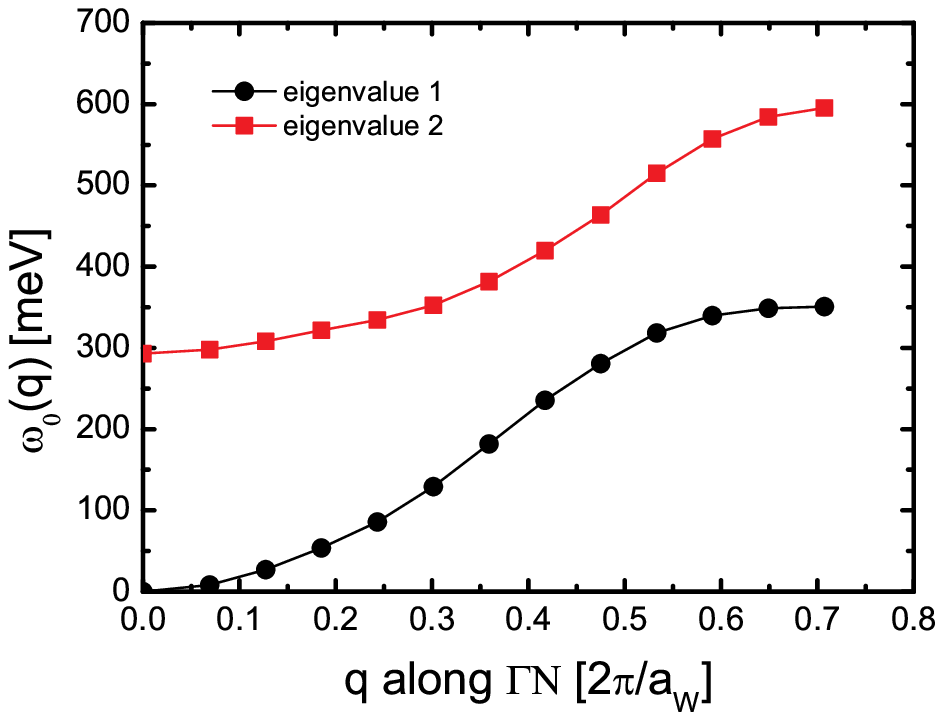} % 6b
  \includegraphics[width=0.49\textwidth]{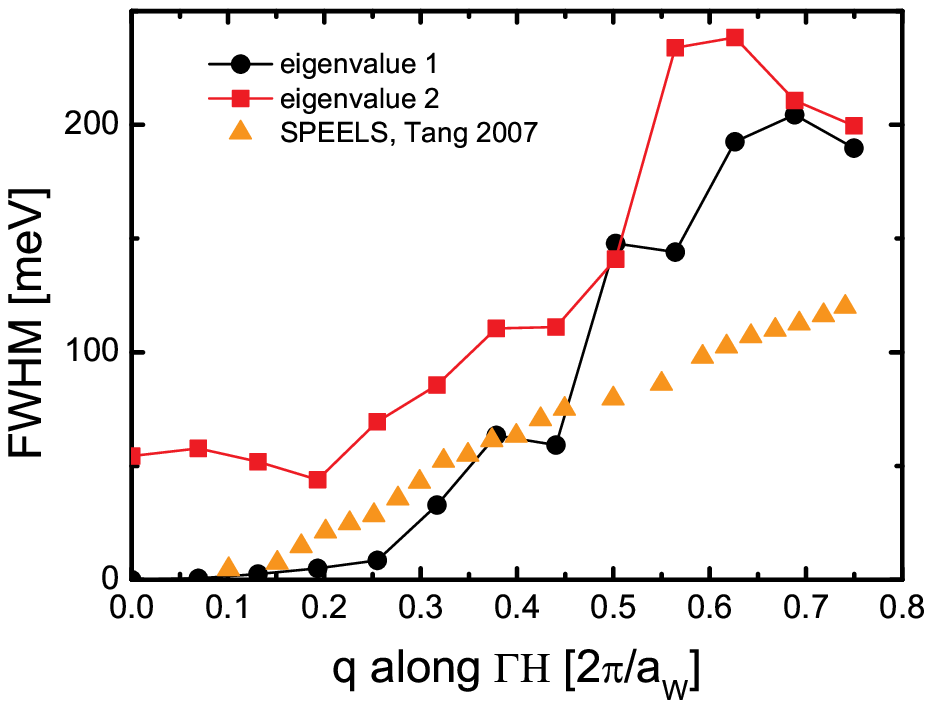}  % 6c
  \includegraphics[width=0.49\textwidth]{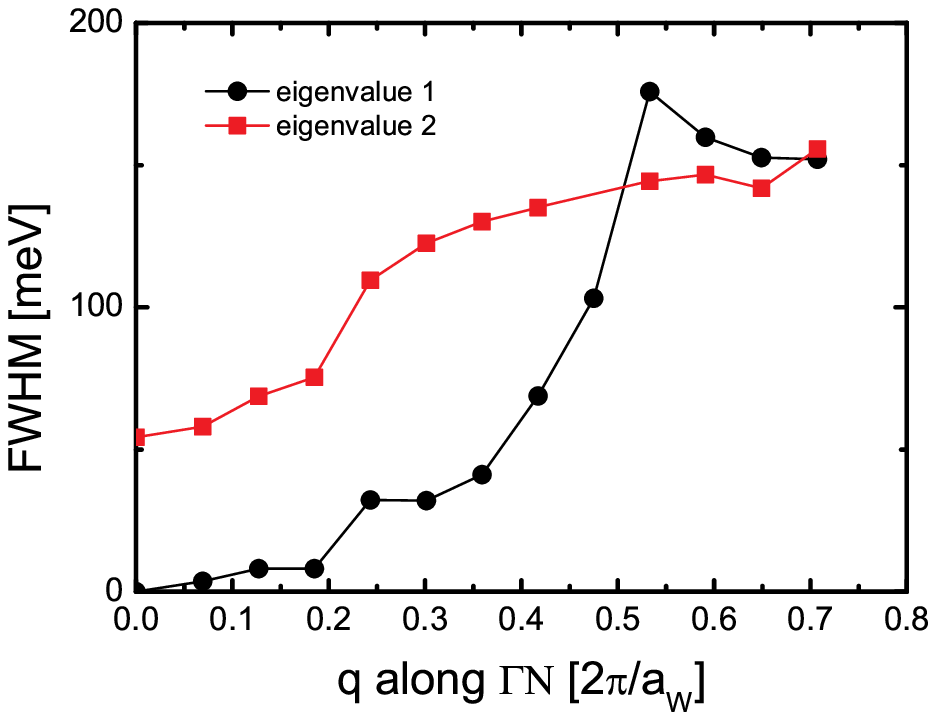}  % 6d
  \caption{Spin-waves of \ml{2} Fe/W(110). Two magnon modes can be discerned in the whole two dimensional Brillouin zone. Experimental points come from Ref.~\cite{Tang2007}.}
  \label{fig:2ML_FeW110}
\end{figure}

The situation is dramatically different in Fe/W(110), as evident from Fig.\ \ref{fig:FeMonolayer110}. The substrate renormalizes magnon energies and strongly enhances the damping. Here, an important role is played by the interface electronic complexes, formed by hybridized surface states of W(110) and electron states of the film \cite{Buczek2011}. The complexes provide an efficient source of Stoner pairs in the region of magnetic overlayer. The strongly damped collective precession has been observed experimentally at the zone boundary in $\unit{1}{ML}$ Fe/W(110) system \cite{Prokop2009}. We emphasize, however, that even for Fe/W(110) the region in the Brillouin zone featuring the spin-wave disappearance effect is small compared to the bulk case. Experiment \cite{Tang2007} shows also that the damping-to-energy ratio in $\unit{2}{ML}$ Fe/W(110) film decreases compared to the single monolayer case. The trend correlates well with our calculations, cf.\ Fig.\ \ref{fig:2ML_FeW110}. We remark that the system is characterized by quite complex structure and in our calculations we took into account the atomic relaxations \cite{Qian1999,Sander1999a,Meyerheim2001}.

In $\unit{2}{ML}$ Fe/W(110) system there are two types of non-equivalent Fe atoms and the loss matrix associated with the enhanced susceptibility features two large eigenvalues corresponding to the acoustic and optical spin-wave modes. The spectra are qualitatively very similar to those of hcp Co presented in Sec.\ \ref{subsubsec:hcpCo}.

The observed magnon energies are roughly $40\%$ smaller than predicted by our theory. The reason of the discrepancy is still not clear, especially in the light of good performance of ALSDA in the bulk bcc iron \cite{Zhang2010}. It is worth to remark here that the LRTDDFT performs much better compared to model Hamiltonians, in which the spin-wave energies of $\unit{1}{ML}$ Fe/W(110) are grossly overestimated \cite{Muniz2002,Costa2008,Prokop2009}. Furthermore, the experiment did not reveal any presence of an optical branch in the $\unit{2}{ML}$ case. It has been conjectured \cite{Muniz2008} that SPEELS could probe only the modes with significant amplitude in the top layer, because of a limited penetration depth of electrons. However, our calculations do not predict the formation of modes localized at the surface of the film. Therefore, the excitation of both types of spin-waves should be expected.

For large wave vectors the both spin-wave peaks in the spectral density are substantially broadened and might appear as a single feature in the spectrum, especially if the finite SPEELS resolution is taken into account. On the other hand, in the center of the zone the optical mode should be discernible. Possibly, the optical mode, being substantially broadened even for $q = 0$, is lost in the background of the signal dominated by the acoustic mode.

We remark that in the case of $\unit{1}{ML}$ Fe/W(110) the MFT yields the spin-wave energies \cite{Buczek2010,Zhang2010} of similar values as LRTDDFT; the same is the case for $\unit{1}{ML}$ Fe/Cu(100) studied by Pajda \el \cite{Pajda2000}. Udvardi \el \cite{Udvardi2003} considered also the relativistic corrections. In the relativistic limit the spin is coupled to the lattice and the Hamiltonian looses its spin-rotational invariance. As a consequence the lowest energy magnon does not have vanishing frequency and the Goldstone mode disappears. Udvardi \el show that in light transition metals the relativistic corrections to the spin-wave spectra are of minor importance on the energy scale of exchange interactions. Simultaneously, they predict the appearance of a series of weakly dispersive spin-wave bands associated with the dynamics of small magnetic moments induced in the Cu substrate. These additional resonances are absent in our data, also in the case of Fe/W(110), where the moment induced in the interface layer of W is sizable ($\approx\unit{0.14}{\Bm}$) and antiferromagnetically aligned with the film magnetization. As it has been already discussed \cite{Sandratskii2007} the appearance of additional spin-wave modes related to the induced moments is an artifact of adiabatic approach where small induced moments are treated as independent adiabatic degrees of freedom. The methods based on the evaluation of magnetic susceptibility are better suited for the investigation of such ``induced'' magnetization dynamics \cite{Buczek2009}.

\subsection{Co}
\label{subsec:Co}

At ambient pressure and low temperature cobalt is a ferromagnet featuring $\varepsilon$ (hcp) structure. \cite{Yoo1998} As the temperature increases a transition to $\gamma$ (fcc) structure occurs at around \unit{750}{\kelvin} and above Curie temperature of around \unit{1400}{\kelvin} the systems becomes paramagnetic. To our knowledge, all previous \textit{ab initio} studies of spin-waves in Co employ adiabatic approximation. \cite{Halilov1998a,Grotheer2001,Pajda2001}

\subsubsection{$\varepsilon$(hcp)-Co}
\label{subsubsec:hcpCo}

Experimental lattice constant of hcp cobalt $a_{\mathrm{hcp}} = \unit{4.738}{\bohr}$ was used. \cite{Taylor1950} The calculated magnetic moment per Co atom reads \unit{1.61}{\Bm}.

\begin{figure}
  \centering
  a)~\includegraphics[width=0.45\textwidth]{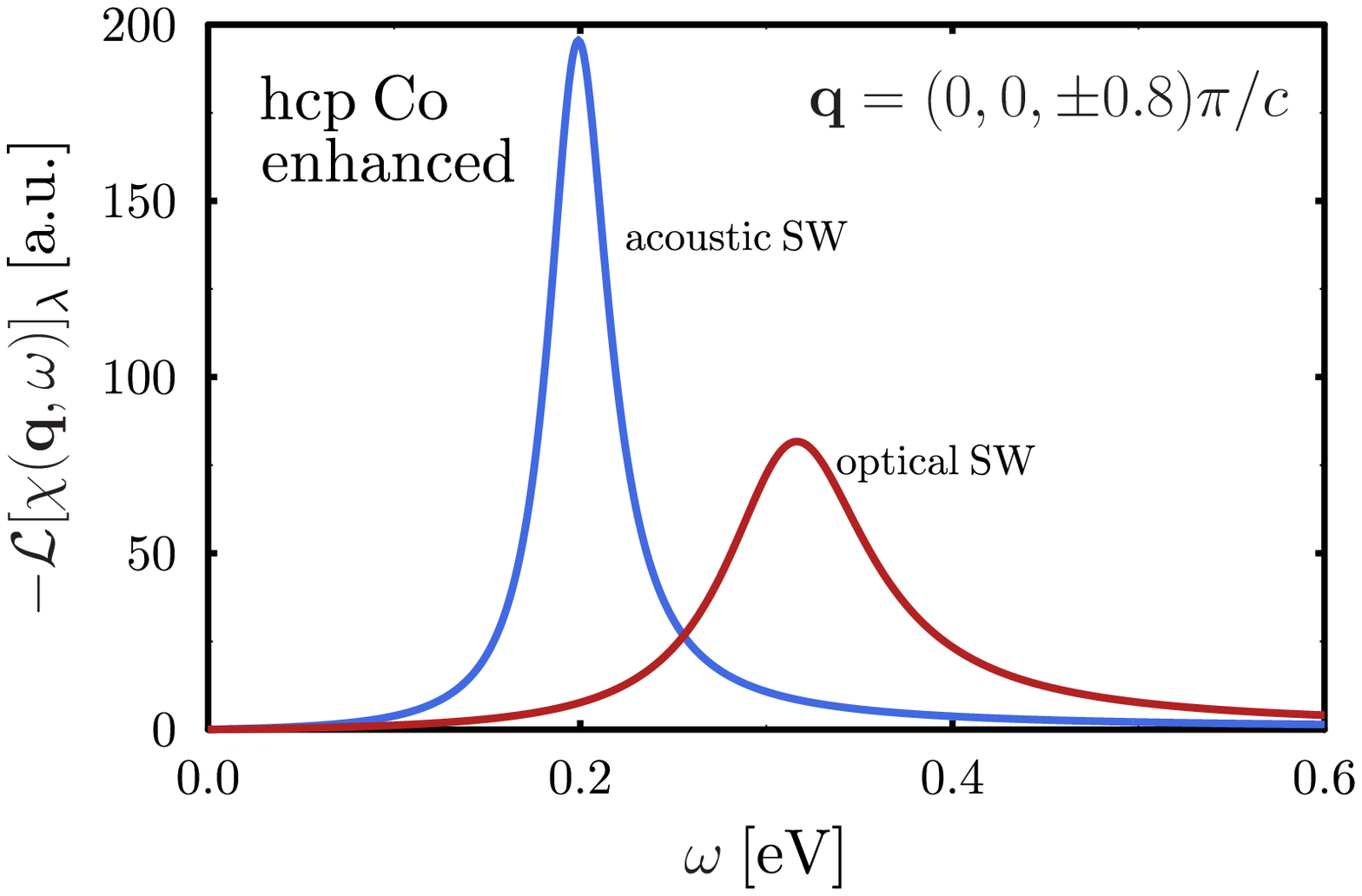}               % 7a
  b)~\includegraphics[width=0.45\textwidth]{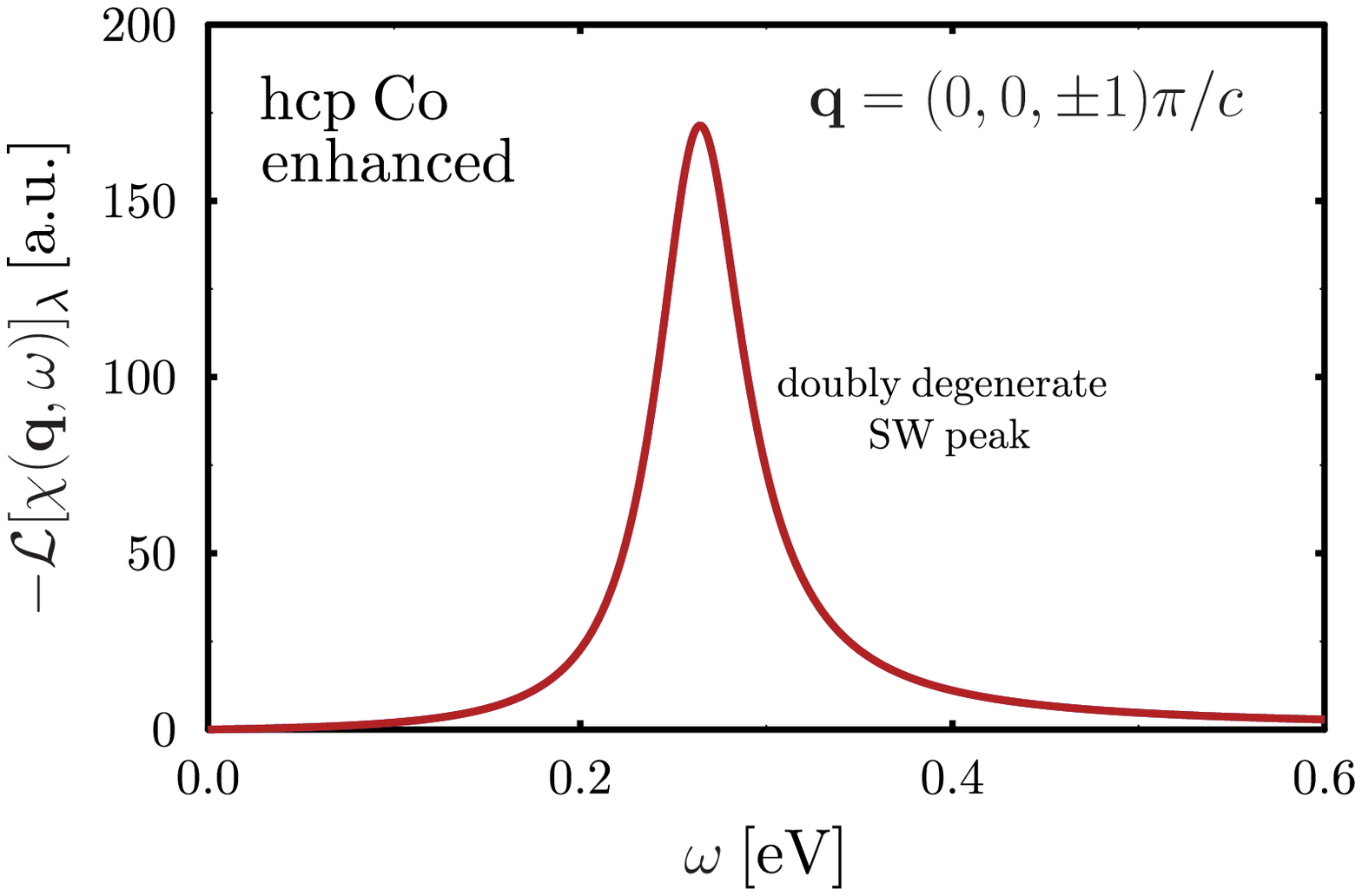}               % 7b
  c)~\includegraphics[width=0.45\textwidth]{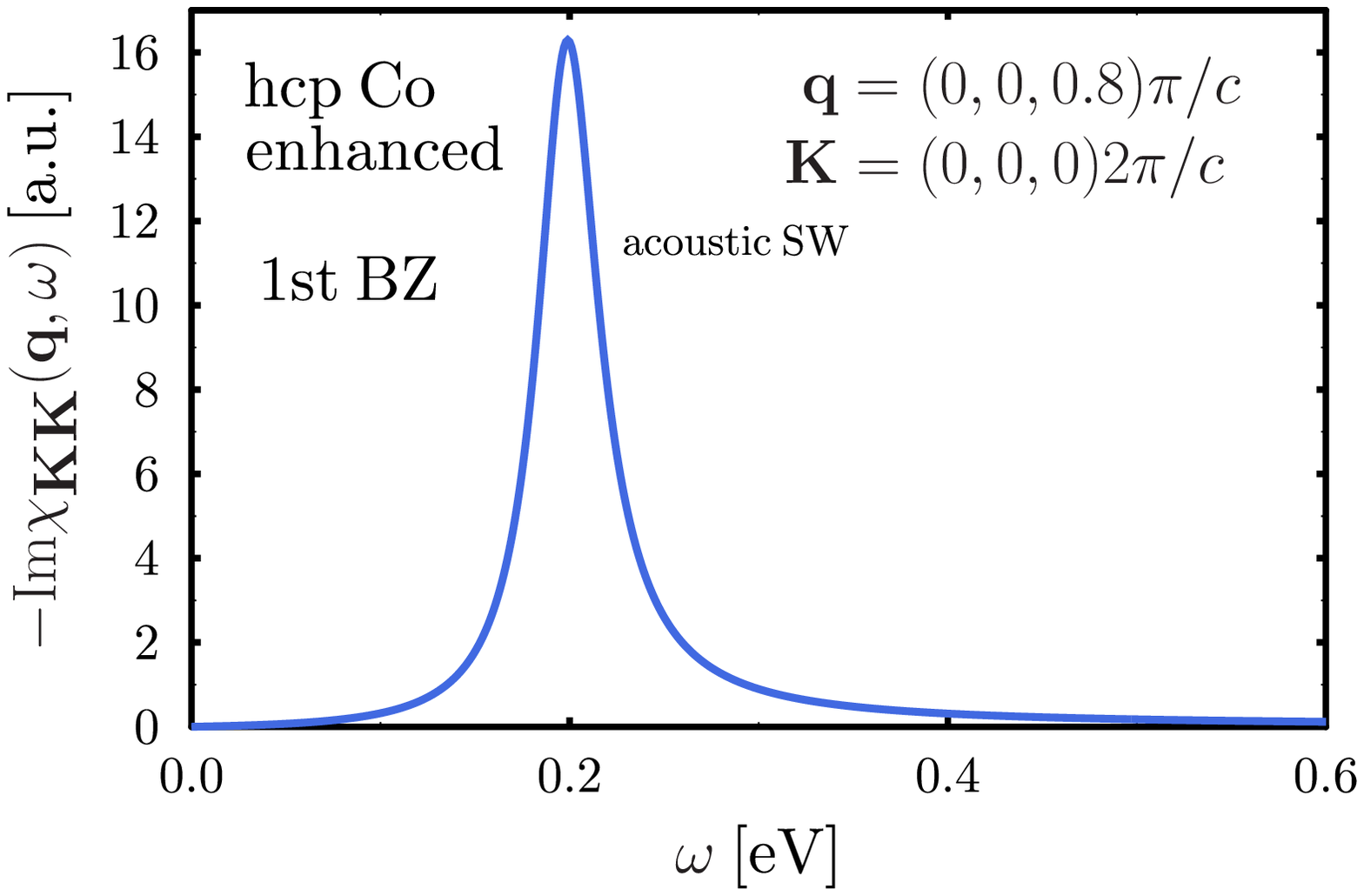}  % 7c
  d)~\includegraphics[width=0.45\textwidth]{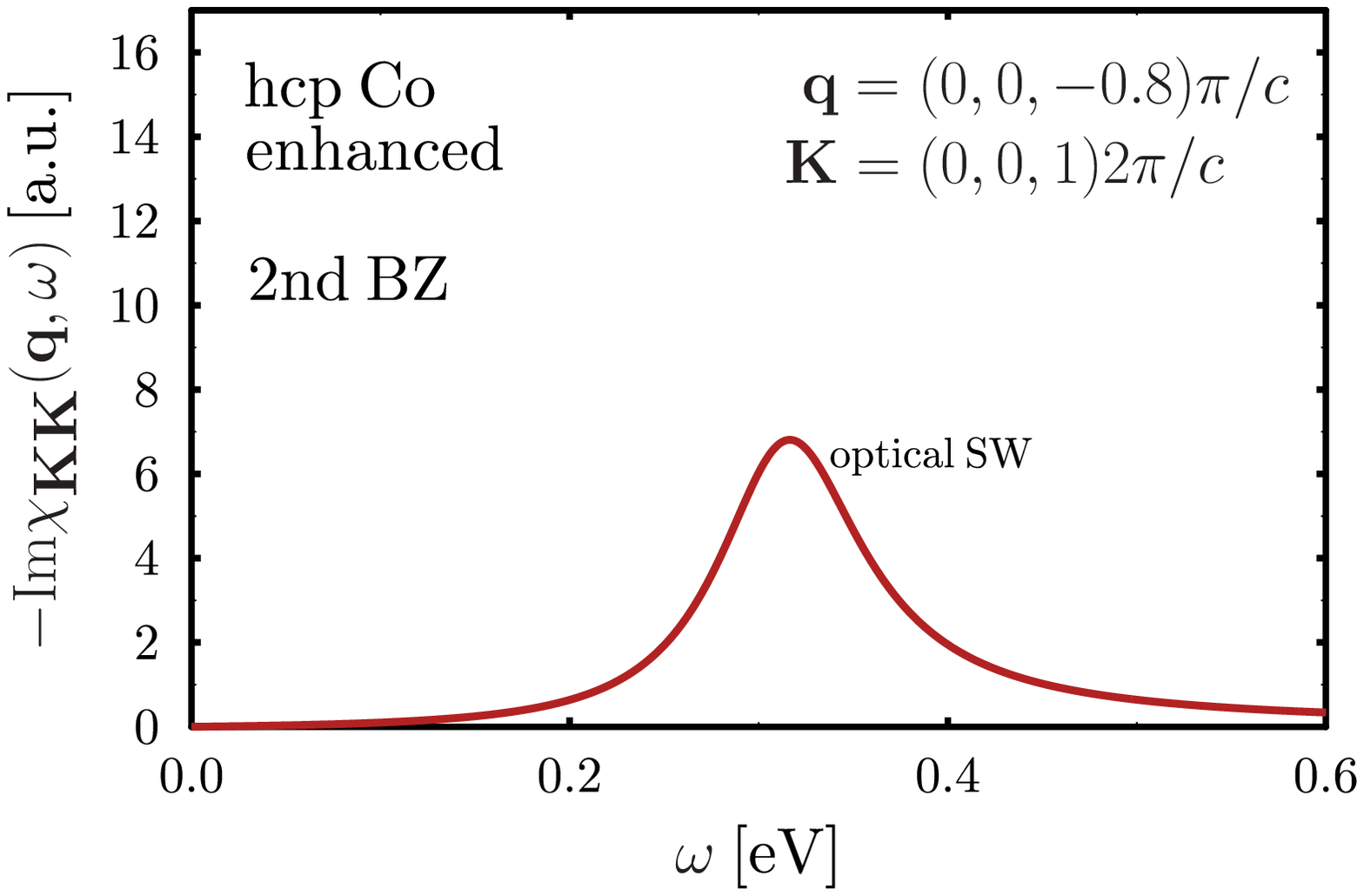} % 7d
  \caption{Examples of spin-flip excitation spectrum in hcp Co, atomic units, for two momenta along $\Gamma\mathrm{A}$ direction. Panels (a) and (b) present the two dominant eigenvalues of the loss matrix, $\Loss\chi\fbr{\vec{q},\omega}$; the eigenvalues correspond to the acoustic and optical spin-wave branches. As expected from symmetry arguments, they become degenerate for $q_{z} = \frac{\pi}{c}$. Panels (c) and (d) show the imaginary part of the Fourier transformed susceptibility, $\Im\chi_{\vec{K}\vec{K}}\fbr{\vec{q},\omega}$, for momenta inside and outside the first Brillouin zone. The momenta differ by a reciprocal lattice vector. $\Im\chi_{\vec{K}\vec{K}}\fbr{\vec{q},\omega}$ is probed in the neutron scattering experiments. By varying the momentum one can access the optical and acoustic spin wave branch.}
  \label{fig:CoExample}
\end{figure}

\begin{figure}
  \centering
  \includegraphics[width=0.45\textwidth]{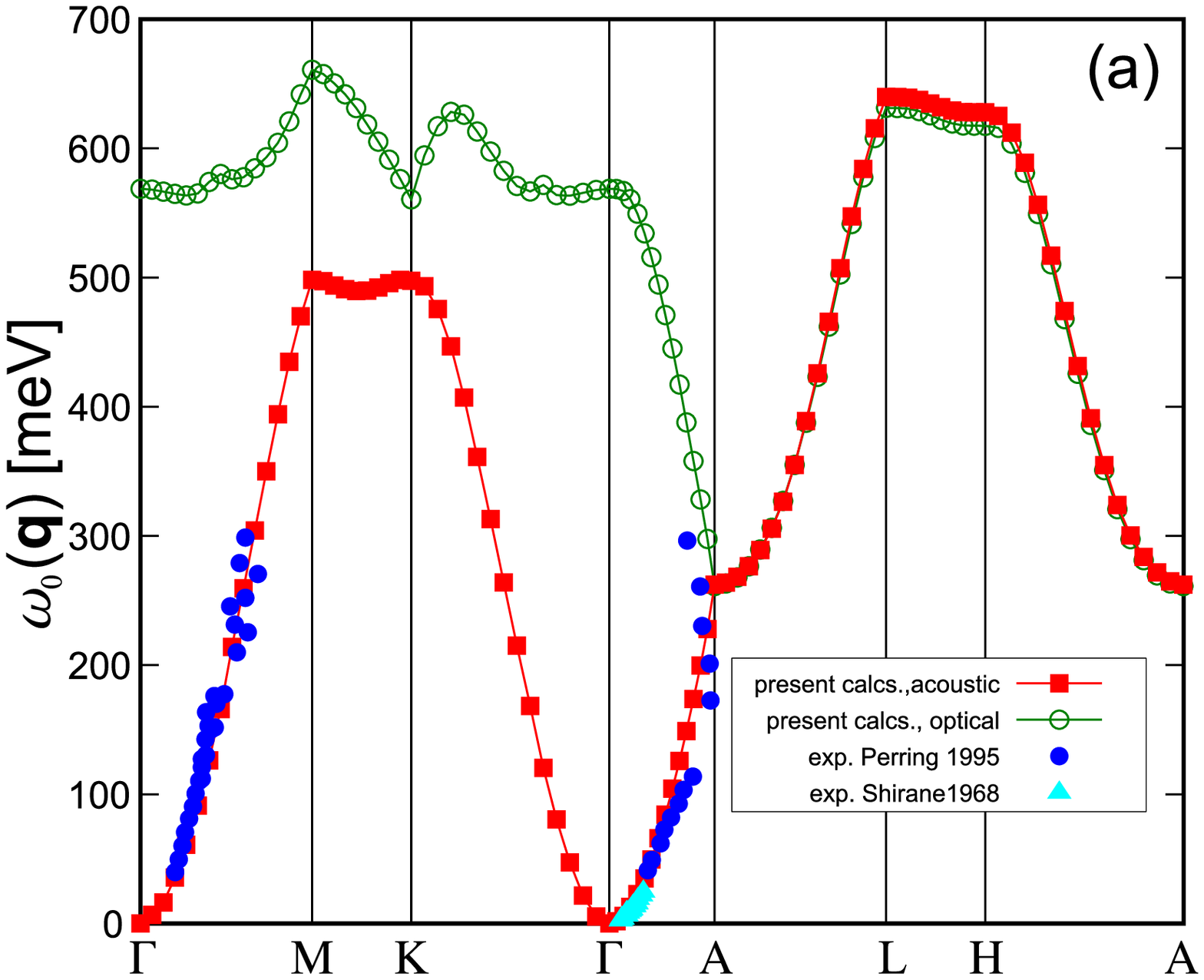}           % 8a
  \includegraphics[width=0.45\textwidth]{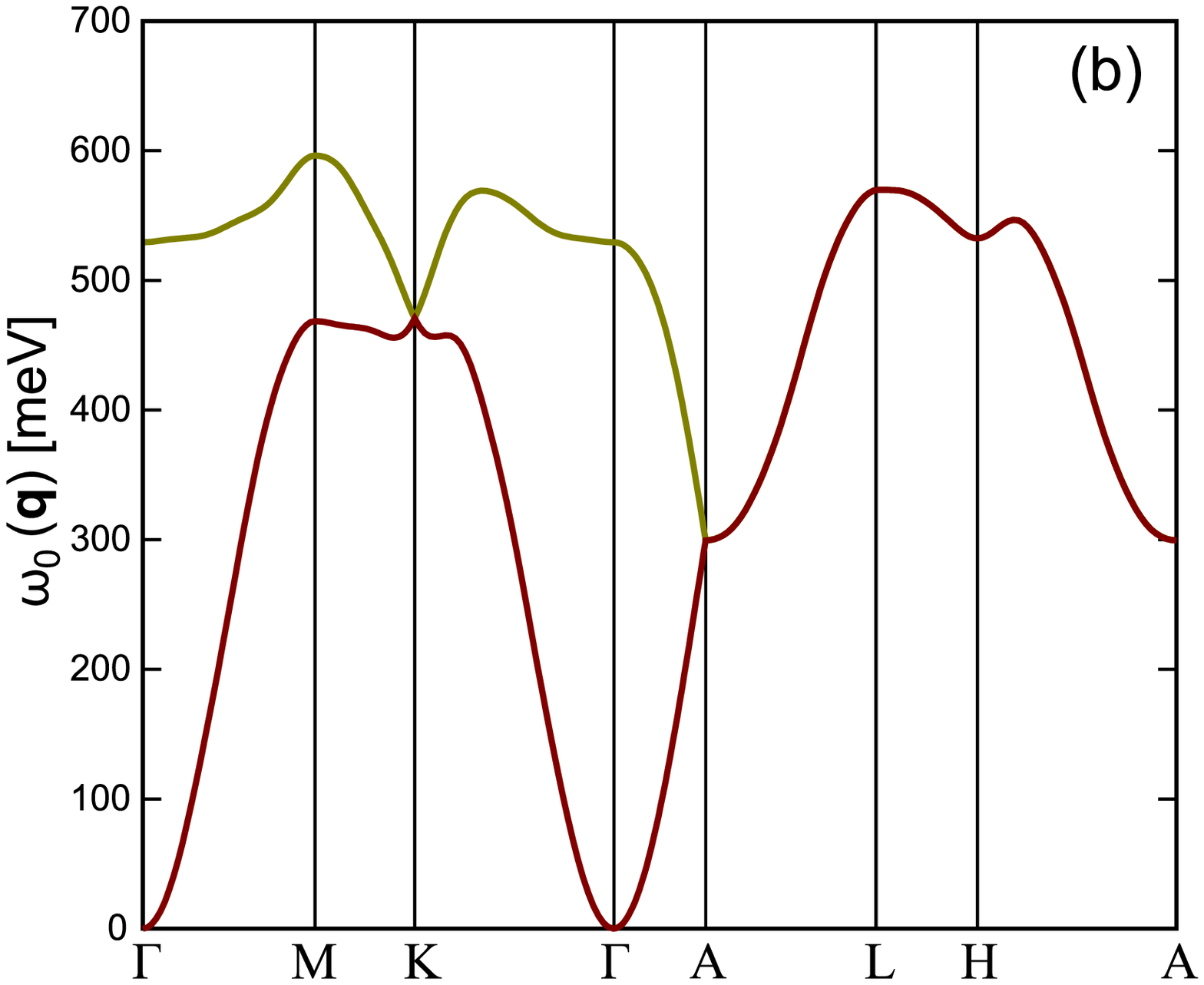} % 8b
  \includegraphics[width=0.45\textwidth]{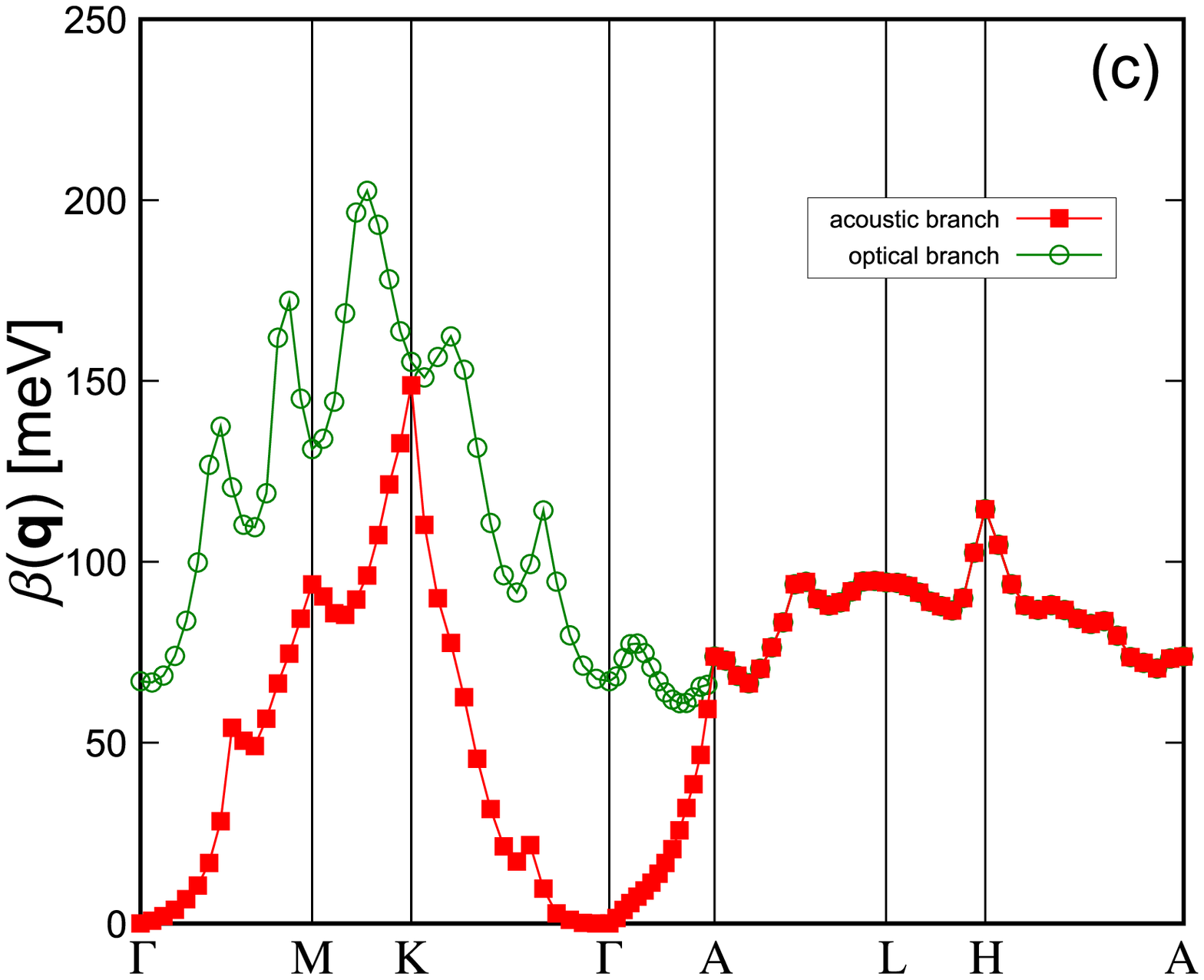}      % 8c
  \caption{Spin-waves of of hcp cobalt. (a) Position of the peak $\omega_{0}\fbr{\vec{q}}$ obtained from the enhanced susceptibility. The energies of spin-waves are degenerated within the numerical error along $\mathrm{ALHA}$ directions. The experimental energies come from Refs.~\cite{Shirane1968,Perring1995}. (b) Adiabatic magnon spectra obtained by means of magnetic force theorem. (c) \textit{half-widths} at half maximum, FWHM$\equiv2\beta\fbr{\vec{q}}$.}
  \label{fig:CoHCPDS}
\end{figure}

The loss matrix associated with the susceptibility features \textit{two} large eigenvalues for $\vec{q}$ in the first Brillouin zone and we interpret them as acoustic and optical spin-wave branches, cf.\ Fig.\ \ref{fig:CoExample}. The positions and widths of the peaks are presented in Fig.~\ref{fig:CoHCPDS} together with adiabatic spectra obtained from magnetic force theorem. Our method, MFT and STSM yield rather similar energies. For $q_{z} = \frac{\pi}{c}$ plane, the both modes are degenerate; it is the consequence of the symmetry of the hcp lattice. Theoretical spin-wave energies correspond very well to the experimental results along $\Gamma\mathrm{M}$ direction, but are larger along $c$-axis ($\Gamma\mathrm{A}$) direction, cf.\ Fig.\ \ref{fig:CoHCPDS}. For small momentum transfers the peak position can be very well described by the biquadratic fit, cf.\ eq.\ \ref{eq:biq_fit}. Table~\ref{tab:Co_hcp_biq_fit} presents the results of the fit for $q\leq0.3 \frac{2\pi}{a_{\mathrm{hcp}}}$ along different directions. Note that for the hcp system the dispersion relation for small $q$ is isotropic in the basal plane. Our parameters $D$ weakly depend on the direction in the reciprocal space because of the limited accuracy of the biquadratic function in the fitted interval of the wave vectors.

\begin{table}[htbp]
  \centering
    \begin{tabular}{@{}ccc@{}} \toprule
        direction                                              &  % 1
        $D$ $[\unit{}{\milli\electronvolt\angstrom\squared}]$  &  % 2
        $\gamma$ $[\angstrom\squared]$                         \\ % 3
    \midrule
    $\Gamma\mathrm{M}$ & 539 & 0.27  \\
    $\Gamma\mathrm{K}$ & 520 & 0.26  \\
    $\Gamma\mathrm{A}$ & 529 & 0.32  \\
    \bottomrule
    \end{tabular}
    \caption{Parameters of the biquadratic fit of Eq.~\eqref{eq:biq_fit} for different direction in Brillouin zone in hcp Co.}
  \label{tab:Co_hcp_biq_fit}
\end{table}

Additionally to the spectrum of the loss matrix, $\Loss\sqbr{\chi\fbr{\vec{q},\omega}}$, we present in the Fig.\ \ref{fig:CoExample} the imaginary part of the Fourier transform of the enhanced susceptibility, $\Im\chi_{\vec{G}\vec{G}}\fbr{\vec{q},\omega}$, for two selected momenta. By varying the momentum transfer in the inelastic scattering experiments, one can modify the intensity of peaks coming from the optical and acoustic spin-waves. \cite{Buczek2010a} In the example of hcp Co, along the $\Gamma\textrm{A}$ direction, only the acoustic magnons are observed in the first Brillouin zone, while the optical ones are detectable only in the second Brillouin zone. Perring \el \cite{Perring1995} succeeded to detect spin-waves for $\vec{q}$ beyond the first Brillouin zone along $\fbr{00\xi}$ direction, thus accessing the optical SW branch of $\Gamma\mathrm{A}$ segment. The formalism of this paper predicts that the two branches exist in the whole Brillouin zone. The acoustic mode corresponds to the moments oscillating in phase and the optical mode in anti-phase. The energy of the optical mode is overestimated in ALSDA. The optical magnons are of much shorter life-time than the acoustic magnons because of the higher density of the Stoner transitions, as expected for higher energies. Characteristic peaks in the FWHM curves correspond to the areas where the spin-wave branch crosses the region of larger density of Stoner excitations. The damping of spin-waves is moderate in the case of Co and peaks have well defined Lorentzian shape. All the majority spin $d$ electrons are occupied and located rather far from the Fermi level. As a consequence, the low energy Stoner excitations involve primarily $s^{\uparrow}$ and $p^{\uparrow}$ to $d^{\downarrow}$ transitions which results in the small Stoner intensity because of weak overlap of the wave-functions of the initial and final states.

Earlier theoretical works on spin susceptibility of hcp cobalt utilize empirical tight binding scheme \cite{Trohidou1991,Bass1992}. They predict correctly the energies of acoustic modes, yielding however too low values of optical modes, in fact in a range where they were not detected despite being experimentally observable. Except for $\Gamma\mathrm{A}$ segment the optical branch energies predicted in our study lie above \unit{0.5}{\electronvolt}, the maximal energy addressed in the calculations mentioned and one guesses that certain complex structure of the spectral density of spin-flip excitations (cf.\ the case of bcc Fe in Fig.\ \ref{fig:FeExample}) might have been erroneously identified as the optical mode.

\subsubsection{$\gamma$(fcc)-Co}
\label{subsubsec:fccCo}

The lattice constant employed in the calculations was determined from the condition of equal atomic volumes for hcp and fcc systems, yielding $a_{\mathrm{fcc}} = \sqrt{2}a_{\mathrm{hcp}}$. Ground state magnetic moment is very close to the one obtained for hcp phase.

\begin{figure}
  \centering
  \includegraphics[width=0.45\textwidth]{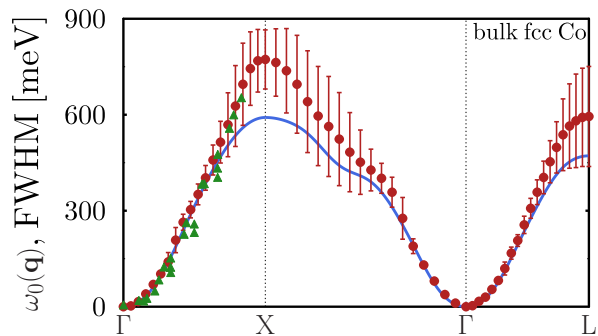} % 9
  \caption{Spin-waves of fcc Co. Solid circles ({\Large\textbullet}) correspond to $\omega_{0}\fbr{\vec{q}}$, while the error bars denote full width at half maximum (FWHM) of the peak. Solid line denotes spin-wave energies obtained from MFT. Solid triangles ($\blacktriangle$) stand for the experimental estimation of the dispersion relation by Balashov. \cite{Balashov2009a}}
  \label{fig:CoFCCDS}
\end{figure}

The loss matrix of the enhanced susceptibility features only one large eigenvalue. The high wave-vectors magnons show slightly higher energy, when compared to the MFT results, cf.\ Fig.\ \ref{fig:CoFCCDS}. The spin-waves are clearly damped, but the spectrum can be well described by a Lorentzian peak for all momenta; no spin-wave disappearance is observed.

Inelastic neutron scattering experiments revealed magnons of slightly smaller energies. \cite{Sinclair1960,Pickart1967} The stiffness constant estimated from our data amounts to $D=\unit{492}{\milli\electronvolt\angstrom^{2}}$, while $D=\unit{369}{\milli\electronvolt\angstrom^{2}}$ \cite{Sinclair1960} and $D=\unit{356}{\milli\electronvolt\angstrom^{2}}$ \cite{Pickart1967} were reported. The results of spin polarized electron energy loss spectroscopy \cite{Vollmer2003} and inelastic scanning tunneling spectroscopy \cite{Balashov2009a} match rather well the neutron scattering data. The difference with our calculations can be traced back to the finite temperature of the experiment. For the data extracted from the film measurements \cite{Vollmer2003,Balashov2009a} the Cu substrate might influence the results, but our calculations reported in the next section exclude this possibility. In the case of bulk (neutron) measurements fcc Co must be alloyed with about 6\% of Fe for the sake of stability, which might further contribute to the quantative differences between the theory and experiment.

Rather limited data exists regarding the life-time of magnons. Vollmer \el \cite{Vollmer2003} were able to perform constant $\vec{q}$ scans and provided an estimate of FHWM about \unit{40-75}{\milli\electronvolt}. This matches quite well our low energy results.

\subsubsection{\ml{1} Co/Cu(100)}
\label{subsec:CoFilms}

\begin{figure}
  \centering
  \includegraphics[width=0.45\textwidth]{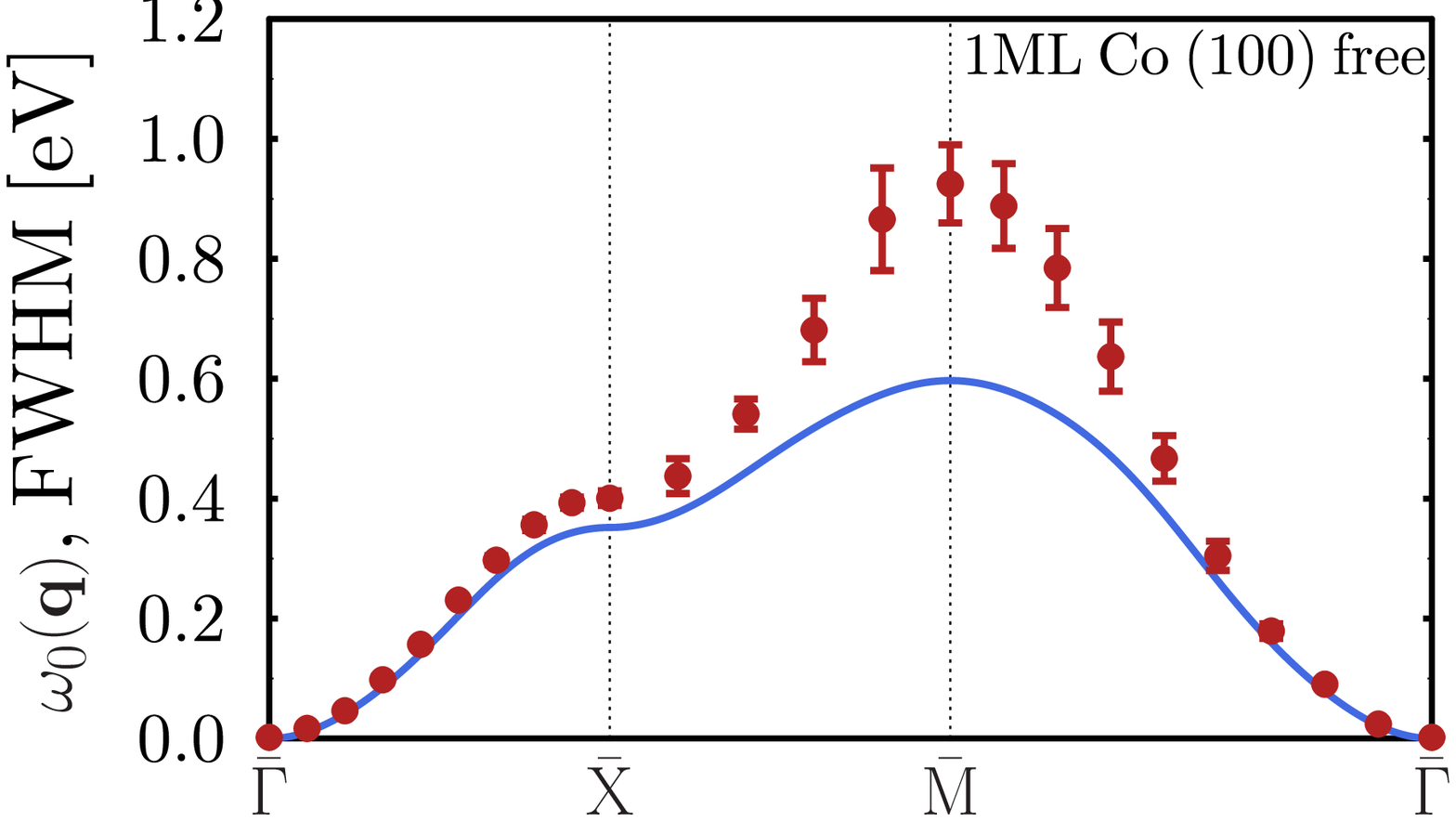} % 10a
  \includegraphics[width=0.45\textwidth]{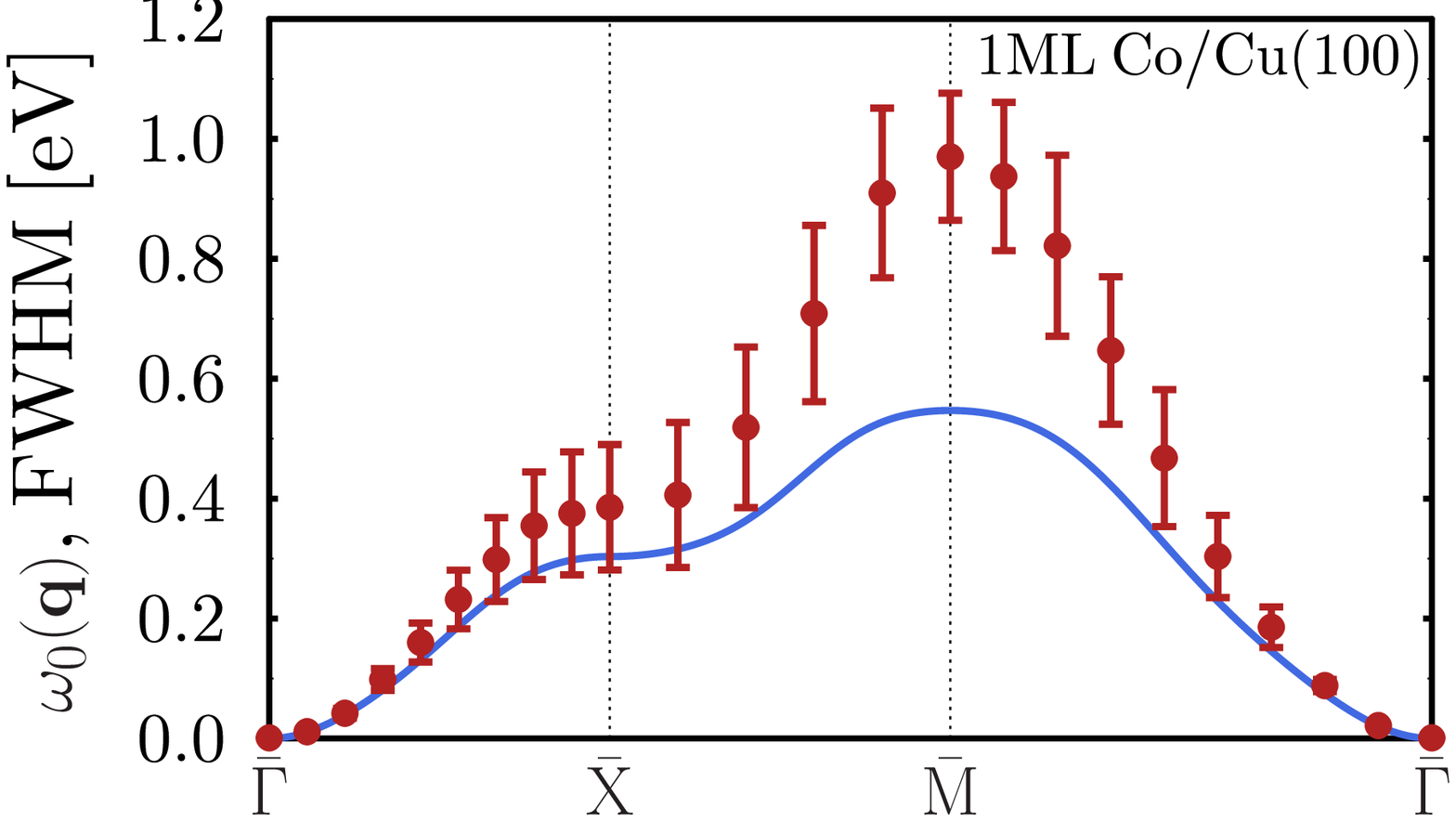}  % 10b
  \caption{Spin-waves of Co(100) monolayer free and supported on Cu(100) surface. Solid circles ({\Large\textbullet}) correspond to $\omega_{0}\fbr{\vec{q}}$, while the error bars denote full width at half maximum (FWHM) of the peak. Solid line denotes spin-wave energies obtained from MFT.}
  \label{fig:CoMonolayer}
\end{figure}

The spin-waves of \ml{1} Co/Cu(100) are clearly of higher energy than in the bulk fcc case, see Fig.\ \ref{fig:CoMonolayer}. Interestingly, the Landau damping is slightly smaller. Compared to the free standing monolayer we observe almost no magnon energy renormalization and moderate increase of the Landau damping. This behavior is qualitatively very similar to the \ml{1} Fe/Cu(100) case.

\subsection{Ni}
\label{subsec:Ni}

\subsubsection{bulk fcc Ni}

\begin{figure}
  \centering
  \includegraphics[width=0.45\textwidth]{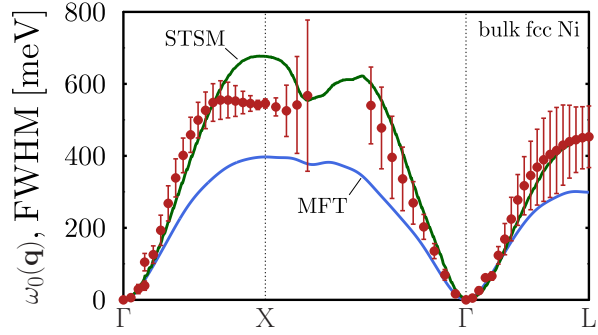} % 11
  \caption{Spin-waves of fcc nickel. Solid circles ({\Large\textbullet}) correspond to $\omega_{0}\fbr{\vec{q}}$, while the error bars denote full width at half maximum (FWHM) of the peak. Solid line denotes spin-wave energies obtained using MFT and STSM; the two methods differ significantly for this system. Most of the spectrum can be described using Lorentzian distribution function, except most energetic spin-waves along $[\xi\xi0]$ direction. An example of such a irregular spectrum is presented in Figure \ref{fig:NiExample}.}
  \label{fig:NiDS}
\end{figure}

The transverse magnetic susceptibility of bulk fcc Ni has been earlier studied using different first-principle approaches \cite{Savrasov1998,Aryasetiawan1999,Karlsson2000,Sasiouglu2010}. All previous calculations and the present study performed by us, cf.\ Fig.\ \ref{fig:NiDS}, yield a very similar picture. The spin dynamics of nickel differs strongly from that of iron, in particular, no spin-wave dissapearance is found. Magnon peaks are well defined in the practically entire Brillouin zone, characterized by a relatively small Landau damping. This is rather surprising, realizing that the Stoner continuum is pronounced in Ni at relatively low energies, corresponding to the  exchange splitting predicted by LSDA to be only about $\unit{0.7}{\electronvolt}$. The effect of the low damping of the Ni magnons has its roots in the fact that the spectral power of the Stoner continuum is pronounced mostly around the exchange splitting energy and is not spread over a wide range of energies as in Fe. Furthermore, all the majority spin $d$ electrons are occupied and located well below the Fermi level. Therefore, the Stoner transitions involving these states are not effective at the energies characteristic to magnon excitations in fcc Ni. In this respect the system is similar to hcp and fcc Co.

\begin{figure}
  \centering
  \includegraphics[width=0.45\textwidth]{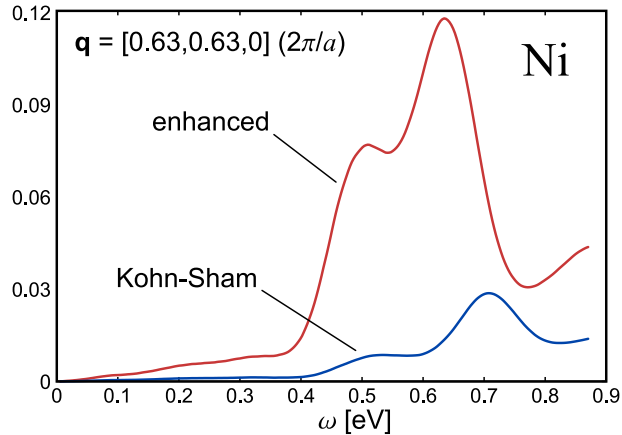} % 12
  \caption{An example of spin-flip excitation spectrum in Ni, $-\Loss\sqbr{\chi\fbr{\vec{q},\omega}}_{\lambda}$, atomic units. Only eigenvalue of the loss matrix dominates. One clearly sees the energy dependence of the Stoner continuum reflected in the behavior of enhanced susceptibility.}
  \label{fig:NiExample}
\end{figure}

\begin{figure}
  \centering
  \includegraphics[width=0.45\textwidth]{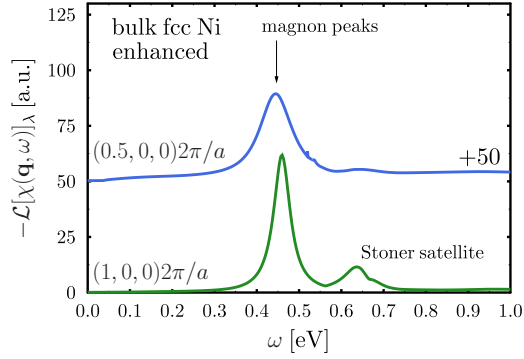} % 13
  \caption{Two examples of spin-flip excitation spectrum in Ni, atomic units, along (100) direction. The non-monotonous dependence of the damping with the magnon momentum is clearly seen. Additionally, close to the $\mathrm{X}$ point in the Brillouin zone a clear coexistence of the spin-wave peak and a Stoner continuum peak can be observed.}
  \label{fig:NiExampleX}
\end{figure}

Another special property of Ni is a pronounced non-monotonous dependence of damping on $\vec{q}$, seen most clearly along $[\xi00]$ in the Brillouin zone, cf.\ Fig.\ \ref{fig:NiExampleX}. At the $\textrm{X}$ point the spin-waves are very weakly Landau damped. This behavior originates again from the very narrow Stoner continuum at the $\textrm{X}$ point; it is almost entirely concentrated around the exchange splitting energy, while at $\Delta\equiv\sqbr{\frac{1}{2}00}$ point it has a significant contribution at smaller energies. This leads to a very interesting effect at the $\textrm{X}$ point where the spin-wave peak (identified by a small eigenvalue of $\onemat - \chi^{\pm}\KS\fbr{0} K\xc$ matrix) appears just below a peak of the Stoner continuum. Existing experimental studies point to a rather monotonous increase of the magnon linewidth with increasing momenta \cite{Mook1969,Mook1979}. However, the momenta close to the zone boundary are difficult to be probed experimentally using thermal neutrons. We conjecture that the coexisting magnon and Stoner peaks were not separately resolved giving rise to one broad spectral feature.

\begin{figure}
  \centering
  a)~\includegraphics[width=0.45\textwidth]{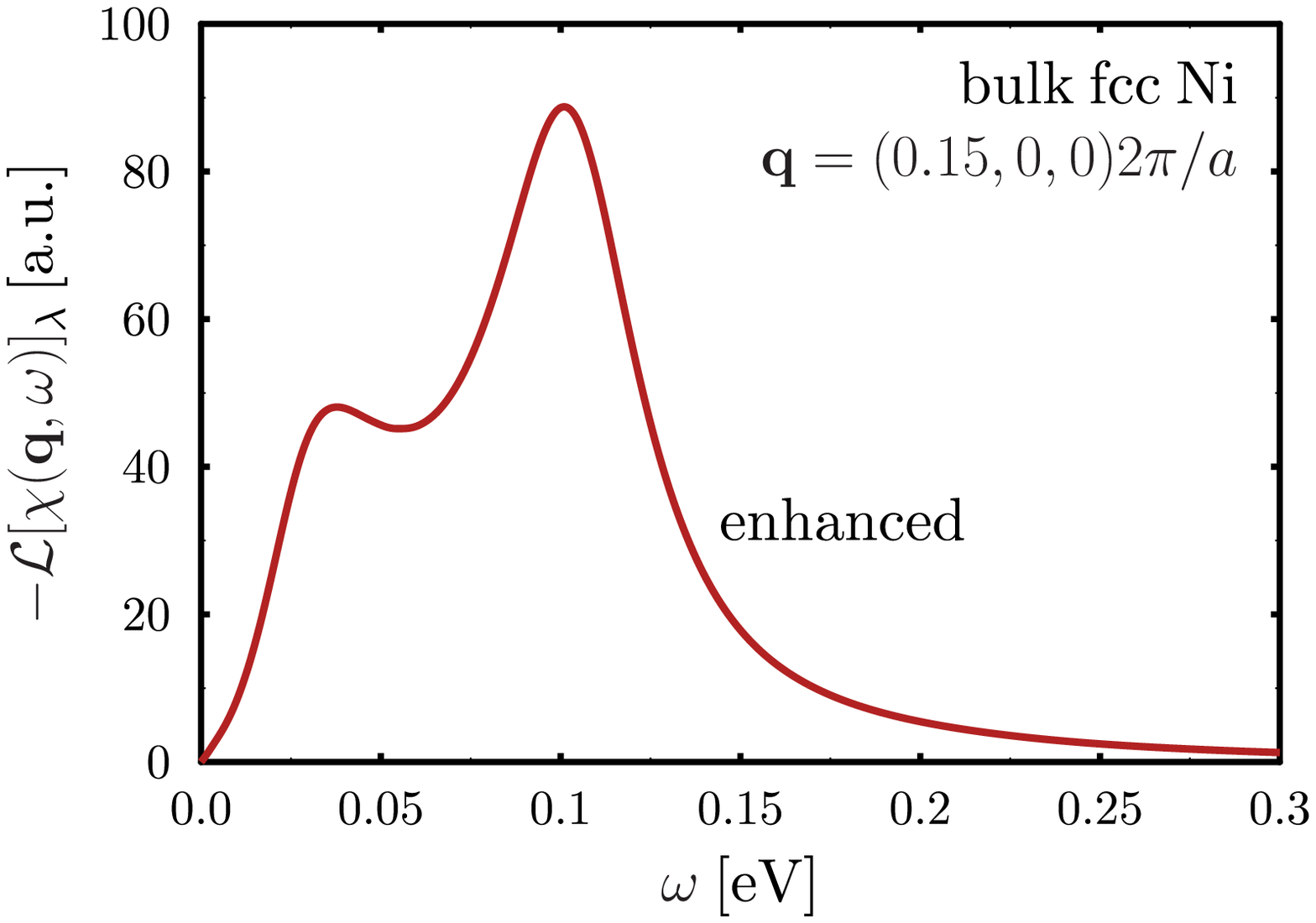}      % 14a
  b)~\includegraphics[width=0.45\textwidth]{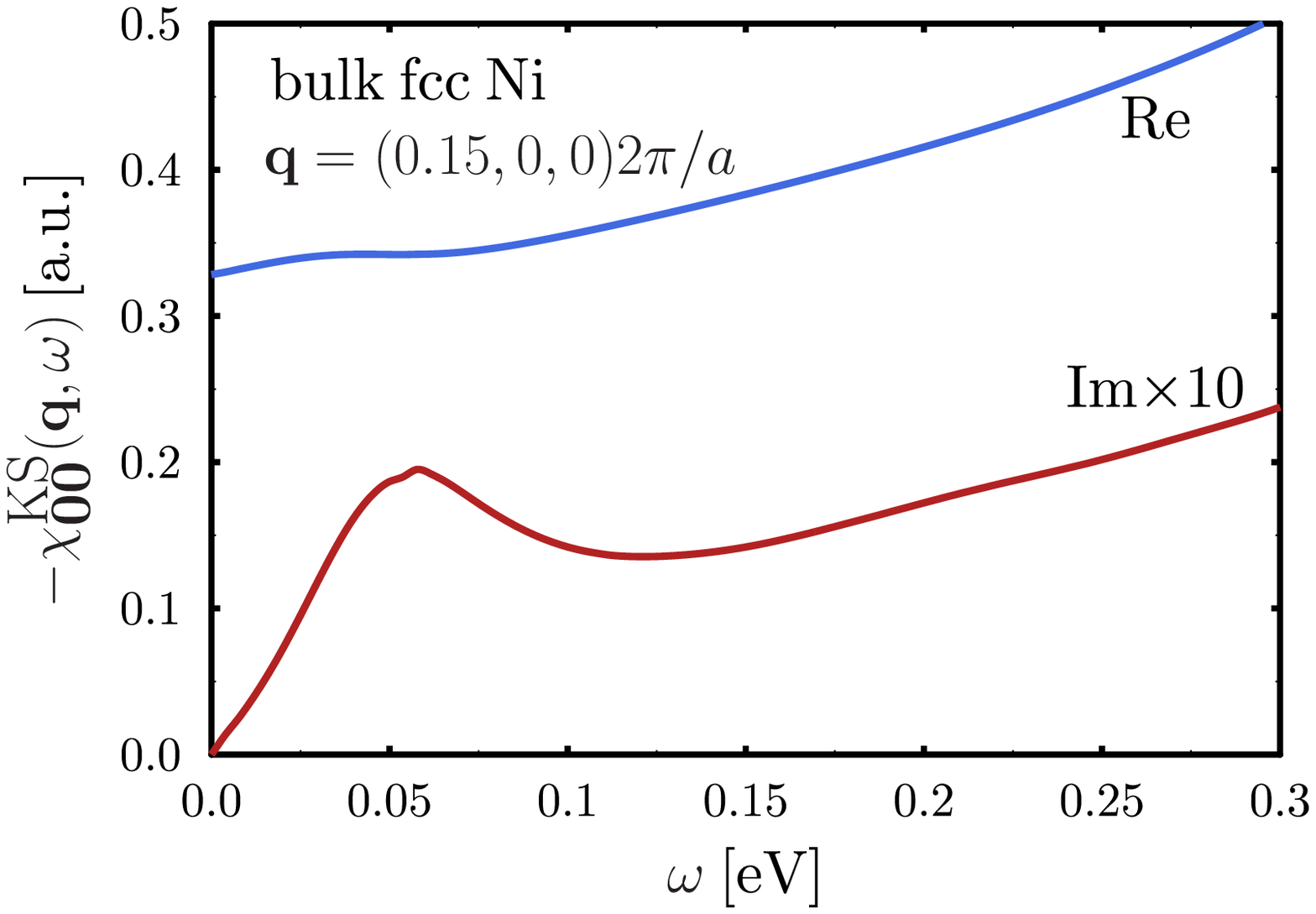} % 14b
  \caption{(a) Double peak feature in bulk fcc Ni. (b) Behavior of the corresponding Kohn-Sham susceptibility.}
  \label{fig:NiDoublePeak}
\end{figure}

Much attention has been paid to the appearance of so called ``optical spin-wave mode'' experimentally detected in fcc Ni \cite{Mook1979,Sasiouglu2010}. In our calculations a low energy double magnon peak structure appears only in a very small part of the Brillouin zone around $\vec{q}=(0.15,0,0)2\pi/a$, cf.\ Fig.\ \ref{fig:NiDoublePeak}a. The feature is strictly absent in the Heisenberg model, as it features only single degree of freedom in monatomic Ni. We note, however, that (i) the loss matrix of the enhanced susceptibility features only one large eigenvalue in this energy region, (ii) both peaks are associated with similar eigenvectors and therefore describing similar magnetization shape. In this respect the double peak structure clearly differs from the ``real'' optical spin-wave mode of hcp Co arising from the presence of two magnetic atoms in the primitive cell. The splitting of the peak was explained by Karlsson \el \cite{Karlsson2000} by analyzing the non-enhanced susceptibility in the region of the double peak structure, cf.\ Fig.\ \ref{fig:NiDoublePeak}b, characterized by an appearance of rather narrow low energy Stoner peak. The corresponding real part of the susceptibility changes non-monotonously with the energy bringing an eigenvalue of $\onemat - \chi^{\pm}\KS\fbr{0} K\xc$ close to zero \textit{twice} on a the energy scale of the peak splitting.

Unfortunately, the agreement with experiment is much worse regarding the magnon energies. The stiffness constant extracted from our data equals \unit{851}{\milli\electronvolt\angstrom^{2}} and is roughly twice as large as the experimentally observed \unit{374-433}{\milli\electronvolt\angstrom^{2}} \cite{Pickart1967,Mook1969}. At larger momenta, the same factor two discrepancy between measured and calculated magnon energies is seen. The problem is associated with the overestimation of the exchange splitting in LDA for Ni \cite{Lichtenstein2001,Sasiouglu2010}.

\begin{figure}
  \centering
  \includegraphics[width=0.45\textwidth]{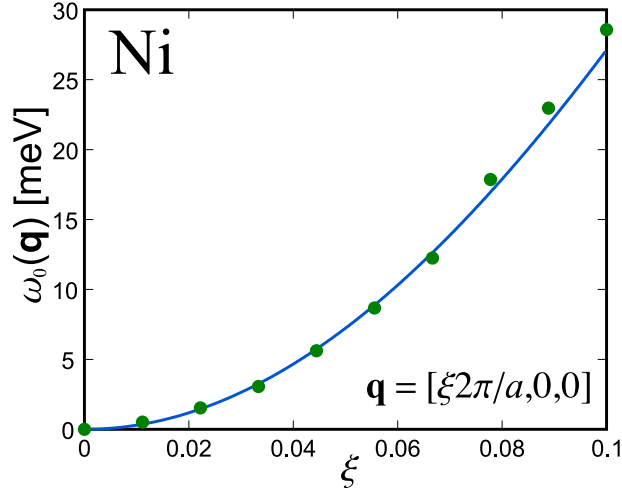} % 15
  \caption{Magnon energies in fcc Ni in the limit of small momenta obtained from dynamic susceptibility ($\blacksquare$) and MFT (line). In the limit both methods yield identical results. This mathematical identity \cite{Bruno2003,Katsnelson2004} provides a neat check for our numerics.}
  \label{fig:NiStiffness}
\end{figure}

The adiabatic STSM and the results based on the dynamic susceptibility calculations, agree very well with each other but yield clearly higher energies than MFT, cf.\ Fig.\ \ref{fig:NiDS}. As it has already been pointed out by Grotheer \el \cite{Grotheer2001}, the systematic error of the MFT method \cite{Bruno2003} is particularly pronounced in Ni, owing to its small exchange splitting. Nevertheless, the spin-wave stiffness obtained from all adiabatic methods is identical to the one obtained from dynamic susceptibility, cf.\ Fig.\ref{fig:NiStiffness}.

\subsubsection{\ml{1} Ni/Cu(100)}
\label{subsec:NiFilms}

\begin{figure}
  \centering
  \includegraphics[width=0.45\textwidth]{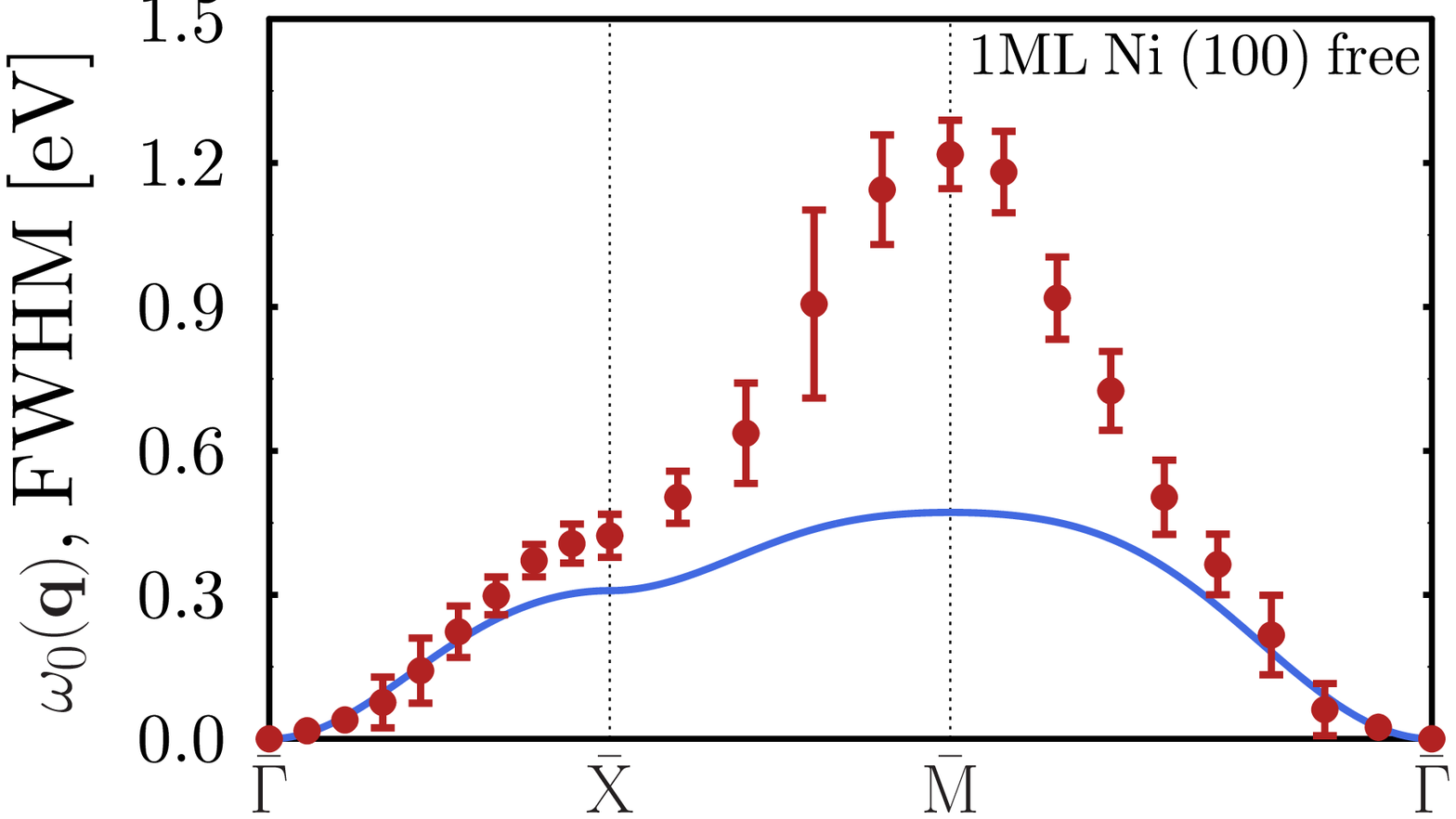} % 16a
  \includegraphics[width=0.45\textwidth]{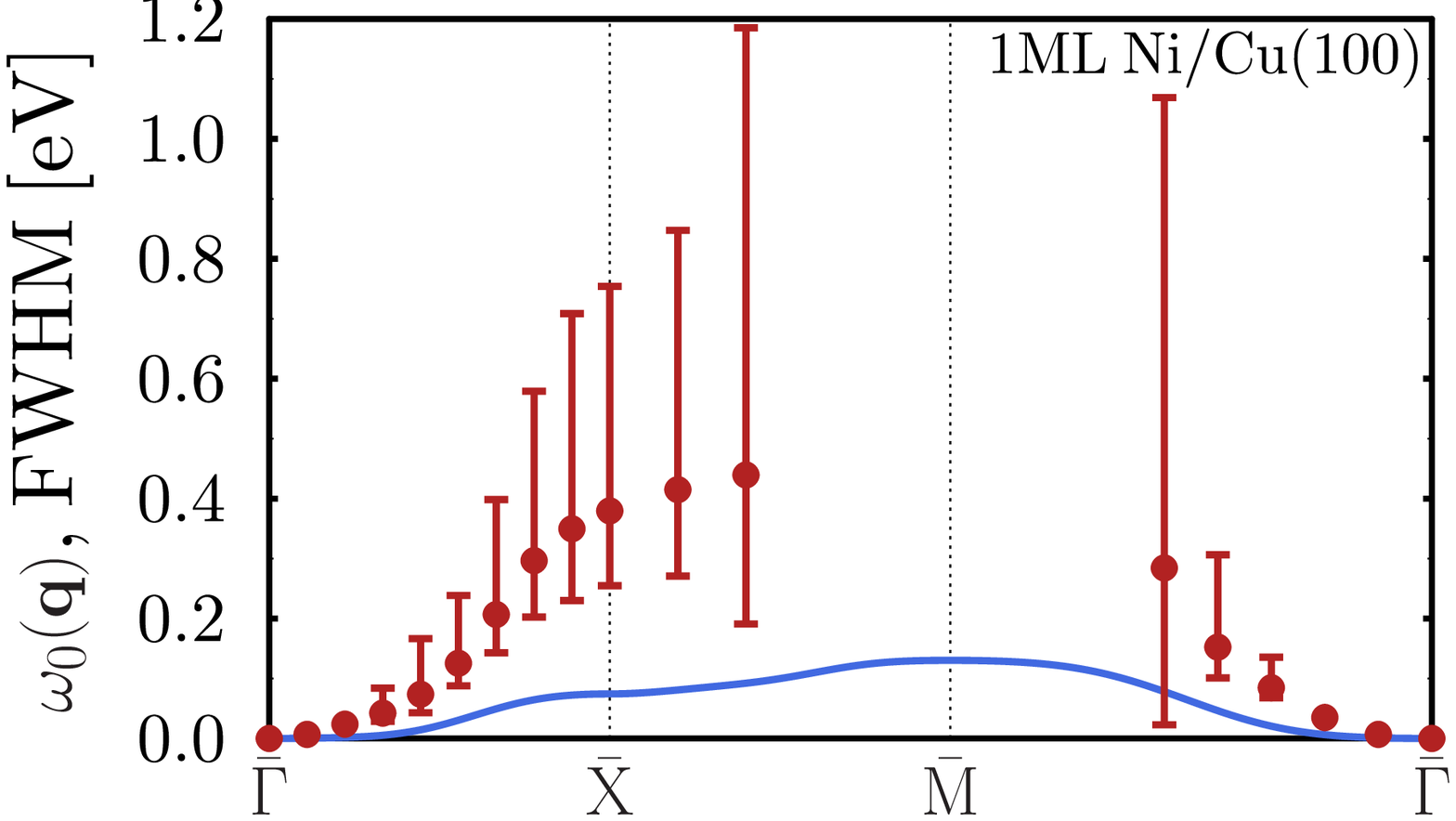}  % 16b
  \caption{Spin-waves of Ni(100) monolayer free and supported on Cu(100) surface. Solid circles ({\Large\textbullet}) correspond to $\omega_{0}\fbr{\vec{q}}$, while the error bars denote FWHM. Solid line denotes spin-wave energies obtained from MFT. Spin-waves with large momenta are not well defined excitations in the case of the absorbed film, while for the free standing film they exist in the whole Brillouin zone.}
  \label{fig:NiMonolayer}
\end{figure}

\begin{figure}
  \centering
  \includegraphics[width=0.45\textwidth]{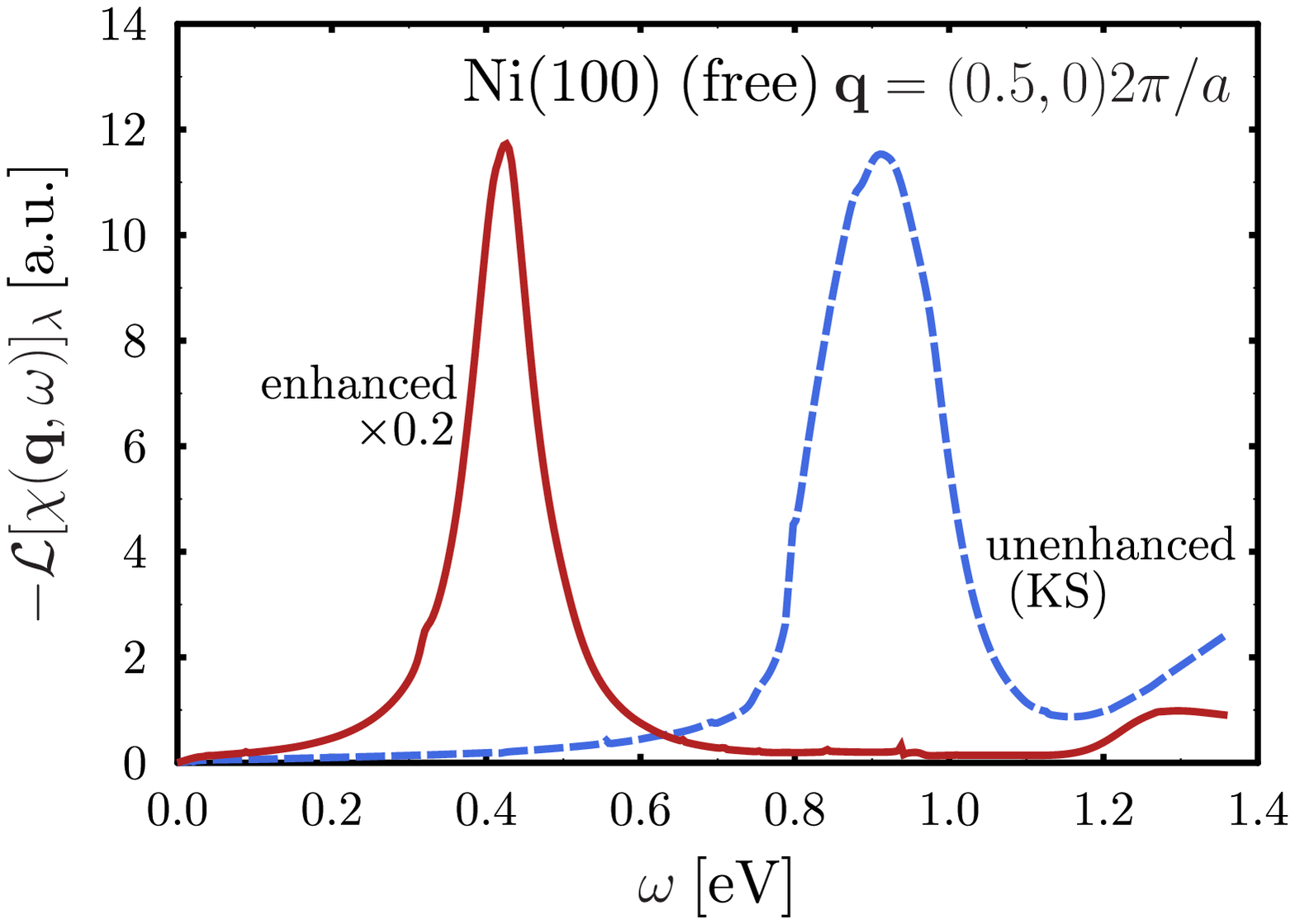}   % 17a
  \includegraphics[width=0.45\textwidth]{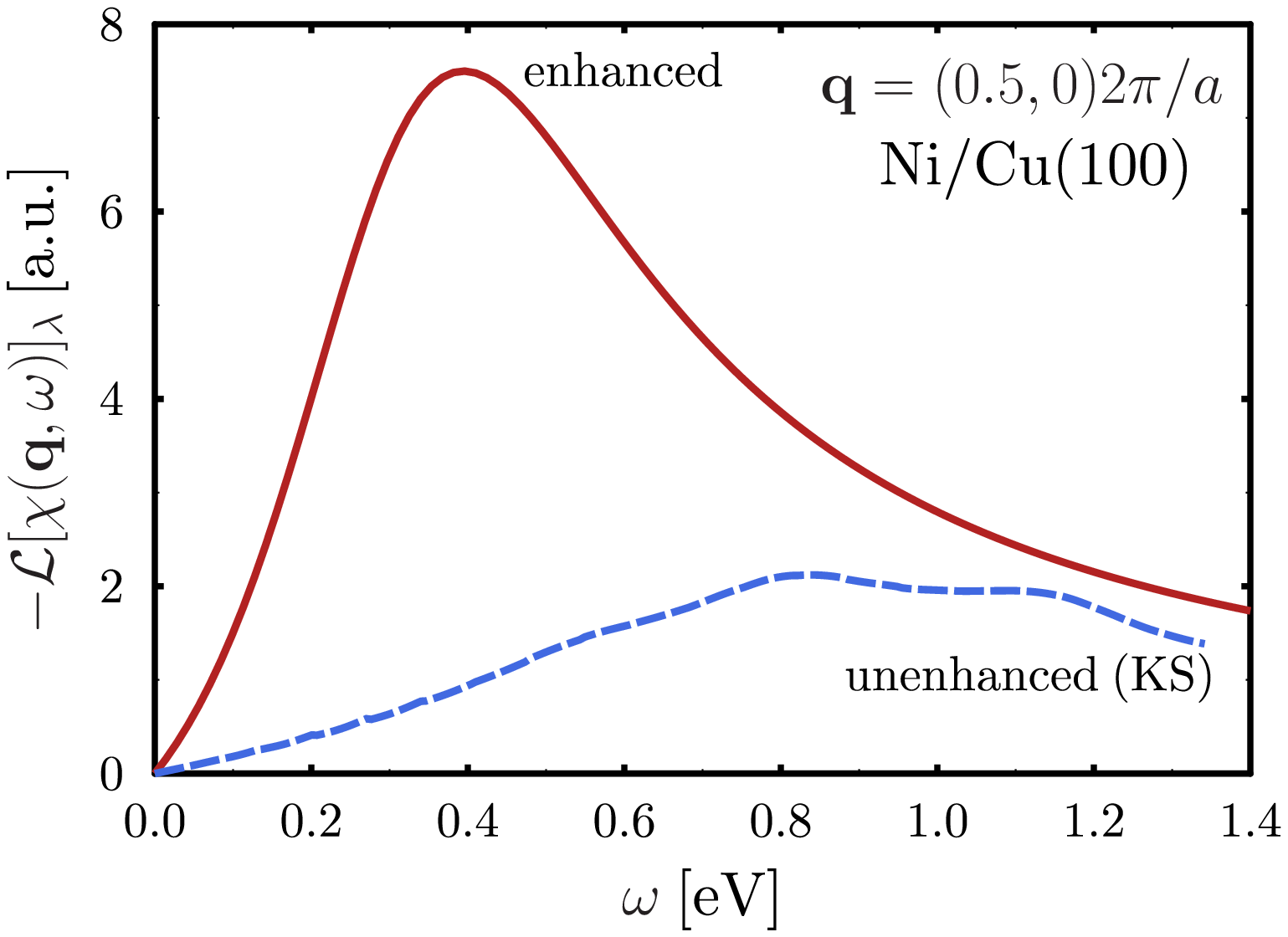} % 17b
  \caption{Example of spectra in $\unit{1}{ML}$ Ni(100) and $\unit{1}{ML}$ Ni/Cu(100), atomic units. $\textrm{\textbf{q}}=(0.5,0)2\pi/a$ corresponds to $\bar{\mathrm{X}}$ point in the Brillouin zone. In the supported monolayer, for large wave-vectors, the Stoner continuum becomes very broad (in contrast to the free standing film) and the Landau damping completely washes out any sharp spin-wave features.}
  \label{fig:NiCuExample}
\end{figure}

\begin{figure}
  \centering
  \vskip10mm
  \includegraphics[width=0.45\textwidth]{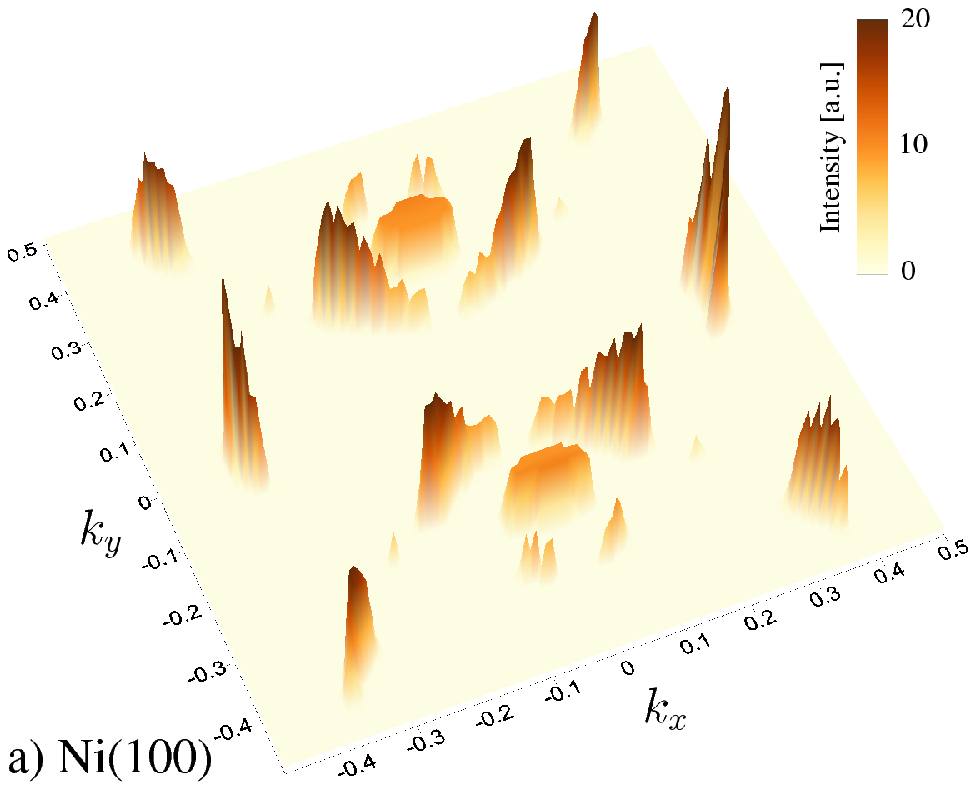}    % 18a
  \includegraphics[width=0.45\textwidth]{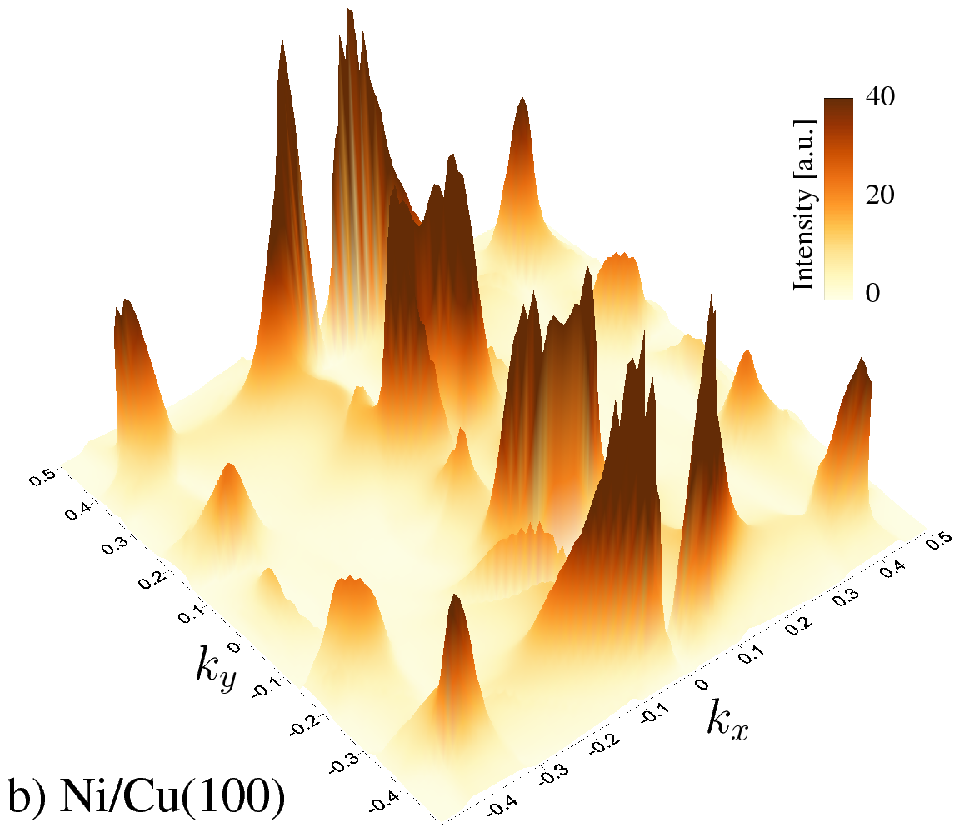} % 18b
  \caption{Landau maps of $\unit{1}{ML}$ Ni(100) and $\unit{1}{ML}$ Ni/Cu(100): intensity of Stoner transitions with momentum $\textrm{\textbf{q}}=(0.5,0)2\pi/a$ ($\mathrm{\bar{X}}$ point) and energy $\omega_{0} = \unit{400}{\milli\electronvolt}$ in the Ni layer resolved for different final $\vec{k}$-vectors in the first Brillouin zone. The Stoner states cause the damping of magnons presented in Fig.\ \ref{fig:NiCuExample}.}
  \label{fig:NiCuLhs}
\end{figure}

Similar to the case of Fe and Co, the magnons in free standing monolayer of Ni (100) are weakly damped, as seen in Fig.\ \ref{fig:NiMonolayer}. Their energies are much higher than in the fcc Ni bulk. However, upon absorption on Cu(100) surface, the spin-dynamics of the systems changes dramatically and in the way that is in strong contrast to the other two transition metals. Spin waves in $\unit{1}{ML}$ Ni/Cu(100) are defined only close to the center of the Brillouin zone and strongly damped for larger momenta. Two examples of spin-flip spectra are shown in Fig.\ \ref{fig:NiCuExample}. The hybridization of the states of the Ni overlayer and the Cu substrate states leads to (i) reduction of the exchange splitting (the Stoner excitations for $q = 0$ are centered around \unit{218}{\milli\electronvolt} vs.\ \unit{700}{\milli\electronvolt} in the bulk), reflected in the smaller magnetic moment ($\unit{0.27}{\Bm}$ vs.\ $\unit{0.62}{\Bm}$ in the bulk), (ii) larger energy width of the Stoner continuum. This two effects result in the enhanced density of low energy Stoner excitations, washing out most of sharp high energy spin-wave features. In \ml{1} Ni/Cu(100) localized moment picture of spin excitations (i.e.\ Heisenberg model) fails altogether. The Landau maps of \ml{1} Ni/Cu(100), cf.\ Fig.\ \ref{fig:NiCuLhs} are qualitatively similar to the to case of the Fe and Co monolayer \cite{Buczek2011} but characterized by much broader hot-spots and more intense diffused background. This property reflects strong hybridization of the electronic states in the film with the substrate electrons.

\section{Concluding remarks}
\label{sec:Conclusions}

The spin-flip dynamics of elementary $3d$ transition-metal ferromagnets is rich and strongly system dependent. The magnons of Co live relatively long for momenta in the whole Brillouin zone. One might be tempted to associate this property with the large exchange splitting of bands, which usually results in the Stoner continuum pronounced at high energies, but such explanation fails badly for Fe, which features even larger exchange splitting and at the same time severe Landau damping of spin-waves. The effect is even more spectacular in Ni, where the band splitting is only a bit larger than the typical magnon energy. Owing to the compactness of the Stoner spectrum and the small density of majority spin $d$ states at the Fermi level, Ni features rather long-living magnons in most of its Brillouin zone. We see that these are fine properties of Stoner continuum which determine the spin-wave attenuation. Obviously, the first principle approaches based on the calculation of transverse magnetic susceptibility are indispensable in the consistent description of spin dynamics in real materials.

This statement holds particularly true for the spin excitations of ultrathin films. Free standing monolayers studied in this paper shows usually well defined long-living spin-waves for all momenta. Upon absorption of the magnetic film on a nonmagnetic substrate the damping generally increases, but the details of the spin-flip dynamics are very sensitive to the details of the electronic hybridization between the electron states of substrate and film. Monolayers of Fe (supported both on Cu(100) and W(110) surfaces) feature relatively long-living, well defined spin-waves in most of the two-dimensional Brillouin zone, in striking contrast to the bulk bcc phase. In \ml{1} Ni/Cu(100) spin-waves exist only for small momenta, the spin-flip spectrum in the rest of the Brillouin zone is dominated by incoherent Stoner excitations. The situation is again opposite to the one in the bulk fcc Ni bulk characterized by generally long-living high energy spin-flip dynamics. The spin dynamics of cobalt assumes an intermediate position and changes weakly upon the transition from the bulk to \ml{1} Co/Cu(100).

Based on the examples above we suggest the following classification of the spin-wave Landau damping in $3d$ magnets. We distinguish three regimes: (A) for small wave vectors the acoustic magnon pole appears outside the energy range of the Stoner continuum and the spin-waves resemble closely atomic-spin like precession assumed in the Heisenberg model. For the elemental transition metal ferromagnets this region contains typically only a very small part of the Brillouin zone, whereas it might span large part of $\vec{q}$ space in the case of half-metals \cite{Buczek2009}; (B) in the energy region of low intensity Stoner excitations ($\Re{\chi\KS} \gg \Im{\chi\KS} \ne 0$) the trend to form the coherent precession of atomic moments is still observed. However, it is opposed by the one-electron Stoner transitions at this energy. This hybridization results in a broadening of the spin-wave peak and, respectively, in finite life time of the magnon. The magnons, however, can still be regarded as well defined excitations. This regime is absent in the uniform electron model, but much of the spin-flip dynamics in Fe, Co and Ni falls into this category; (C) finally, ``spin-wave disappearance'' regime corresponds to the situation where the expected position of the magnon pole lies in the energy region with high density of Stoner states. In this case $\Re{\chi\KS} \approx \Im{\chi\KS}$ and no well defined resonance peak can form since the phase relationship between the external field and the induced exchange-correlation field is destroyed due to the intense excitations of incoherent Stoner pairs.

In different systems the relative importance of different spin-wave damping regimes may vary. In this respect already the comparison of the bulk $3d$ metals brings interesting observations. In bcc Fe, along certain directions in the Brillouin zone, ``spin-wave disappearance'' regime sets in very abruptly. In fcc Ni and fcc and hcp Co regimes A and B dominate and practically no spin-wave disappearance is seen. The case of ultrathin films can differ strongly compared to the corresponding bulk system under the influence of the dimensionality and the presence of non-magnetic substrate.

\begin{acknowledgments}
PB acknowledges stimulating discussions with
Vladimir P.\ Antropov,
Patrick Bruno,
E.K.U.\ Gross,
Bal\'azs Gy\H{o}rffy,
Josef Kudrnovsk\'y,
Julie Staunton,
Zdzis\l{}awa Szotek,
Walter Temmerman, and
Herman Winter.
We thank Udo Schmidt for technical assistance. This work is supported by the Sonderforschungsbereich SFB 762, ``Functionality of Oxidic Interfaces'', and DFG priority programs SPP 1386, ``Nanostrukturierte Thermoelektrika'' and SPP 1538, ``Spin Caloric Transport''. Part of the calculations was performed at the Rechenzentrum Garching of the Max Planck Society (Germany).
\end{acknowledgments}

\appendix

\section{Representation of the spatial dependency of susceptibilities}
\label{app:Basis}

The susceptibilities depend on two spatial arguments and are represented in a separable basis as follows
\begin{align}
  \chi\fbr{\vec{x},\vec{x}'} = \sum_{\lambda\lambda'} \chi_{\lambda\lambda'}
    \varphi_{\lambda}\fbr{\vec{x}}\varphi_{\lambda'}\fbr{\vec{x}'}^{\ast},
\end{align}
where the functions $\varphi_{\lambda}$ form a complete basis. $\chi_{\lambda\lambda'}$ is the \textit{susceptibility matrix} in this basis.

In this work, we mostly used the following function set, referred to as \textit{Y-Ch basis}, to represent the susceptibilities. The angular dependence in an atomic (Voronoi) cell $s$ is cast into real spherical harmonics $\Ylm_{lm}\fbr{\hat{\vec{r}}}$ whereas the radial dependence is represented by Chebyshev polynomials $\Ch_{\mu}\fbr{\xi}$
\begin{align}
  \varphi_{\lambda}\fbr{\vec{x}} \equiv \frac{1}{r_{s}} \Ch_{\mu}\fbr{\xi_{\sqbr{0,R_{s}}}\fbr{r_{s}}} \Ylm_{lm}\fbr{\hat{\vec{r}}_{s}} \Theta_{s}\fbr{\vec{x}}.
\label{Eq:basis}
\end{align}
Here, $\vec{r}_{s} \equiv \vec{x} - \vec{s}_{s}$, where $\vec{s}_{s}$ is the center of the Voronoi cell containing point $\vec{x}$. $\Theta_{s}\fbr{\vec{x}}$ is the shape function equal $1$ when $\vec{x}$ is inside cell $s$ and $0$ otherwise. In the atomic sphere approximation, the Voronoi cells are substituted by atomic spheres with radii $R_{s}$. $r_{s} \equiv \abs{\vec{r}_{s}}$ and $\hat{\vec{r}}_{s} \equiv \vec{r}_{s}/r_{s}$. $\xi_{\sqbr{a,b}}\fbr{r}$ is an invertible function mapping interval $\sqbr{a,b}$ into interval $\sqbr{-1,1}$. The additional multiplier $r^{-1}$ in (\ref{Eq:basis}) improves the convergence properties of the basis and is convenient when solving the susceptibility Dyson equation. The composite index $\lambda=s\mu lm$ determines a supermatrix structure of the susceptibility.

The Y-Ch basis offers a complete, accurate, and efficient representation of the spatial dependencies of the susceptibilities. Compared to other approaches \cite{Kotani2008,Lounis2010,Lounis2011} no assumptions regarding the nature of orbitals responsible for magnetism are necessary: \textit{all} orbitals are included on an equal footing. Also the full spatial dependence of the exchange-correlation kernel is taken into account. The number of necessary Chebyshev polynomials per site per spherical harmonic needed for accurate representation of the susceptibilities for systems considered in this work varies between 8 and 16. The basis functions are localized on atomic sites and, unlike the plane wave basis, can be used equally well for representing the spatial dependencies in the periodic solids, at surfaces and interfaces and in finite clusters of atoms.

In the case when the cells are approximated with spheres, the basis functions are orthonormal regarding $slm$ indices but not the Chebyshev index $\mu$. Prior to eigenvalue analysis it is convenient to transform the susceptibility matrices into an orthonormal representation. We use the L\"owdin transformation \cite{Lowdin1950,Mayer2002} based on the matrix square root algorithm of Denman and Beavers \cite{Denman1976}.

For systems featuring a discrete translational invariance, it is convenient to express the quantities using the following mixed $\vec{r}$-$\vec{q}$ representation. We define the lattice Fourier transformation
\begin{align}
  f(\vec{r},\vec{q}) = \sum_{\vec{R}} f(\vec{r}+\vec{R})\ee{-\ii\vec{q}\cdot\vec{R}},
\end{align}
where the summation proceeds over the crystal lattice and $\vec{r}$ belongs to the Wigner-Seitz cell $\WS$ of the crystal. $\vec{q}$ is a vector in the first Brillouin zone $\Omega_{\mathrm{BZ}}$. The retarded susceptibility $\chi^{ij}(\vec{r},\vec{r}',\vec{q})$ relates the components of the external field and the induced density (the frequency argument has been suppressed)
\begin{align}
  \delta n^{i}(\vec{r},\vec{q}) =
    \sum_{j}\int_{\WS}d\vec{r}'\chi^{ij}(\vec{r},\vec{r}',\vec{q})\Xi^{j}(\vec{r}',\vec{q}),
\end{align}
where
\begin{align}
  \chi(\vec{r},\vec{r}',\vec{q}) = \sum_{\vec{R}} \chi(\vec{r} + \vec{R},\vec{r'})\ee{-\ii\vec{q}\cdot\vec{R}}.
  \label{eq:latticeFTchi}
\end{align}
The $\fbr{\vec{r},\vec{r}'}$ dependency above is given in Y-Ch basis.

The inelastic scattering experiments probe the imaginary part of the Fourier transformed susceptibility \cite{VanHove1954}, obtained by projecting it on the plane waves
\begin{align}
  \varphi_{\vec{K}}\fbr{\vec{r}} \equiv \frac{1}{\WS^{1/2}} \ee{\ii\vec{K}\cdot\vec{r}},
\end{align}
where $\vec{r}\in\WS$ and $\vec{K}$ is the reciprocal lattice vector. The Fourier transformation is defined now as
\begin{align}
  \chi_{\vec{K}\vec{K}'}\fbr{\vec{q}} \equiv \WS^{-1}
  \iint_{\WS^{2}} d \vec{r} d \vec{r}'
    \ee{ - \ii\fbr{\vec{q} + \vec{K}}\cdot\vec{r}}
    \chi\fbr{\vec{r},\vec{r}',\vec{q}}
    \ee{   \ii\fbr{\vec{q} + \vec{K}'}\cdot\vec{r}'}.
\end{align}
$\vec{K},\vec{K}'$ are reciprocal lattice vectors and $\vec{q}\in\BZ$. The definition is consistent in the sense that in uniform systems one obtains
\begin{align}
  \chi_{\vec{K}\vec{K}'}\fbr{\vec{q}} = \tilde{\chi}\fbr{\vec{q} + \vec{K}} \delta_{\vec{K}\vec{K}'}, \quad
  \tilde{\chi}\fbr{\vec{q}} = \int d \vec{x} \chi\fbr{\vec{x}} \ee{ - \ii\vec{q}\cdot\vec{x}}.
\end{align}

\section{Evaluation of the integral in Eq.~\eqref{eq:Schmalian} and analytic continuation}
\label{app:ComplexIntegration}

Our purpose is to transform the integral in Eq.\ \eqref{eq:Schmalian} in such a way that in the complex integration contours as little as possible evaluations of KKRGF are performed close to the real axis.

This can be easily achieved in the case when $\omega = 0$ and $\gamma = 2n\pi\kB T\equiv \omb{n}, n\in\Zahl$, as equation \eqref{eq:Schmalian} reduces to the Matsubara susceptibility evaluated at Bosonic frequencies \cite{Fetter1971,Staunton1999}
\begin{align}
  \chi^{ij}\KS&\fbr{\vec{x},\vec{x}',\ii\omb{n}} = \frac{1}{\beta} \sum_{n\in\Zahl}
    S_{ij}\fbr{\vec{x},\vec{x}',\theta_{m},\theta_{m-n}},
\label{eq:BosonicKS}
\end{align}
where $\theta_{n} \equiv  \mu + \ii\omf{n}$, $\mu$ stands for the chemical potential and $\omf{n}\equiv\fbr{2n + 1}\pi\kB T$ is the fermionic Matsubara frequency. We remark that the temperature is introduced everywhere in this work for the computational expediency only, in order to smooth the discontinuity of the Fermi-Dirac distribution. $T$ does not influence the results as long as it remains much smaller than the characteristic band width. \cite{Wildberger1995}

\begin{figure}
  \centering
  \includegraphics[width=0.4\textwidth]{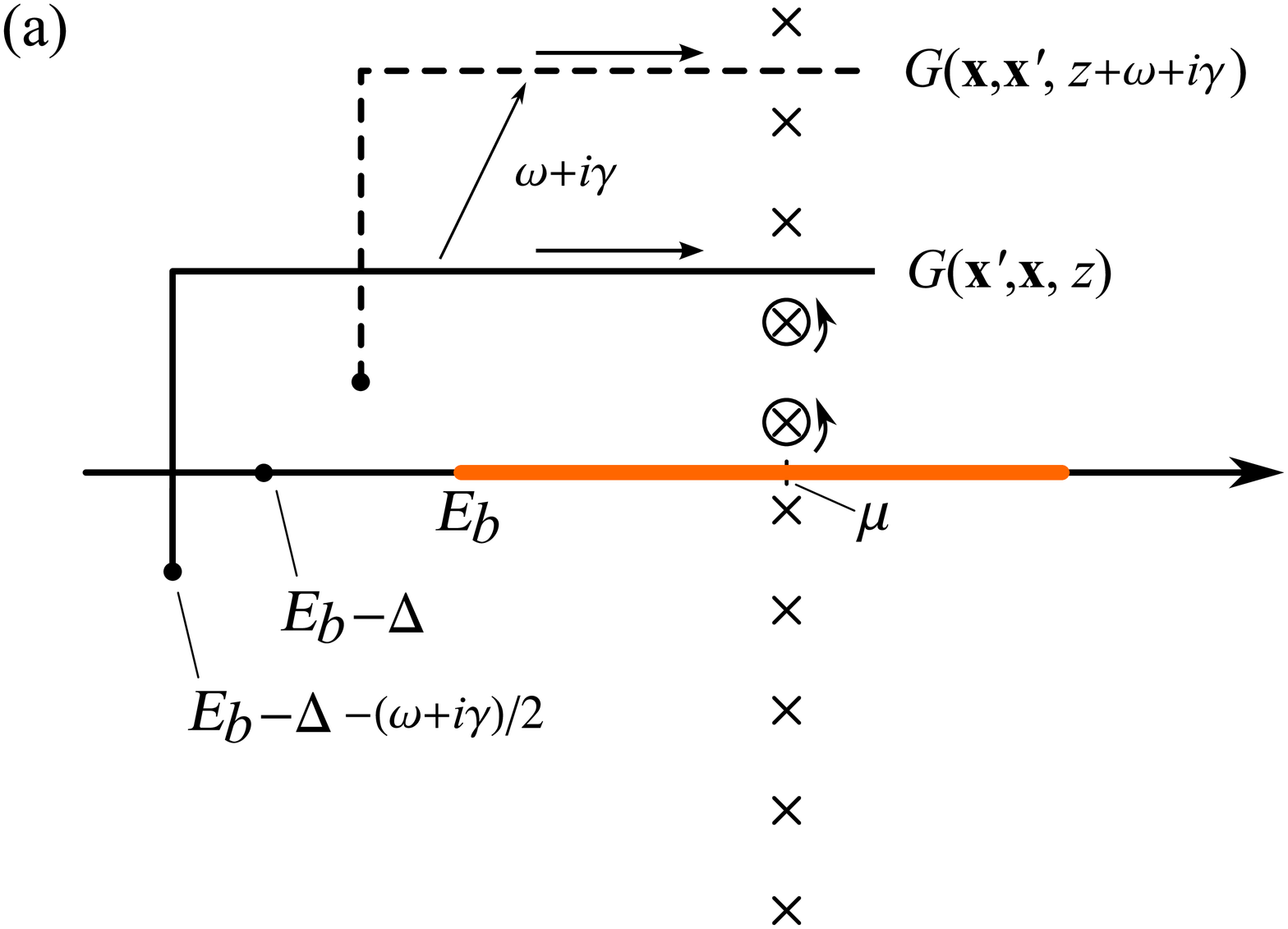} % 19a
  \includegraphics[width=0.4\textwidth]{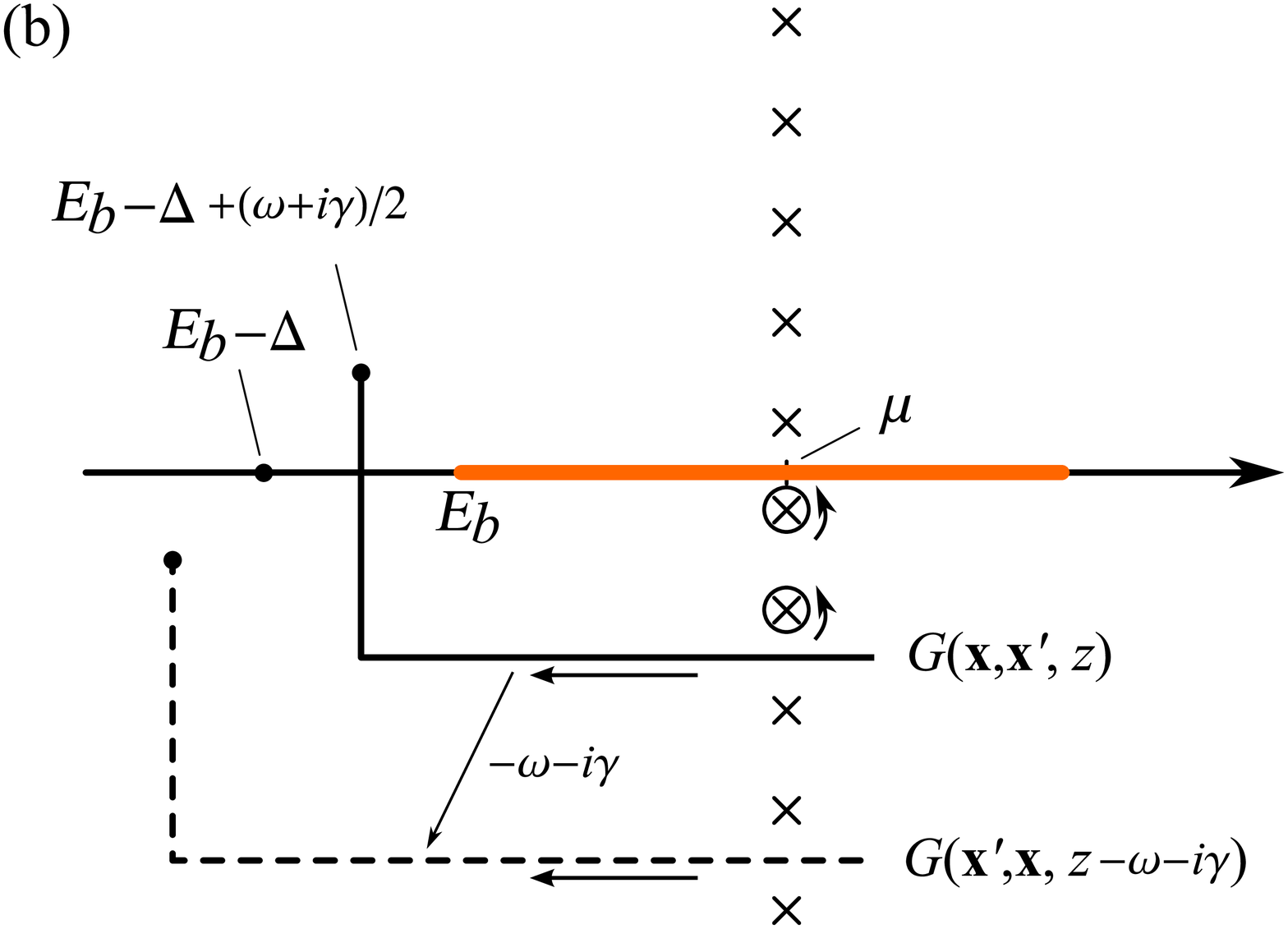} % 19b
  \caption{Evaluating of the first (a) and the fourth term (b) in Eq.~\eqref{eq:Schmalian} around the valence band. Crosses ($\pmb{\times}$) denote the positions of the Fermionic poles $\theta_{n}$. Part of the integration is transformed into the sums over Fermionic frequencies (small circular contours around selected $\theta_{n}$). The integration weighted with the Fermi-Dirac distribution function \cite{Wildberger1995} can be approximated close to the chemical potential $\mu$ using Sommerfeld expansion. This way of integration allows to avoid evaluation of KKRGF close to its singularities originating from the valence states; the valence band is marked with thick orange line. The sum of the integrals a, b and the one given by Eq.~\eqref{eq:belowEBcontrib} gives joint contribution of the first and fourth term in Eq.~\eqref{eq:Schmalian}.}
  \label{fig:term_14}
\end{figure}

The case $\omega\neq0$ is more difficult to handle. Here we use the procedure that we refer to as \textit{nearly real axis approach}. \cite{Schmalian1996} We take advantage of the periodicity of $\fFD\fbr{z}$ on the imaginary plane and consider only the case when $\gamma = \omb{2M},\Zahl\ni M>0$. The first and the last terms of Eq.~\eqref{eq:Schmalian} can be computed rather straightforwardly, since the Green's functions involved are evaluated on the same complex semi-planes (upper in the case of the first term and the lower for the fourth term). The integration contours can be deformed in order to perform the integration away from the singularities corresponding to valence states, as presented in Fig.~\ref{fig:term_14}. For energies $\eps$ well below the bottom of the valence band $E_{b}$ and above the core states $E_{c}$ the respective contours can be taken parallel to the real axis; in this case their joint contribution reads
\begin{align}
  \frac{\ii}{2\pi} \int_{E_{c} + \Delta}^{E_{b} - \Delta} d\eps
  \fbr{\fFD\fbr{\eps - \frac{\omega}{2} - \ii\frac{\gamma}{2}} - \fFD\fbr{\eps + \frac{\omega}{2} + \ii\frac{\gamma}{2}}}
     S_{ij}\fbr{\vec{x},\vec{x}',\eps + \frac{\omega}{2} + \ii\frac{\gamma}{2},\eps - \frac{\omega}{2} - \ii\frac{\gamma}{2}},
  \label{eq:belowEBcontrib}
\end{align}
where $\Delta>\abs{\omega/2}$. In the vicinity of the core states the integration contours must be deformed again so the KKRGF evaluation is performed away from the singularities. The contribution given by Eq.~\eqref{eq:belowEBcontrib} vanishes rigorously for $\omega = 0$ and $\gamma = \omb{n}$. In the case of finite $\omega$ it can still be safely neglected, providing that $\kB T$ and $\omega$ are small.

\begin{figure}
  \centering
  \includegraphics[width=0.4\textwidth]{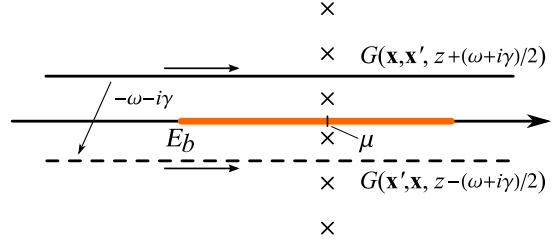} % 20
  \caption{The structure of the integral in Eq.~\eqref{eq:CetralTerm}.}
  \label{fig:term_23}
\end{figure}

The second and the third terms cancel each other only for $\omega=0$ and $\gamma = \omb{n}$. In the $\omega\neq0$ case they are quite intricate to compute, since they involve Green's functions evaluated simultaneously on both complex semi-planes. To minimize the computational effort we rewrite the joint contribution of the second and third terms as follows
\begin{align}
  \frac{\ii}{2\pi} \int_{-\infty}^{\infty} d\eps
    \fbr{\fFD\fbr{\eps + \frac{\omega}{2}} - \fFD\fbr{\eps - \frac{\omega}{2}}}
    S_{ij}\fbr{\vec{x},\vec{x}',\eps + \frac{\omega}{2} + \ii\frac{\gamma}{2},\eps - \frac{\omega}{2} - \ii\frac{\gamma}{2}}\nno\\
  +\frac{1}{\beta}\sum_{m = - M}^{-1} S_{ij}\fbr{\vec{x},\vec{x}',\theta_{m} + \omega + \ii\gamma,\theta_{m}}
  +\frac{1}{\beta}\sum_{m = 0}^{M - 1}S_{ij}\fbr{\vec{x},\vec{x}',\theta_{m},\theta_{m} - \omega - \ii\gamma}.
\label{eq:CetralTerm}
\end{align}
The structure of the integral is presented in Fig.~\ref{fig:term_23}; it has been ``symmetrized'' with respect to the real axis. The evaluations of both GFs are performed now as far as possible from the Kohn-Sham poles. When resorting to Sommerfeld expansion, the integration is reduced to a finite range $\eps\in\sqbr{-\frac{\omega}{2},\frac{\omega}{2}}$. The two Matsubara sums are computed directly term by term. The integration involves now evaluating the Green's function at distance $\gamma/2$ from the real axis and for sufficiently large $M$ rapid convergence can be achieved.

\begin{figure}
  \centering
  \includegraphics[width=0.4\textwidth]{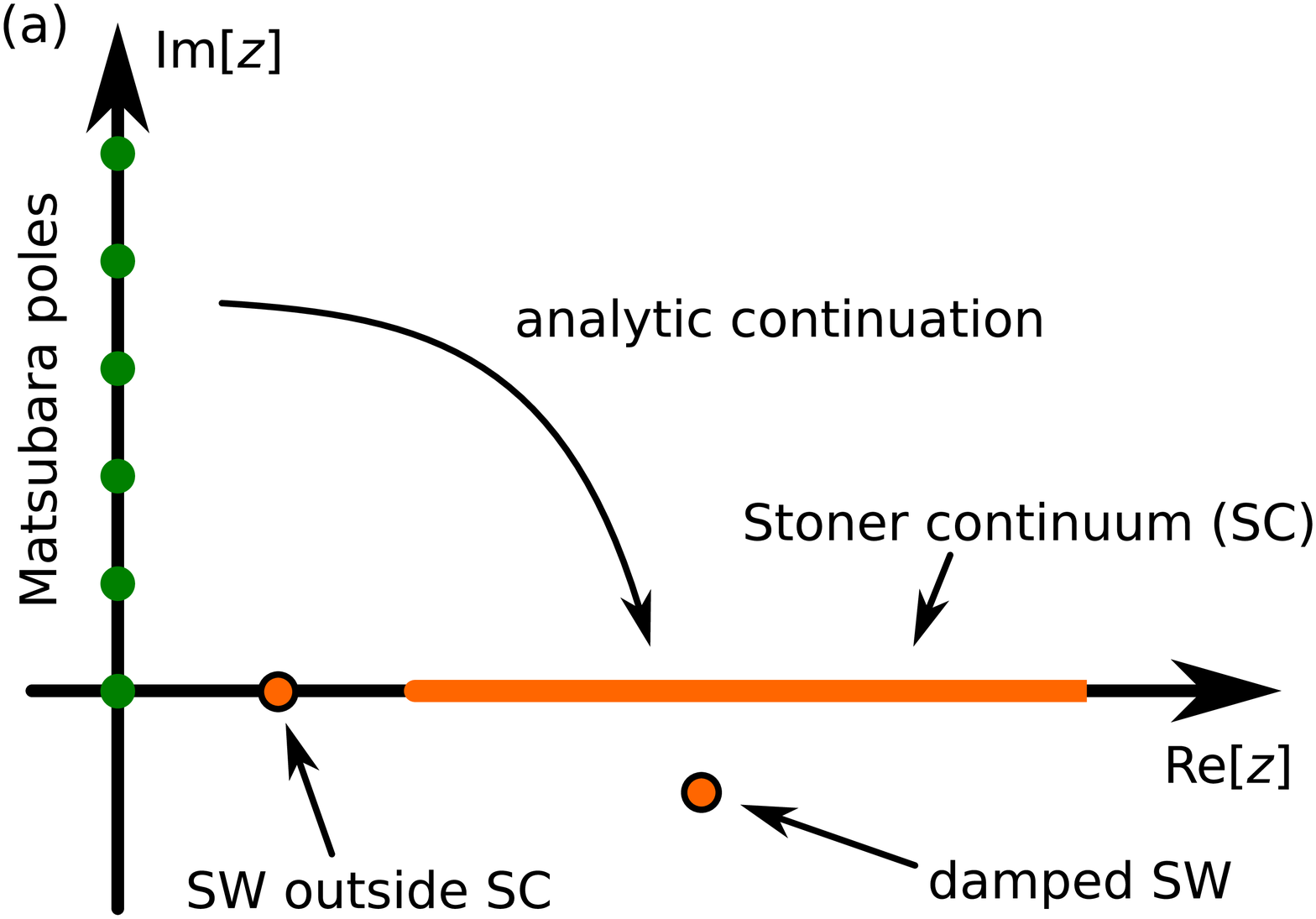} % 21a
  \includegraphics[width=0.4\textwidth]{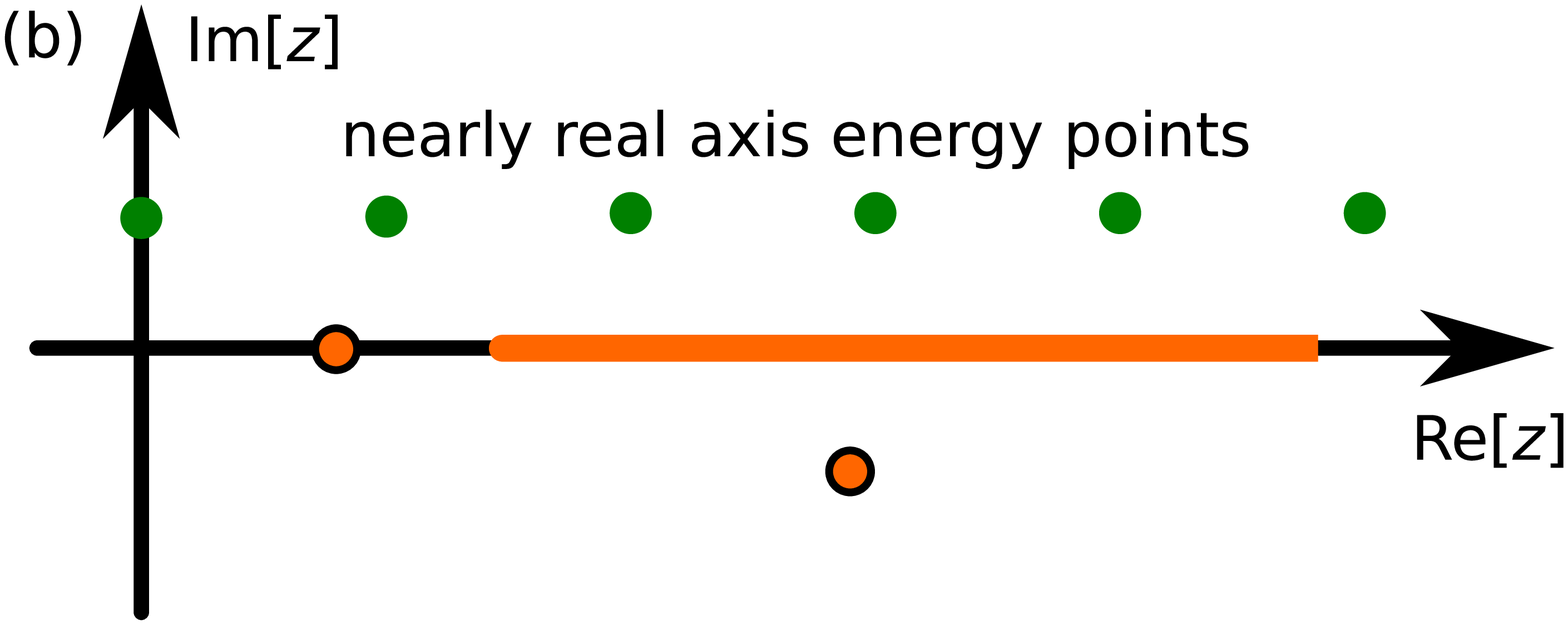} % 21b
  \caption{The analytic structure of transverse susceptibility and schematic presentations of different analytic continuation schemes. For particular $\vec{q}$ the singularities of KS susceptibility form Stoner continuum (SC). The additional singularities introduced into the enhanced susceptibility by the Dyson equation, i.e.\ spin-waves, can appear outside the continuum; such magnons cannot decay via Landau mechanism. On the contrary, when the SW pole appears in the Stoner continuum it acquires a finite life time manifested by an apparent shift of the pole into the lower complex semi-plane. The analytic continuation in the case of the temperature susceptibility (a) involves the reconstruction of the real time dynamics based on the values given in points located on the imaginary axis; it is in general unstable. (b) Nearly real axis calculations lead to much better stability and accuracy of the analytic continuation.}
  \label{fig:AC}
\end{figure}

It is well known that the analytic continuation poses a tricky and potentially unstable numerical problem. \cite{Beach2000} In the context of this work one faces in general two contradicting requirements. The evaluation of temperature susceptibility given by eq.\ \eqref{eq:BosonicKS}, corresponding to the purely imaginary frequency in eq.\ \eqref{eq:Schmalian}, is much easier to implement and numerically much faster. Unfortunately, the subsequent analytic continuation becomes pathologically unstable, since the distance between the points of the complex plane where the susceptibility is actually evaluated and the points on the real axis where we want to determine it by means of the analytical continuation is large, cf.\ Fig.\ \ref{fig:AC}a. In fact, the continuation cannot be performed without assuming certain explicit analytical form of the susceptibility, as e.g.\ in Ref.\ \cite{Staunton1999}. It is much more beneficial to work with $\omega \ne 0$, Fig.\ \ref{fig:AC}b. In this case, the computational effort increases with decreasing $\gamma$, since much denser sampling in the Brillouin zone integration is necessary. Smaller $\gamma$, however, stabilizes greatly the subsequent analytic continuation procedure.

To perform numerical analytic continuation, in most cases we employ a rational function (Pad\'e) approximation \cite{Eschrig1986,Lee1996,Beach2000}, where a complex function $f\fbr{z}$ is represented by a ratio of two polynomials. Alternatively we resort to the method of Haas, Velick\'y and Ehrenreich. \cite{Hass1984,Gray1986}

\section{Explicit form of products of two KKR Green's functions}
\label{app:KKRGFProduct}

By solving the multiple scattering problem in the atomic sphere approximation, the Korringa-Kohn-Rostoker (KKR) Green's function (GF), eq.\ \eqref{eq:BareGreensFunction}, is obtained in the following representation
\begin{align}
  G_{\sig}\fbr{\vec{x},\vec{x}',z} = \sqrt{z} \sum_{L}
  H_{\sig L}^{m}\fbr{\vec{r}_>,z} R_{\sig L}^{m}\fbr{\vec{r}_<,z}
  \delta_{mn} +
  \sum_{LL'} R_{\sig L}^{m}\fbr{\vec{r}_{m},z}
  G_{\sig LL'}^{mn}\fbr{z} R_{\sig L'}^{n}\fbr{\vec{r}'_{n},z},
\label{eq:KKRGF}
\end{align}
where $\sig = \fbr{\alpha,\beta}$ is the spin index. In the case of collinear magnets, only the two diagonal spin components of the GF are non-zero ($\alpha = \beta = \;\uparrow,\downarrow$). $\vec{r}_{m} \equiv \vec{x} - s_{m}$, where $s_{m}$ is the position of the atomic site closest to $\vec{x}$, $\vec{r}_{<}$ denotes the one of two vectors $\vec{r}_{m},\vec{r}_{m}'$ being shorter and $\vec{r}_{>}$ the longer one, $R_{L}^{m}\fbr{\vec{r},z} = R_{L}^{m}\fbr{r,z}Y_{L}\fbr{\hat{\vec{r}}}$ and $H_{L}^{m}\fbr{\vec{r},z} = R_{L}^{m}\fbr{r}Y_L\fbr{\hat{\vec{r}}}$ are regular and irregular solutions of the radial Schr\"odinger equation for atomic site $m$. The first term in Eq.~\eqref{eq:KKRGF} represents the single-scattering GF, while the second term describes multiple-scattering processes via the back-scattering operator $G_{\sig LL'}^{\alpha\beta}\fbr{z}$, which can be computed from the algebraic KKR Dyson equation
\begin{align}
  G_{\sig LL'}^{mn}\fbr{z} = g_{LL'}^{mn}\fbr{z} +
  \sum\limits_{k, L''} g_{LL''}^{mk} \fbr{z}
  t_{\sig L''}^{k}\fbr{z}  G_{\sig L''L'}^{kn}\fbr{z},
 \label{eq:KKRequation}
\end{align}
where $g_{LL'}^{mn}$ are the KKR structure constants and $t_{\sig L}^{m}\fbr{z}$ is the single-site scattering matrix. We note that the computational method presented in this paper is trivially generalizable to the full potential treatment.

In periodic systems, the product of two Green's functions appearing in Eq.~\eqref{eq:Schmalian} involves additionally the convolution over the Brillouin zone
\begin{align}
  S_{\sigma_{1}\sigma_{2}}\fbr{\vec{r},\vec{r}',\vec{q},z_{1},z_{2}} = 
    \sigma^{i}_{\alpha\beta} \sigma^{j}_{\gamma\delta}
    \int_{\Omega_{\mathrm{BZ}}} \frac{d^{D} \vec{k}}{\Omega_{\mathrm{BZ}}}
    G_{\sigma_{1}}\fbr{\vec{r} ,\vec{r}',\vec{k}          ,z_{1}}
    G_{\sigma_{2}}\fbr{\vec{r}',\vec{r} ,\vec{k} - \vec{q},z_{2}}.
\label{eq:Sijq}
\end{align}
$D$ stands for the dimensionality of the periodic lattice. The number of necessary integration $\vec{k}-$points decreases rapidly as one moves away from the singularities of KKRGF, i.e.\ Kohn-Sham energies located on the real axis.

Upon substituting the KKR form of $G$ in equation \eqref{eq:Sijq} we obtain
\begin{align}
\frac{1}{\BZ}&\int_{\BZ}d^{D}\vec{k}
      \bareG_{\sig_{1}}\fbr{\vec{r} ,\vec{r}',\vec{k}          ,z_{1}}
      \bareG_{\sig_{2}}\fbr{\vec{r}',\vec{r} ,\vec{k} - \vec{q},z_{2}} \nno \\
  &=\sum_{L_{1}L_{2}L_{3}L_{4}}
    Y_{L_{1}}\fbr{\hat{\vec{r}} _{m}}
    Y_{L_{2}}\fbr{\hat{\vec{r}}'_{n}}
    Y_{L_{3}}\fbr{\hat{\vec{r}}'_{n}}
    Y_{L_{4}}\fbr{\hat{\vec{r}} _{m}} \times \nno \\
    ( &^{\sig_{1}\sig_{2}}C_{L_{1}L_{2}L_{3}L_{4}}^{mn}\fbr{z_{1},z_{2},\vec{q}}
        R^{m}_{\sig_{1}L_{1}}\fbr{r_{m},z_{1}}
        R^{n}_{\sig_{1}L_{2}}\fbr{r'_{n},z_{1}}
        R^{n}_{\sig_{2}L_{3}}\fbr{r'_{n},z_{2}}
        R^{m}_{\sig_{2}L_{4}}\fbr{r_{m},z_{2}} + \nno \\
      &\delta_{mn}\delta_{L_{1}L_{2}}\sqrt{z_{1}}
        R^{m}_{\sig_{1}L_{1}}\fbr{r_{<},z_{1}}
        H^{m}_{\sig_{1}L_{1}}\fbr{r_{>},z_{1}}
        B_{\sig_{2}L_{3}L_{4}}^{m}\fbr{z_{2}}
        R^{m}_{\sig_{2}L_{3}}\fbr{r'_{m},z_{2}}
        R^{m}_{\sig_{2}L_{4}}\fbr{r _{m},z_{2}} + \nno \\
      &\delta_{mn}\delta_{L_{3}L_{4}}
        B_{\sig_{1}L_{1}L_{2}}^{m}\fbr{z_{1}}
        R^{m}_{\sig_{1}L_{1}}\fbr{r _{m},z_{1}}
        R^{m}_{\sig_{1}L_{2}}\fbr{r'_{m},z_{1}}
        \sqrt{z_{2}}
        R^{m}_{\sig_{2}L_{3}}\fbr{r_{<},z_{2}}
        H^{m}_{\sig_{2}L_{3}}\fbr{r_{>},z_{2}} + \nno \\
      &\delta_{mn}\delta_{L_{1}L_{2}}\delta_{L_{3}L_{4}}\sqrt{z_{1}z_{2}}
        R^{m}_{\sig_{1}L_{1}}\fbr{r_{<},z_{1}}
        H^{m}_{\sig_{1}L_{1}}\fbr{r_{>},z_{1}}
        R^{m}_{\sig_{2}L_{3}}\fbr{r_{<},z_{2}}
        H^{m}_{\sig_{2}L_{3}}\fbr{r_{>},z_{2}} ).
\end{align}
The first term comes from the convolution of two backscattering operators
\begin{align}
  ^{\sig_{1}\sig_{2}}C_{L_{1}L_{2}L_{3}L_{4}}^{mn}\fbr{z_{1},z_{2},\vec{q}} =
    \frac{1}{\BZ}&\int_{\BZ}d^{D}\vec{k}
      G^{mn}_{\sig_{1}L_{1}L_{2}}\fbr{z_{1},\vec{k}}
      G^{nm}_{\sig_{2}L_{3}L_{4}}\fbr{z_{2},\vec{k} - \vec{q}},
\end{align}
while the next two terms involve only diagonal part of it
\begin{align}
  B_{\sig LL'}^{m}\fbr{z} =
    \frac{1}{\BZ}&\int_{\BZ}d^{D}\vec{k}
      G^{mm}_{\sig LL'}\fbr{z,\vec{k}}.
\end{align}

By means of Gaunt coefficients the products of four spherical harmonics are reduced to pairs $Y_{L}\fbr{\hat{\vec{r}}}Y_{L'}\fbr{\hat{\vec{r}}'}$. The remaining radial dependence is approximated using Chebyshev polynomials. This gives the representation of the susceptibility in the Y-Ch basis.

\input{bibliography}
\end{document}

%% file: headers.tex
\usepackage[T1]{fontenc}

\usepackage[thinspace,squaren,pstricks,italian,derivedinbase,derived]{SIunits}
\addunit{\bohr}{au}
\addunit{\rydberg}{Ry}
\addunit{\chwilka}{ch}
\addunit{\kubo}{Kb}

\usepackage[latin2]{inputenc}
\usepackage[dvips]{epsfig}
\usepackage{amsfonts}
\usepackage{amsmath}
\usepackage{amssymb}
\usepackage{upgreek}  % upgreek fonts
\usepackage{mathrsfs}
\usepackage{booktabs} % nice tables
\usepackage{setspace}

%% file: commands.tex
\newcommand{\pder}[2]
{
{\frac{\partial #1}{\partial #2}}
}
\renewcommand{\vec}[1]
{
{\mathbf #1}
}
\newcommand{\abs}[1]
{
{\left| #1 \right|} 
}
\newcommand{\kB}
{
{k_{\textrm{B}}}
}
\newcommand{\Bm}
{
{\mu_{\textrm{B}}}
}

\newcommand{\nno}
{
{\nonumber}
}
\newcommand{\eps}
{
{\epsilon}
}
\newcommand{\sig}
{
{\sigma}
}
\newcommand{\onemat}
{
{\sf I}     %sf version 
}
\newcommand{\Heav}
{
{\theta}
}
\renewcommand{\Re}%[1]
{
{\mathrm{Re}}%\sqbr{#1}}
}
\renewcommand{\Im}%[1]
{
{\mathrm{Im}}%\sqbr{#1}}
}
\newcommand{\Loss}%[1]
{
{\mathcal{L}}%\sqbr{#1}}
}
\newcommand{\ii}
{
{{\mathfrak{i}}} %Euler version
}
\newcommand{\ee}[1]
{
{\mathfrak{e}^{#1}}   %bold version
}
\newcommand{\av}[1]
{
{\left\langle #1 \right\rangle} 
}
\newcommand{\sqbr}[1]
{
{\left[ #1 \right]}
}
\newcommand{\fbr}[1]
{
{\left( #1 \right)} 
}
\newcommand{\Complex}
{
{\mathbb{C}} 
}
\newcommand{\Reals}
{
{\mathbb{R}} 
}
\newcommand{\Zahl}
{
{\mathbb{Z}} 
}
\newcommand{\omb}[1]
{
{\omega_{#1}^{b}}
}
\newcommand{\omf}[1]
{
{\omega_{#1}^{f}}
}
\newcommand{\zoneSymbol}
{
\Omega
}
\newcommand{\WS}
{
\zoneSymbol_{\textrm{WS}}
}
\newcommand{\BZ}
{
\zoneSymbol_{\textrm{BZ}}
}
\newcommand{\bareG}
{
{G}
}
\newcommand{\comm}[3]
{
{\sqbr{#1,#2}_{#3}}
}
\newcommand{\dOp}
{{c}}
\newcommand{\cOp}
{
c^{\dagger}
}
\newcommand{\fop}
{
{\widehat{\uppsi}}
}
\newcommand{\fopc}
{
{\widehat{\uppsi}^{\dagger}}
}
\newcommand{\Psigma}
{
{\pmb{\sigma}}
}
\newcommand{\fFD}
{
{f_{T}}
}
\newcommand{\zp}
{
{0^{+}}
}

\newcommand{\bra}[1]
{{\left\langle #1 \right|}}
\newcommand{\ket}[1]
{{\left| #1 \right\rangle}}
\newcommand{\Ylm}
{
{\mathcal{Y}}
}
\newcommand{\Ch}
{
{\mathfrak{Ch}}
}
\newcommand{\xc}
{
_{\mathrm{xc}}
}
\newcommand{\KS}
{
_{\mathrm{KS}}
}
\newcommand{\GS}
{
_{\mathrm{GS}}
}
\newcommand{\ex}
{
\mathrm{ex}
}
\newcommand{\el}
{%
\textit{et~al}.\
}
\newcommand{\ml}[1]{$\textrm{#1}\usk\textrm{ML}$}

%% file: text.bbl
\begin{thebibliography}{133}
\expandafter\ifx\csname natexlab\endcsname\relax\def\natexlab#1{#1}\fi
\expandafter\ifx\csname bibnamefont\endcsname\relax
  \def\bibnamefont#1{#1}\fi
\expandafter\ifx\csname bibfnamefont\endcsname\relax
  \def\bibfnamefont#1{#1}\fi
\expandafter\ifx\csname citenamefont\endcsname\relax
  \def\citenamefont#1{#1}\fi
\expandafter\ifx\csname url\endcsname\relax
  \def\url#1{\texttt{#1}}\fi
\expandafter\ifx\csname urlprefix\endcsname\relax\def\urlprefix{URL }\fi
\providecommand{\bibinfo}[2]{#2}
\providecommand{\eprint}[2][]{\url{#2}}

\bibitem[{\citenamefont{Moriya}(1985)}]{Moriya1985}
\bibinfo{author}{\bibfnamefont{T.}~\bibnamefont{Moriya}},
  \emph{\bibinfo{title}{Spin fluctuations in itinerant electron magnetism}},
  vol.~\bibinfo{volume}{56} of \emph{\bibinfo{series}{Springer series in
  solid-state sciences}} (\bibinfo{publisher}{Springer},
  \bibinfo{address}{Berlin}, \bibinfo{year}{1985}).

\bibitem[{\citenamefont{Pajda et~al.}(2001)\citenamefont{Pajda, Kudrnovsk\'y,
  Turek, Drchal, and Bruno}}]{Pajda2001}
\bibinfo{author}{\bibfnamefont{M.}~\bibnamefont{Pajda}},
  \bibinfo{author}{\bibfnamefont{J.}~\bibnamefont{Kudrnovsk\'y}},
  \bibinfo{author}{\bibfnamefont{I.}~\bibnamefont{Turek}},
  \bibinfo{author}{\bibfnamefont{V.}~\bibnamefont{Drchal}}, \bibnamefont{and}
  \bibinfo{author}{\bibfnamefont{P.}~\bibnamefont{Bruno}},
  \bibinfo{journal}{Phys. Rev. B} \textbf{\bibinfo{volume}{64}},
  \bibinfo{pages}{174402} (\bibinfo{year}{2001}),
  \urlprefix\url{http://link.aps.org/abstract/PRB/v64/e174402}.

\bibitem[{\citenamefont{Rusz et~al.}(2005)\citenamefont{Rusz, Turek, and
  Divi\v{s}}}]{Rusz2005}
\bibinfo{author}{\bibfnamefont{J.}~\bibnamefont{Rusz}},
  \bibinfo{author}{\bibfnamefont{I.}~\bibnamefont{Turek}}, \bibnamefont{and}
  \bibinfo{author}{\bibfnamefont{M.}~\bibnamefont{Divi\v{s}}},
  \bibinfo{journal}{Phys. Rev. B} \textbf{\bibinfo{volume}{71}},
  \bibinfo{pages}{174408} (\bibinfo{year}{2005}),
  \urlprefix\url{http://prola.aps.org/abstract/PRB/v71/i17/e174408}.

\bibitem[{\citenamefont{Doniach and Engelsberg}(1966)}]{Doniach1966}
\bibinfo{author}{\bibfnamefont{S.}~\bibnamefont{Doniach}} \bibnamefont{and}
  \bibinfo{author}{\bibfnamefont{S.}~\bibnamefont{Engelsberg}},
  \bibinfo{journal}{Phys. Rev. Lett.} \textbf{\bibinfo{volume}{17}},
  \bibinfo{pages}{750} (\bibinfo{year}{1966}),
  \urlprefix\url{http://link.aps.org/abstract/PRL/v17/p750}.

\bibitem[{\citenamefont{Vignale and Singwi}(1985)}]{Vignale1985}
\bibinfo{author}{\bibfnamefont{G.}~\bibnamefont{Vignale}} \bibnamefont{and}
  \bibinfo{author}{\bibfnamefont{K.~S.} \bibnamefont{Singwi}},
  \bibinfo{journal}{Phys. Rev. B} \textbf{\bibinfo{volume}{32}},
  \bibinfo{pages}{2824} (\bibinfo{year}{1985}),
  \urlprefix\url{http://link.aps.org/abstract/PRB/v32/p2824}.

\bibitem[{\citenamefont{Hertz and Edwards}(1972)}]{Hertz1972}
\bibinfo{author}{\bibfnamefont{J.~A.} \bibnamefont{Hertz}} \bibnamefont{and}
  \bibinfo{author}{\bibfnamefont{D.~M.} \bibnamefont{Edwards}},
  \bibinfo{journal}{Phys. Rev. Lett.} \textbf{\bibinfo{volume}{28}},
  \bibinfo{pages}{1334} (\bibinfo{year}{1972}),
  \urlprefix\url{http://link.aps.org/abstract/PRL/v28/p1334}.

\bibitem[{\citenamefont{Hertz and Edwards}(1973)}]{Hertz1973}
\bibinfo{author}{\bibfnamefont{J.~A.} \bibnamefont{Hertz}} \bibnamefont{and}
  \bibinfo{author}{\bibfnamefont{D.~M.} \bibnamefont{Edwards}},
  \bibinfo{journal}{Journal of Physics F: Metal Physics}
  \textbf{\bibinfo{volume}{3}}, \bibinfo{pages}{2174} (\bibinfo{year}{1973}),
  ISSN \bibinfo{issn}{0305-4608}.

\bibitem[{\citenamefont{Edwards and Hertz}(1973)}]{Edwards1973}
\bibinfo{author}{\bibfnamefont{D.~M.} \bibnamefont{Edwards}} \bibnamefont{and}
  \bibinfo{author}{\bibfnamefont{J.~A.} \bibnamefont{Hertz}},
  \bibinfo{journal}{Journal of Physics F: Metal Physics}
  \textbf{\bibinfo{volume}{3}}, \bibinfo{pages}{2191} (\bibinfo{year}{1973}),
  ISSN \bibinfo{issn}{0305-4608}.

\bibitem[{\citenamefont{Kleinman}(1978)}]{Kleinman1978}
\bibinfo{author}{\bibfnamefont{L.}~\bibnamefont{Kleinman}},
  \bibinfo{journal}{Phys. Rev. B} \textbf{\bibinfo{volume}{17}},
  \bibinfo{pages}{3666} (\bibinfo{year}{1978}),
  \urlprefix\url{http://link.aps.org/abstract/PRB/v17/p3666}.

\bibitem[{\citenamefont{Sch\"afer et~al.}(2004)\citenamefont{Sch\"afer,
  Schrupp, Rotenberg, Rossnagel, Koh, Blaha, and Claessen}}]{Schafer2004}
\bibinfo{author}{\bibfnamefont{J.}~\bibnamefont{Sch\"afer}},
  \bibinfo{author}{\bibfnamefont{D.}~\bibnamefont{Schrupp}},
  \bibinfo{author}{\bibfnamefont{E.}~\bibnamefont{Rotenberg}},
  \bibinfo{author}{\bibfnamefont{K.}~\bibnamefont{Rossnagel}},
  \bibinfo{author}{\bibfnamefont{H.}~\bibnamefont{Koh}},
  \bibinfo{author}{\bibfnamefont{P.}~\bibnamefont{Blaha}}, \bibnamefont{and}
  \bibinfo{author}{\bibfnamefont{R.}~\bibnamefont{Claessen}},
  \bibinfo{journal}{Phys. Rev. Lett.} \textbf{\bibinfo{volume}{92}},
  \bibinfo{pages}{097205} (\bibinfo{year}{2004}),
  \urlprefix\url{http://link.aps.org/abstract/PRL/v92/e097205}.

\bibitem[{\citenamefont{Hofmann et~al.}(2009)\citenamefont{Hofmann, Cui,
  Schafer, Meyer, Hopfner, Blumenstein, Paul, Patthey, Rotenberg, Bunemann
  et~al.}}]{Hofmann2009}
\bibinfo{author}{\bibfnamefont{A.}~\bibnamefont{Hofmann}},
  \bibinfo{author}{\bibfnamefont{X.~Y.} \bibnamefont{Cui}},
  \bibinfo{author}{\bibfnamefont{J.}~\bibnamefont{Schafer}},
  \bibinfo{author}{\bibfnamefont{S.}~\bibnamefont{Meyer}},
  \bibinfo{author}{\bibfnamefont{P.}~\bibnamefont{Hopfner}},
  \bibinfo{author}{\bibfnamefont{C.}~\bibnamefont{Blumenstein}},
  \bibinfo{author}{\bibfnamefont{M.}~\bibnamefont{Paul}},
  \bibinfo{author}{\bibfnamefont{L.}~\bibnamefont{Patthey}},
  \bibinfo{author}{\bibfnamefont{E.}~\bibnamefont{Rotenberg}},
  \bibinfo{author}{\bibfnamefont{J.}~\bibnamefont{Bunemann}},
  \bibnamefont{et~al.}, \bibinfo{journal}{Phys. Rev. Lett.}
  \textbf{\bibinfo{volume}{102}}, \bibinfo{pages}{187204}
  (\bibinfo{year}{2009}),
  \urlprefix\url{http://dx.doi.org/10.1103/PhysRevLett.102.187204}.

\bibitem[{\citenamefont{Schmidt et~al.}(2010)\citenamefont{Schmidt, Pickel,
  Donath, Buczek, Ernst, Zhukov, Echenique, Sandratskii, Chulkov, and
  Weinelt}}]{Schmidt2010a}
\bibinfo{author}{\bibfnamefont{A.~B.} \bibnamefont{Schmidt}},
  \bibinfo{author}{\bibfnamefont{M.}~\bibnamefont{Pickel}},
  \bibinfo{author}{\bibfnamefont{M.}~\bibnamefont{Donath}},
  \bibinfo{author}{\bibfnamefont{P.}~\bibnamefont{Buczek}},
  \bibinfo{author}{\bibfnamefont{A.}~\bibnamefont{Ernst}},
  \bibinfo{author}{\bibfnamefont{V.~P.} \bibnamefont{Zhukov}},
  \bibinfo{author}{\bibfnamefont{P.~M.} \bibnamefont{Echenique}},
  \bibinfo{author}{\bibfnamefont{L.~M.} \bibnamefont{Sandratskii}},
  \bibinfo{author}{\bibfnamefont{E.~V.} \bibnamefont{Chulkov}},
  \bibnamefont{and} \bibinfo{author}{\bibfnamefont{M.}~\bibnamefont{Weinelt}},
  \bibinfo{journal}{Phys. Rev. Lett.} \textbf{\bibinfo{volume}{105}},
  \bibinfo{pages}{197401} (\bibinfo{year}{2010}),
  \urlprefix\url{http://link.aps.org/doi/10.1103/PhysRevLett.105.197401}.

\bibitem[{\citenamefont{Gokhale et~al.}(1992)\citenamefont{Gokhale, Ormeci, and
  Mills}}]{Gokhale1992}
\bibinfo{author}{\bibfnamefont{M.~P.} \bibnamefont{Gokhale}},
  \bibinfo{author}{\bibfnamefont{A.}~\bibnamefont{Ormeci}}, \bibnamefont{and}
  \bibinfo{author}{\bibfnamefont{D.~L.} \bibnamefont{Mills}},
  \bibinfo{journal}{Physical Review B (Condensed Matter)}
  \textbf{\bibinfo{volume}{46}}, \bibinfo{pages}{8978} (\bibinfo{year}{1992}),
  \urlprefix\url{http://link.aps.org/abstract/PRB/v46/p8978}.

\bibitem[{\citenamefont{Plihal and Mills}(1995)}]{Plihal1995}
\bibinfo{author}{\bibfnamefont{M.}~\bibnamefont{Plihal}} \bibnamefont{and}
  \bibinfo{author}{\bibfnamefont{D.~L.} \bibnamefont{Mills}},
  \bibinfo{journal}{Physical Review B (Condensed Matter)}
  \textbf{\bibinfo{volume}{52}}, \bibinfo{pages}{12813} (\bibinfo{year}{1995}),
  \urlprefix\url{http://link.aps.org/abstract/PRB/v52/p12813}.

\bibitem[{\citenamefont{Hong and Mills}(1999)}]{Hong1999}
\bibinfo{author}{\bibfnamefont{J.}~\bibnamefont{Hong}} \bibnamefont{and}
  \bibinfo{author}{\bibfnamefont{D.~L.} \bibnamefont{Mills}},
  \bibinfo{journal}{Phys. Rev. B} \textbf{\bibinfo{volume}{59}},
  \bibinfo{pages}{13840} (\bibinfo{year}{1999}),
  \urlprefix\url{http://link.aps.org/abstract/PRB/v59/p13840}.

\bibitem[{\citenamefont{Hong and Mills}(2000)}]{Hong2000}
\bibinfo{author}{\bibfnamefont{J.}~\bibnamefont{Hong}} \bibnamefont{and}
  \bibinfo{author}{\bibfnamefont{D.~L.} \bibnamefont{Mills}},
  \bibinfo{journal}{Phys. Rev. B} \textbf{\bibinfo{volume}{62}},
  \bibinfo{pages}{5589} (\bibinfo{year}{2000}),
  \urlprefix\url{http://link.aps.org/abstract/PRB/v62/p5589}.

\bibitem[{\citenamefont{Zhukov et~al.}(2004{\natexlab{a}})\citenamefont{Zhukov,
  Chulkov, and Echenique}}]{Zhukov2004}
\bibinfo{author}{\bibfnamefont{V.~P.} \bibnamefont{Zhukov}},
  \bibinfo{author}{\bibfnamefont{E.~V.} \bibnamefont{Chulkov}},
  \bibnamefont{and} \bibinfo{author}{\bibfnamefont{P.~M.}
  \bibnamefont{Echenique}}, \bibinfo{journal}{Phys. Rev. Lett.}
  \textbf{\bibinfo{volume}{93}}, \bibinfo{pages}{096401}
  (\bibinfo{year}{2004}{\natexlab{a}}),
  \urlprefix\url{http://link.aps.org/abstract/PRL/v93/e096401}.

\bibitem[{\citenamefont{Zhukov et~al.}(2004{\natexlab{b}})\citenamefont{Zhukov,
  Chulkov, and Echenique}}]{Zhukov2004a}
\bibinfo{author}{\bibfnamefont{V.~P.} \bibnamefont{Zhukov}},
  \bibinfo{author}{\bibfnamefont{E.~V.} \bibnamefont{Chulkov}},
  \bibnamefont{and} \bibinfo{author}{\bibfnamefont{P.~M.}
  \bibnamefont{Echenique}}, \bibinfo{journal}{Journal of Magnetism and Magnetic
  Materials} \textbf{\bibinfo{volume}{272-276}}, \bibinfo{pages}{466}
  (\bibinfo{year}{2004}{\natexlab{b}}),
  \urlprefix\url{http://www.sciencedirect.com/science/article/B6TJJ-4BM8JNB-1/%
2/adda00211726600793dc144c4db09d15}.

\bibitem[{\citenamefont{Zhukov et~al.}(2005)\citenamefont{Zhukov, Chulkov, and
  Echenique}}]{Zhukov2005}
\bibinfo{author}{\bibfnamefont{V.~P.} \bibnamefont{Zhukov}},
  \bibinfo{author}{\bibfnamefont{E.~V.} \bibnamefont{Chulkov}},
  \bibnamefont{and} \bibinfo{author}{\bibfnamefont{P.~M.}
  \bibnamefont{Echenique}}, \bibinfo{journal}{Phys. Rev. B}
  \textbf{\bibinfo{volume}{72}}, \bibinfo{pages}{155109}
  (\bibinfo{year}{2005}),
  \urlprefix\url{http://link.aps.org/abstract/PRB/v72/e155109}.

\bibitem[{\citenamefont{Scalapino}(1995)}]{Scalapino1995}
\bibinfo{author}{\bibfnamefont{D.~J.} \bibnamefont{Scalapino}},
  \bibinfo{journal}{Physics Reports} \textbf{\bibinfo{volume}{250}},
  \bibinfo{pages}{329} (\bibinfo{year}{1995}),
  \urlprefix\url{http://www.sciencedirect.com/science/article/B6TVP-3YGTS3W-J/%
1/8f0b85026c3ce23b78a8890d6c06558a}.

\bibitem[{\citenamefont{Eschrig}(2006)}]{Eschrig2006}
\bibinfo{author}{\bibfnamefont{M.}~\bibnamefont{Eschrig}},
  \bibinfo{journal}{Advances in Physics} \textbf{\bibinfo{volume}{55}},
  \bibinfo{pages}{47} (\bibinfo{year}{2006}),
  \urlprefix\url{http://www.informaworld.com/10.1080/00018730600645636}.

\bibitem[{\citenamefont{Mazin et~al.}(2008)\citenamefont{Mazin, Singh,
  Johannes, and Du}}]{Mazin2008}
\bibinfo{author}{\bibfnamefont{I.~I.} \bibnamefont{Mazin}},
  \bibinfo{author}{\bibfnamefont{D.~J.} \bibnamefont{Singh}},
  \bibinfo{author}{\bibfnamefont{M.~D.} \bibnamefont{Johannes}},
  \bibnamefont{and} \bibinfo{author}{\bibfnamefont{M.~H.} \bibnamefont{Du}},
  \bibinfo{journal}{Phys. Rev. Lett.} \textbf{\bibinfo{volume}{101}},
  \bibinfo{pages}{057003} (\bibinfo{year}{2008}),
  \urlprefix\url{http://link.aps.org/doi/10.1103/PhysRevLett.101.057003}.

\bibitem[{\citenamefont{Mazin and Johannes}(2009)}]{Mazin2009}
\bibinfo{author}{\bibfnamefont{I.~I.} \bibnamefont{Mazin}} \bibnamefont{and}
  \bibinfo{author}{\bibfnamefont{M.~D.} \bibnamefont{Johannes}},
  \bibinfo{journal}{Nat. Phys.} \textbf{\bibinfo{volume}{5}},
  \bibinfo{pages}{141} (\bibinfo{year}{2009}), ISSN \bibinfo{issn}{1745-2473},
  \urlprefix\url{http://dx.doi.org/10.1038/nphys1160}.

\bibitem[{\citenamefont{\v{Z}uti\'c et~al.}(2004)\citenamefont{\v{Z}uti\'c,
  Fabian, and Das~Sarma}}]{Zutic2004}
\bibinfo{author}{\bibfnamefont{I.}~\bibnamefont{\v{Z}uti\'c}},
  \bibinfo{author}{\bibfnamefont{J.}~\bibnamefont{Fabian}}, \bibnamefont{and}
  \bibinfo{author}{\bibfnamefont{S.}~\bibnamefont{Das~Sarma}},
  \bibinfo{journal}{Rev. Mod. Phys.} \textbf{\bibinfo{volume}{76}},
  \bibinfo{pages}{323} (\bibinfo{year}{2004}),
  \urlprefix\url{http://link.aps.org/abstract/RMP/v76/p323}.

\bibitem[{\citenamefont{Kaka et~al.}(2005)\citenamefont{Kaka, Pufall, Rippard,
  Silva, Russek, and Katine}}]{Kaka2005}
\bibinfo{author}{\bibfnamefont{S.}~\bibnamefont{Kaka}},
  \bibinfo{author}{\bibfnamefont{M.~R.} \bibnamefont{Pufall}},
  \bibinfo{author}{\bibfnamefont{W.~H.} \bibnamefont{Rippard}},
  \bibinfo{author}{\bibfnamefont{T.~J.} \bibnamefont{Silva}},
  \bibinfo{author}{\bibfnamefont{S.~E.} \bibnamefont{Russek}},
  \bibnamefont{and} \bibinfo{author}{\bibfnamefont{J.~A.}
  \bibnamefont{Katine}}, \bibinfo{journal}{Nature}
  \textbf{\bibinfo{volume}{437}}, \bibinfo{pages}{389} (\bibinfo{year}{2005}),
  ISSN \bibinfo{issn}{0028-0836},
  \urlprefix\url{http://dx.doi.org/10.1038/nature04035}.

\bibitem[{\citenamefont{Neusser and Grundler}(2009)}]{Neusser2009}
\bibinfo{author}{\bibfnamefont{S.}~\bibnamefont{Neusser}} \bibnamefont{and}
  \bibinfo{author}{\bibfnamefont{D.}~\bibnamefont{Grundler}},
  \bibinfo{journal}{Advanced Materials} \textbf{\bibinfo{volume}{21}},
  \bibinfo{pages}{2927} (\bibinfo{year}{2009}),
  \urlprefix\url{http://dx.doi.org/10.1002/adma.200900809}.

\bibitem[{\citenamefont{Kajiwara et~al.}(2010)\citenamefont{Kajiwara, Harii,
  Takahashi, Ohe, Uchida, Mizuguchi, Umezawa, Kawai, Ando, Takanashi
  et~al.}}]{Kajiwara2010}
\bibinfo{author}{\bibfnamefont{Y.}~\bibnamefont{Kajiwara}},
  \bibinfo{author}{\bibfnamefont{K.}~\bibnamefont{Harii}},
  \bibinfo{author}{\bibfnamefont{S.}~\bibnamefont{Takahashi}},
  \bibinfo{author}{\bibfnamefont{J.}~\bibnamefont{Ohe}},
  \bibinfo{author}{\bibfnamefont{K.}~\bibnamefont{Uchida}},
  \bibinfo{author}{\bibfnamefont{M.}~\bibnamefont{Mizuguchi}},
  \bibinfo{author}{\bibfnamefont{H.}~\bibnamefont{Umezawa}},
  \bibinfo{author}{\bibfnamefont{H.}~\bibnamefont{Kawai}},
  \bibinfo{author}{\bibfnamefont{K.}~\bibnamefont{Ando}},
  \bibinfo{author}{\bibfnamefont{K.}~\bibnamefont{Takanashi}},
  \bibnamefont{et~al.}, \bibinfo{journal}{Nature}
  \textbf{\bibinfo{volume}{464}}, \bibinfo{pages}{262} (\bibinfo{year}{2010}),
  ISSN \bibinfo{issn}{0028-0836},
  \urlprefix\url{http://dx.doi.org/10.1038/nature08876}.

\bibitem[{\citenamefont{Khitun et~al.}(2010)\citenamefont{Khitun, Bao, and
  Wang}}]{Khitun2010}
\bibinfo{author}{\bibfnamefont{A.}~\bibnamefont{Khitun}},
  \bibinfo{author}{\bibfnamefont{M.}~\bibnamefont{Bao}}, \bibnamefont{and}
  \bibinfo{author}{\bibfnamefont{K.~L.} \bibnamefont{Wang}},
  \bibinfo{journal}{Journal of Physics D: Applied Physics}
  \textbf{\bibinfo{volume}{43}}, \bibinfo{pages}{264005}
  (\bibinfo{year}{2010}), ISSN \bibinfo{issn}{0022-3727},
  \urlprefix\url{http://stacks.iop.org/0022-3727/43/i=26/a=264005}.

\bibitem[{\citenamefont{van Vleck and van Kranendonk}(1958)}]{Vleck1958}
\bibinfo{author}{\bibfnamefont{J.~H.} \bibnamefont{van Vleck}}
  \bibnamefont{and} \bibinfo{author}{\bibfnamefont{J.}~\bibnamefont{van
  Kranendonk}}, \bibinfo{journal}{Rev. Mod. Phys.}
  \textbf{\bibinfo{volume}{30}}, \bibinfo{pages}{1} (\bibinfo{year}{1958}),
  \urlprefix\url{http://link.aps.org/abstract/RMP/v30/p1}.

\bibitem[{\citenamefont{Shirane et~al.}(1965)\citenamefont{Shirane, Nathans,
  Steinsvoll, Alperin, and Pickart}}]{Shirane1965}
\bibinfo{author}{\bibfnamefont{G.}~\bibnamefont{Shirane}},
  \bibinfo{author}{\bibfnamefont{R.}~\bibnamefont{Nathans}},
  \bibinfo{author}{\bibfnamefont{O.}~\bibnamefont{Steinsvoll}},
  \bibinfo{author}{\bibfnamefont{H.~A.} \bibnamefont{Alperin}},
  \bibnamefont{and} \bibinfo{author}{\bibfnamefont{S.~J.}
  \bibnamefont{Pickart}}, \bibinfo{journal}{Phys. Rev. Lett.}
  \textbf{\bibinfo{volume}{15}}, \bibinfo{pages}{146} (\bibinfo{year}{1965}),
  \urlprefix\url{http://link.aps.org/abstract/PRL/v15/p146}.

\bibitem[{\citenamefont{Cooke et~al.}(1980)\citenamefont{Cooke, Davis, and
  Lynn}}]{Cooke1980}
\bibinfo{author}{\bibfnamefont{J.~F.} \bibnamefont{Cooke}},
  \bibinfo{author}{\bibfnamefont{H.~L.} \bibnamefont{Davis}}, \bibnamefont{and}
  \bibinfo{author}{\bibfnamefont{J.~W.} \bibnamefont{Lynn}},
  \bibinfo{journal}{Phys. Rev. B} \textbf{\bibinfo{volume}{21}},
  \bibinfo{pages}{4118} (\bibinfo{year}{1980}).

\bibitem[{\citenamefont{Vollmer et~al.}(2003)\citenamefont{Vollmer, Etzkorn,
  Kumar, Ibach, and Kirschner}}]{Vollmer2003}
\bibinfo{author}{\bibfnamefont{R.}~\bibnamefont{Vollmer}},
  \bibinfo{author}{\bibfnamefont{M.}~\bibnamefont{Etzkorn}},
  \bibinfo{author}{\bibfnamefont{P.~S.~A.} \bibnamefont{Kumar}},
  \bibinfo{author}{\bibfnamefont{H.}~\bibnamefont{Ibach}}, \bibnamefont{and}
  \bibinfo{author}{\bibfnamefont{J.}~\bibnamefont{Kirschner}},
  \bibinfo{journal}{Physical Review Letters} \textbf{\bibinfo{volume}{91}},
  \bibinfo{eid}{147201} (pages~\bibinfo{numpages}{4}) (\bibinfo{year}{2003}),
  \urlprefix\url{http://link.aps.org/abstract/PRL/v91/e147201}.

\bibitem[{\citenamefont{Kittel}(2005)}]{Kittel2005}
\bibinfo{author}{\bibfnamefont{C.}~\bibnamefont{Kittel}},
  \emph{\bibinfo{title}{Introduction to solid state physics}}
  (\bibinfo{publisher}{Wiley}, \bibinfo{year}{2005}).

\bibitem[{\citenamefont{Landau}(1946)}]{Landau1946}
\bibinfo{author}{\bibfnamefont{L.~D.} \bibnamefont{Landau}},
  \bibinfo{journal}{J. Phys. USSR} \textbf{\bibinfo{volume}{10}},
  \bibinfo{pages}{25} (\bibinfo{year}{1946}).

\bibitem[{\citenamefont{Fetter and Walecka}(1971)}]{Fetter1971}
\bibinfo{author}{\bibfnamefont{A.~L.} \bibnamefont{Fetter}} \bibnamefont{and}
  \bibinfo{author}{\bibfnamefont{J.~D.} \bibnamefont{Walecka}},
  \emph{\bibinfo{title}{Quantum theory of many-particle systems}},
  International series in pure and applied physics
  (\bibinfo{publisher}{McGraw-Hill}, \bibinfo{address}{New York, NY},
  \bibinfo{year}{1971}).

\bibitem[{\citenamefont{Gross and Kohn}(1985)}]{Gross1985}
\bibinfo{author}{\bibfnamefont{E.~K.~U.} \bibnamefont{Gross}} \bibnamefont{and}
  \bibinfo{author}{\bibfnamefont{W.}~\bibnamefont{Kohn}},
  \bibinfo{journal}{Physical Review Letters} \textbf{\bibinfo{volume}{55}},
  \bibinfo{pages}{2850} (\bibinfo{year}{1985}),
  \urlprefix\url{http://link.aps.org/abstract/PRL/v55/p2850}.

\bibitem[{\citenamefont{Savrasov}(1998)}]{Savrasov1998}
\bibinfo{author}{\bibfnamefont{S.~Y.} \bibnamefont{Savrasov}},
  \bibinfo{journal}{Physical Review Letters} \textbf{\bibinfo{volume}{81}},
  \bibinfo{pages}{2570} (\bibinfo{year}{1998}).

\bibitem[{\citenamefont{Staunton et~al.}(1999)\citenamefont{Staunton, Poulter,
  Ginatempo, Bruno, and Johnson}}]{Staunton1999}
\bibinfo{author}{\bibfnamefont{J.~B.} \bibnamefont{Staunton}},
  \bibinfo{author}{\bibfnamefont{J.}~\bibnamefont{Poulter}},
  \bibinfo{author}{\bibfnamefont{B.}~\bibnamefont{Ginatempo}},
  \bibinfo{author}{\bibfnamefont{E.}~\bibnamefont{Bruno}}, \bibnamefont{and}
  \bibinfo{author}{\bibfnamefont{D.~D.} \bibnamefont{Johnson}},
  \bibinfo{journal}{Physical Review Letters} \textbf{\bibinfo{volume}{82}},
  \bibinfo{pages}{3340} (\bibinfo{year}{1999}).

\bibitem[{\citenamefont{Buczek et~al.}(2009)\citenamefont{Buczek, Ernst, Bruno,
  and Sandratskii}}]{Buczek2009}
\bibinfo{author}{\bibfnamefont{P.}~\bibnamefont{Buczek}},
  \bibinfo{author}{\bibfnamefont{A.}~\bibnamefont{Ernst}},
  \bibinfo{author}{\bibfnamefont{P.}~\bibnamefont{Bruno}}, \bibnamefont{and}
  \bibinfo{author}{\bibfnamefont{L.~M.} \bibnamefont{Sandratskii}},
  \bibinfo{journal}{Phys. Rev. Lett.} \textbf{\bibinfo{volume}{102}},
  \bibinfo{pages}{247206} (\bibinfo{year}{2009}),
  \urlprefix\url{http://dx.doi.org/10.1103/PhysRevLett.102.247206}.

\bibitem[{\citenamefont{Buczek et~al.}(2010{\natexlab{a}})\citenamefont{Buczek,
  Ernst, and Sandratskii}}]{Buczek2010d}
\bibinfo{author}{\bibfnamefont{P.}~\bibnamefont{Buczek}},
  \bibinfo{author}{\bibfnamefont{A.}~\bibnamefont{Ernst}}, \bibnamefont{and}
  \bibinfo{author}{\bibfnamefont{L.~M.} \bibnamefont{Sandratskii}},
  \bibinfo{journal}{Phys. Rev. Lett.} \textbf{\bibinfo{volume}{105}},
  \bibinfo{pages}{097205} (\bibinfo{year}{2010}{\natexlab{a}}),
  \urlprefix\url{http://link.aps.org/doi/10.1103/PhysRevLett.105.097205}.

\bibitem[{\citenamefont{Buczek et~al.}(2011)\citenamefont{Buczek, Ernst, and
  Sandratskii}}]{Buczek2011}
\bibinfo{author}{\bibfnamefont{P.}~\bibnamefont{Buczek}},
  \bibinfo{author}{\bibfnamefont{A.}~\bibnamefont{Ernst}}, \bibnamefont{and}
  \bibinfo{author}{\bibfnamefont{L.~M.} \bibnamefont{Sandratskii}},
  \bibinfo{journal}{Phys. Rev. Lett.} \textbf{\bibinfo{volume}{106}},
  \bibinfo{pages}{157204} (\bibinfo{year}{2011}),
  \urlprefix\url{http://link.aps.org/doi/10.1103/PhysRevLett.106.157204}.

\bibitem[{\citenamefont{K{\"u}bler}(2000)}]{Kuebler2000}
\bibinfo{author}{\bibfnamefont{J.~K.} \bibnamefont{K{\"u}bler}},
  \emph{\bibinfo{title}{Theory of itinerant electron magnetism}}, vol.
  \bibinfo{volume}{106} of \emph{\bibinfo{series}{International series of
  monographs on physics}} (\bibinfo{publisher}{Clarendon Press},
  \bibinfo{address}{Oxford}, \bibinfo{year}{2000}).

\bibitem[{\citenamefont{Lichtenstein et~al.}(2001)\citenamefont{Lichtenstein,
  Katsnelson, and Kotliar}}]{Lichtenstein2001}
\bibinfo{author}{\bibfnamefont{A.~I.} \bibnamefont{Lichtenstein}},
  \bibinfo{author}{\bibfnamefont{M.~I.} \bibnamefont{Katsnelson}},
  \bibnamefont{and} \bibinfo{author}{\bibfnamefont{G.}~\bibnamefont{Kotliar}},
  \bibinfo{journal}{Phys. Rev. Lett.} \textbf{\bibinfo{volume}{87}},
  \bibinfo{pages}{067205} (\bibinfo{year}{2001}),
  \urlprefix\url{http://link.aps.org/abstract/PRL/v87/e067205}.

\bibitem[{\citenamefont{Tang et~al.}(1998)\citenamefont{Tang, Plihal, and
  Mills}}]{Tang1998}
\bibinfo{author}{\bibfnamefont{H.}~\bibnamefont{Tang}},
  \bibinfo{author}{\bibfnamefont{M.}~\bibnamefont{Plihal}}, \bibnamefont{and}
  \bibinfo{author}{\bibfnamefont{D.~L.} \bibnamefont{Mills}},
  \bibinfo{journal}{Journal of Magnetism and Magnetic Materials}
  \textbf{\bibinfo{volume}{187}}, \bibinfo{pages}{23} (\bibinfo{year}{1998}),
  \urlprefix\url{http://www.sciencedirect.com/science?_ob=ArticleURL&_udi=B6TJ%
J-3VF49HW-7B&_user=28881&_handle=V-WA-A-W-AE-MsSAYVA-UUA-U-AAVABDYZYC-AABYECEV%
YC-ZWDBEUWAD-AE-U&_fmt=summary&_coverDate=08%2F04%2F1998&_rdoc=4&_orig=browse&%
_srch=%23toc%235312%231998%23998129998%2344352!&_cdi=5312&view=c&_acct=C000002%
678&_version=1&_urlVersion=0&_userid=28881&md5=07a27fe2c118a0be83c62b156e5aeb6%
5}.

\bibitem[{\citenamefont{Muniz and Mills}(2002)}]{Muniz2002}
\bibinfo{author}{\bibfnamefont{R.~B.} \bibnamefont{Muniz}} \bibnamefont{and}
  \bibinfo{author}{\bibfnamefont{D.~L.} \bibnamefont{Mills}},
  \bibinfo{journal}{Physical Review B (Condensed Matter and Materials Physics)}
  \textbf{\bibinfo{volume}{66}}, \bibinfo{eid}{174417}
  (pages~\bibinfo{numpages}{15}) (\bibinfo{year}{2002}),
  \urlprefix\url{http://link.aps.org/abstract/PRB/v66/e174417}.

\bibitem[{\citenamefont{Costa et~al.}(2003)\citenamefont{Costa, Muniz, and
  Mills}}]{Costa2003}
\bibinfo{author}{\bibfnamefont{A.~T.} \bibnamefont{Costa}},
  \bibinfo{author}{\bibfnamefont{R.~B.} \bibnamefont{Muniz}}, \bibnamefont{and}
  \bibinfo{author}{\bibfnamefont{D.~L.} \bibnamefont{Mills}},
  \bibinfo{journal}{Phys. Rev. B} \textbf{\bibinfo{volume}{68}},
  \bibinfo{pages}{224435} (\bibinfo{year}{2003}),
  \urlprefix\url{http://link.aps.org/abstract/PRB/v68/e224435}.

\bibitem[{\citenamefont{Muniz et~al.}(2003)\citenamefont{Muniz, Costa, and
  Mills}}]{Muniz2003}
\bibinfo{author}{\bibfnamefont{R.~B.} \bibnamefont{Muniz}},
  \bibinfo{author}{\bibfnamefont{A.~T.} \bibnamefont{Costa}}, \bibnamefont{and}
  \bibinfo{author}{\bibfnamefont{D.~L.} \bibnamefont{Mills}},
  \bibinfo{journal}{Journal of Physics: Condensed Matter}
  \textbf{\bibinfo{volume}{15}}, \bibinfo{pages}{S495} (\bibinfo{year}{2003}),
  \urlprefix\url{http://stacks.iop.org/0953-8984/15/S495}.

\bibitem[{\citenamefont{Costa et~al.}(2004)\citenamefont{Costa, Muniz, and
  Mills}}]{Costa2004a}
\bibinfo{author}{\bibfnamefont{A.~T.} \bibnamefont{Costa}},
  \bibinfo{author}{\bibfnamefont{R.~B.} \bibnamefont{Muniz}}, \bibnamefont{and}
  \bibinfo{author}{\bibfnamefont{D.~L.} \bibnamefont{Mills}},
  \bibinfo{journal}{Phys. Rev. B} \textbf{\bibinfo{volume}{70}},
  \bibinfo{pages}{054406} (\bibinfo{year}{2004}),
  \urlprefix\url{http://link.aps.org/doi/10.1103/PhysRevB.70.054406}.

\bibitem[{\citenamefont{{Costa, Jr.} et~al.}(2004)\citenamefont{{Costa, Jr.},
  Muniz, and Mills}}]{Costa2004}
\bibinfo{author}{\bibfnamefont{A.~T.} \bibnamefont{{Costa, Jr.}}},
  \bibinfo{author}{\bibfnamefont{R.~B.} \bibnamefont{Muniz}}, \bibnamefont{and}
  \bibinfo{author}{\bibfnamefont{D.~L.} \bibnamefont{Mills}},
  \bibinfo{journal}{Physical Review B (Condensed Matter and Materials Physics)}
  \textbf{\bibinfo{volume}{69}}, \bibinfo{eid}{064413}
  (pages~\bibinfo{numpages}{5}) (\bibinfo{year}{2004}),
  \urlprefix\url{http://link.aps.org/abstract/PRB/v69/e064413}.

\bibitem[{\citenamefont{Costa et~al.}(2006)\citenamefont{Costa, Muniz, and
  Mills}}]{Costa2006}
\bibinfo{author}{\bibfnamefont{A.~T.} \bibnamefont{Costa}},
  \bibinfo{author}{\bibfnamefont{R.~B.} \bibnamefont{Muniz}}, \bibnamefont{and}
  \bibinfo{author}{\bibfnamefont{D.~L.} \bibnamefont{Mills}},
  \bibinfo{journal}{Phys. Rev. B} \textbf{\bibinfo{volume}{74}},
  \bibinfo{pages}{214403} (\bibinfo{year}{2006}),
  \urlprefix\url{http://link.aps.org/doi/10.1103/PhysRevB.74.214403}.

\bibitem[{\citenamefont{Halilov et~al.}(1998)\citenamefont{Halilov, Eschrig,
  Perlov, and Oppeneer}}]{Halilov1998a}
\bibinfo{author}{\bibfnamefont{S.~V.} \bibnamefont{Halilov}},
  \bibinfo{author}{\bibfnamefont{H.}~\bibnamefont{Eschrig}},
  \bibinfo{author}{\bibfnamefont{A.~Y.} \bibnamefont{Perlov}},
  \bibnamefont{and} \bibinfo{author}{\bibfnamefont{P.~M.}
  \bibnamefont{Oppeneer}}, \bibinfo{journal}{Phys. Rev. B}
  \textbf{\bibinfo{volume}{58}}, \bibinfo{pages}{293} (\bibinfo{year}{1998}),
  \urlprefix\url{http://link.aps.org/abstract/PRB/v58/p293}.

\bibitem[{\citenamefont{Antropov et~al.}(1999)\citenamefont{Antropov, Harmon,
  and Smirnov}}]{Antropov1999}
\bibinfo{author}{\bibfnamefont{V.~P.} \bibnamefont{Antropov}},
  \bibinfo{author}{\bibfnamefont{B.~N.} \bibnamefont{Harmon}},
  \bibnamefont{and} \bibinfo{author}{\bibfnamefont{A.~N.}
  \bibnamefont{Smirnov}}, \bibinfo{journal}{Journal of Magnetism and Magnetic
  Materials} \textbf{\bibinfo{volume}{200}}, \bibinfo{pages}{148}
  (\bibinfo{year}{1999}),
  \urlprefix\url{http://www.sciencedirect.com/science/article/B6TJJ-3Y2NV45-R/%
2/fb00e680a9a2f00946af94a6fca5c716}.

\bibitem[{\citenamefont{Liechtenstein et~al.}(1987)\citenamefont{Liechtenstein,
  Katsnelson, Antropov, and Gubanova}}]{Liechtenstein1987}
\bibinfo{author}{\bibfnamefont{A.~I.} \bibnamefont{Liechtenstein}},
  \bibinfo{author}{\bibfnamefont{M.~I.} \bibnamefont{Katsnelson}},
  \bibinfo{author}{\bibfnamefont{V.~P.} \bibnamefont{Antropov}},
  \bibnamefont{and} \bibinfo{author}{\bibfnamefont{V.~A.}
  \bibnamefont{Gubanova}}, \bibinfo{journal}{Journal of Magnetism and Magnetic
  Materials} \textbf{\bibinfo{volume}{67}}, \bibinfo{pages}{65}
  (\bibinfo{year}{1987}).

\bibitem[{\citenamefont{Sandratskii}(1991)}]{Sandratskii1991}
\bibinfo{author}{\bibfnamefont{L.~M.} \bibnamefont{Sandratskii}},
  \bibinfo{journal}{Journal of Physics: Condensed Matter}
  \textbf{\bibinfo{volume}{3}}, \bibinfo{pages}{8565} (\bibinfo{year}{1991}),
  \urlprefix\url{http://stacks.iop.org/0953-8984/3/8565}.

\bibitem[{\citenamefont{Sandratskii}(1998)}]{Sandratskii1998}
\bibinfo{author}{\bibfnamefont{L.~M.} \bibnamefont{Sandratskii}},
  \bibinfo{journal}{Advances in Physics} \textbf{\bibinfo{volume}{47}},
  \bibinfo{pages}{91 } (\bibinfo{year}{1998}).

\bibitem[{\citenamefont{Grotheer et~al.}(2001)\citenamefont{Grotheer, Ederer,
  and F\"ahnle}}]{Grotheer2001}
\bibinfo{author}{\bibfnamefont{O.}~\bibnamefont{Grotheer}},
  \bibinfo{author}{\bibfnamefont{C.}~\bibnamefont{Ederer}}, \bibnamefont{and}
  \bibinfo{author}{\bibfnamefont{M.}~\bibnamefont{F\"ahnle}},
  \bibinfo{journal}{Phys. Rev. B} \textbf{\bibinfo{volume}{63}},
  \bibinfo{pages}{100401} (\bibinfo{year}{2001}).

\bibitem[{\citenamefont{Bruno}(2003)}]{Bruno2003}
\bibinfo{author}{\bibfnamefont{P.}~\bibnamefont{Bruno}},
  \bibinfo{journal}{Phys. Rev. Lett.} \textbf{\bibinfo{volume}{90}},
  \bibinfo{pages}{087205} (\bibinfo{year}{2003}).

\bibitem[{\citenamefont{Callaway and Wang}(1975)}]{Callaway1975}
\bibinfo{author}{\bibfnamefont{J.}~\bibnamefont{Callaway}} \bibnamefont{and}
  \bibinfo{author}{\bibfnamefont{C.~S.} \bibnamefont{Wang}},
  \bibinfo{journal}{Journal of Physics F: Metal Physics}
  \textbf{\bibinfo{volume}{5}}, \bibinfo{pages}{2119} (\bibinfo{year}{1975}),
  \urlprefix\url{http://stacks.iop.org/0305-4608/5/2119}.

\bibitem[{\citenamefont{Stenzel and Winter}(1985)}]{Stenzel1985}
\bibinfo{author}{\bibfnamefont{E.}~\bibnamefont{Stenzel}} \bibnamefont{and}
  \bibinfo{author}{\bibfnamefont{H.}~\bibnamefont{Winter}},
  \bibinfo{journal}{Journal of Physics F: Metal Physics}
  \textbf{\bibinfo{volume}{15}}, \bibinfo{pages}{1571} (\bibinfo{year}{1985}),
  \urlprefix\url{http://stacks.iop.org/0305-4608/15/1571}.

\bibitem[{\citenamefont{Winter et~al.}(1988)\citenamefont{Winter, Stenzel,
  Szotek, and Temmerman}}]{Winter1988}
\bibinfo{author}{\bibfnamefont{H.}~\bibnamefont{Winter}},
  \bibinfo{author}{\bibfnamefont{E.}~\bibnamefont{Stenzel}},
  \bibinfo{author}{\bibfnamefont{Z.}~\bibnamefont{Szotek}}, \bibnamefont{and}
  \bibinfo{author}{\bibfnamefont{W.~M.} \bibnamefont{Temmerman}},
  \bibinfo{journal}{Journal of Physics F: Metal Physics}
  \textbf{\bibinfo{volume}{18}}, \bibinfo{pages}{485} (\bibinfo{year}{1988}),
  \urlprefix\url{http://stacks.iop.org/0305-4608/18/485}.

\bibitem[{\citenamefont{Staunton et~al.}(1986)\citenamefont{Staunton, Gyorffy,
  Stocks, and Wadsworth}}]{Staunton1986}
\bibinfo{author}{\bibfnamefont{J.}~\bibnamefont{Staunton}},
  \bibinfo{author}{\bibfnamefont{B.~L.} \bibnamefont{Gyorffy}},
  \bibinfo{author}{\bibfnamefont{G.~M.} \bibnamefont{Stocks}},
  \bibnamefont{and}
  \bibinfo{author}{\bibfnamefont{J.}~\bibnamefont{Wadsworth}},
  \bibinfo{journal}{Journal of Physics F: Metal Physics}
  \textbf{\bibinfo{volume}{16}}, \bibinfo{pages}{1761} (\bibinfo{year}{1986}),
  \urlprefix\url{http://stacks.iop.org/0305-4608/16/1761}.

\bibitem[{\citenamefont{Staunton et~al.}(2000)\citenamefont{Staunton, Poulter,
  Ginatempo, Bruno, and Johnson}}]{Staunton2000}
\bibinfo{author}{\bibfnamefont{J.~B.} \bibnamefont{Staunton}},
  \bibinfo{author}{\bibfnamefont{J.}~\bibnamefont{Poulter}},
  \bibinfo{author}{\bibfnamefont{B.}~\bibnamefont{Ginatempo}},
  \bibinfo{author}{\bibfnamefont{E.}~\bibnamefont{Bruno}}, \bibnamefont{and}
  \bibinfo{author}{\bibfnamefont{D.~D.} \bibnamefont{Johnson}},
  \bibinfo{journal}{Physical Review B (Condensed Matter and Materials Physics)}
  \textbf{\bibinfo{volume}{62}}, \bibinfo{pages}{1075} (\bibinfo{year}{2000}).

\bibitem[{\citenamefont{Buczek}(2009)}]{Buczek2009a}
\bibinfo{author}{\bibfnamefont{P.~A.} \bibnamefont{Buczek}}, Ph.D. thesis,
  \bibinfo{school}{Martin Luther Universit\"at Halle-Wittenberg}
  (\bibinfo{year}{2009}),
  \urlprefix\url{http://digital.bibliothek.uni-halle.de/ulbhalhs/urn/urn:nbn:d%
e:gbv:3:4-552}.

\bibitem[{\citenamefont{Lounis et~al.}(2010)\citenamefont{Lounis, Costa, Muniz,
  and Mills}}]{Lounis2010}
\bibinfo{author}{\bibfnamefont{S.}~\bibnamefont{Lounis}},
  \bibinfo{author}{\bibfnamefont{A.~T.} \bibnamefont{Costa}},
  \bibinfo{author}{\bibfnamefont{R.~B.} \bibnamefont{Muniz}}, \bibnamefont{and}
  \bibinfo{author}{\bibfnamefont{D.~L.} \bibnamefont{Mills}},
  \bibinfo{journal}{Phys. Rev. Lett.} \textbf{\bibinfo{volume}{105}},
  \bibinfo{pages}{187205} (\bibinfo{year}{2010}),
  \urlprefix\url{http://link.aps.org/doi/10.1103/PhysRevLett.105.187205}.

\bibitem[{\citenamefont{Lounis et~al.}(2011)\citenamefont{Lounis, Costa, Muniz,
  and Mills}}]{Lounis2011}
\bibinfo{author}{\bibfnamefont{S.}~\bibnamefont{Lounis}},
  \bibinfo{author}{\bibfnamefont{A.~T.} \bibnamefont{Costa}},
  \bibinfo{author}{\bibfnamefont{R.~B.} \bibnamefont{Muniz}}, \bibnamefont{and}
  \bibinfo{author}{\bibfnamefont{D.~L.} \bibnamefont{Mills}},
  \bibinfo{journal}{Phys. Rev. B} \textbf{\bibinfo{volume}{83}},
  \bibinfo{pages}{035109} (\bibinfo{year}{2011}),
  \urlprefix\url{http://link.aps.org/doi/10.1103/PhysRevB.83.035109}.

\bibitem[{\citenamefont{Kotani and van Schilfgaarde}(2008)}]{Kotani2008}
\bibinfo{author}{\bibfnamefont{T.}~\bibnamefont{Kotani}} \bibnamefont{and}
  \bibinfo{author}{\bibfnamefont{M.}~\bibnamefont{van Schilfgaarde}},
  \bibinfo{journal}{Journal of Physics: Condensed Matter}
  \textbf{\bibinfo{volume}{20}}, \bibinfo{pages}{295214}
  (\bibinfo{year}{2008}), ISSN \bibinfo{issn}{0953-8984},
  \urlprefix\url{http://stacks.iop.org/0953-8984/20/i=29/a=295214}.

\bibitem[{\citenamefont{Hedin}(1965)}]{Hedin1965}
\bibinfo{author}{\bibfnamefont{L.}~\bibnamefont{Hedin}},
  \bibinfo{journal}{Phys. Rev.} \textbf{\bibinfo{volume}{139}},
  \bibinfo{pages}{A796} (\bibinfo{year}{1965}).

\bibitem[{\citenamefont{Aryasetiawan and Karlsson}(1999)}]{Aryasetiawan1999}
\bibinfo{author}{\bibfnamefont{F.}~\bibnamefont{Aryasetiawan}}
  \bibnamefont{and} \bibinfo{author}{\bibfnamefont{K.}~\bibnamefont{Karlsson}},
  \bibinfo{journal}{Physical Review B (Condensed Matter and Materials Physics)}
  \textbf{\bibinfo{volume}{60}}, \bibinfo{pages}{7419} (\bibinfo{year}{1999}),
  \urlprefix\url{http://link.aps.org/abstract/PRB/v60/p7419}.

\bibitem[{\citenamefont{Karlsson and Aryasetiawan}(2000)}]{Karlsson2000}
\bibinfo{author}{\bibfnamefont{K.}~\bibnamefont{Karlsson}} \bibnamefont{and}
  \bibinfo{author}{\bibfnamefont{F.}~\bibnamefont{Aryasetiawan}},
  \bibinfo{journal}{Phys. Rev. B} \textbf{\bibinfo{volume}{62}},
  \bibinfo{pages}{3006} (\bibinfo{year}{2000}),
  \urlprefix\url{http://link.aps.org/abstract/PRB/v62/p3006}.

\bibitem[{\citenamefont{\c{S}a\c{s}io\u{g}lu
  et~al.}(2010)\citenamefont{\c{S}a\c{s}io\u{g}lu, Schindlmayr, Friedrich,
  Freimuth, and Bl\"ugel}}]{Sasiouglu2010}
\bibinfo{author}{\bibfnamefont{E.}~\bibnamefont{\c{S}a\c{s}io\u{g}lu}},
  \bibinfo{author}{\bibfnamefont{A.}~\bibnamefont{Schindlmayr}},
  \bibinfo{author}{\bibfnamefont{C.}~\bibnamefont{Friedrich}},
  \bibinfo{author}{\bibfnamefont{F.}~\bibnamefont{Freimuth}}, \bibnamefont{and}
  \bibinfo{author}{\bibfnamefont{S.}~\bibnamefont{Bl\"ugel}},
  \bibinfo{journal}{Phys. Rev. B} \textbf{\bibinfo{volume}{81}},
  \bibinfo{pages}{054434} (\bibinfo{year}{2010}),
  \urlprefix\url{http://link.aps.org/doi/10.1103/PhysRevB.81.054434}.

\bibitem[{\citenamefont{Costa et~al.}(2008)\citenamefont{Costa, Muniz, Cao, Wu,
  and Mills}}]{Costa2008}
\bibinfo{author}{\bibfnamefont{A.~T.} \bibnamefont{Costa}},
  \bibinfo{author}{\bibfnamefont{R.~B.} \bibnamefont{Muniz}},
  \bibinfo{author}{\bibfnamefont{J.~X.} \bibnamefont{Cao}},
  \bibinfo{author}{\bibfnamefont{R.~Q.} \bibnamefont{Wu}}, \bibnamefont{and}
  \bibinfo{author}{\bibfnamefont{D.~L.} \bibnamefont{Mills}},
  \bibinfo{journal}{Phys. Rev. B} \textbf{\bibinfo{volume}{78}},
  \bibinfo{pages}{054439} (\bibinfo{year}{2008}),
  \urlprefix\url{http://link.aps.org/abstract/PRB/v78/e054439}.

\bibitem[{\citenamefont{Goldstone}(1961)}]{Goldstone1961}
\bibinfo{author}{\bibfnamefont{J.}~\bibnamefont{Goldstone}},
  \bibinfo{journal}{Il Nuovo Cimento (1955-1965)}
  \textbf{\bibinfo{volume}{19}}, \bibinfo{pages}{154} (\bibinfo{year}{1961}),
  \urlprefix\url{http://dx.doi.org/10.1007/BF02812722}.

\bibitem[{\citenamefont{Goldstone et~al.}(1962)\citenamefont{Goldstone, Salam,
  and Weinberg}}]{Goldstone1962}
\bibinfo{author}{\bibfnamefont{J.}~\bibnamefont{Goldstone}},
  \bibinfo{author}{\bibfnamefont{A.}~\bibnamefont{Salam}}, \bibnamefont{and}
  \bibinfo{author}{\bibfnamefont{S.}~\bibnamefont{Weinberg}},
  \bibinfo{journal}{Phys. Rev.} \textbf{\bibinfo{volume}{127}},
  \bibinfo{pages}{965} (\bibinfo{year}{1962}),
  \urlprefix\url{http://link.aps.org/doi/10.1103/PhysRev.127.965}.

\bibitem[{\citenamefont{Runge and Gross}(1984)}]{Runge1984}
\bibinfo{author}{\bibfnamefont{E.}~\bibnamefont{Runge}} \bibnamefont{and}
  \bibinfo{author}{\bibfnamefont{E.~K.~U.} \bibnamefont{Gross}},
  \bibinfo{journal}{Physical Review Letters} \textbf{\bibinfo{volume}{52}},
  \bibinfo{pages}{997} (\bibinfo{year}{1984}),
  \urlprefix\url{http://link.aps.org/abstract/PRL/v52/p997}.

\bibitem[{\citenamefont{von Barth and Hedin}(1972)}]{vonBarth1972}
\bibinfo{author}{\bibfnamefont{U.}~\bibnamefont{von Barth}} \bibnamefont{and}
  \bibinfo{author}{\bibfnamefont{L.}~\bibnamefont{Hedin}},
  \bibinfo{journal}{Journal of Physics C: Solid State Physics}
  \textbf{\bibinfo{volume}{5}}, \bibinfo{pages}{1629} (\bibinfo{year}{1972}),
  \urlprefix\url{http://stacks.iop.org/0022-3719/5/1629}.

\bibitem[{\citenamefont{Vignale et~al.}(1997)\citenamefont{Vignale, Ullrich,
  and Conti}}]{Vignale1997}
\bibinfo{author}{\bibfnamefont{G.}~\bibnamefont{Vignale}},
  \bibinfo{author}{\bibfnamefont{C.~A.} \bibnamefont{Ullrich}},
  \bibnamefont{and} \bibinfo{author}{\bibfnamefont{S.}~\bibnamefont{Conti}},
  \bibinfo{journal}{Phys. Rev. Lett.} \textbf{\bibinfo{volume}{79}},
  \bibinfo{pages}{4878} (\bibinfo{year}{1997}),
  \urlprefix\url{http://link.aps.org/abstract/PRL/v79/p4878}.

\bibitem[{\citenamefont{Onida et~al.}(2002)\citenamefont{Onida, Reining, and
  Rubio}}]{Onida2002}
\bibinfo{author}{\bibfnamefont{G.}~\bibnamefont{Onida}},
  \bibinfo{author}{\bibfnamefont{L.}~\bibnamefont{Reining}}, \bibnamefont{and}
  \bibinfo{author}{\bibfnamefont{A.}~\bibnamefont{Rubio}},
  \bibinfo{journal}{Rev. Mod. Phys.} \textbf{\bibinfo{volume}{74}},
  \bibinfo{pages}{601} (\bibinfo{year}{2002}),
  \urlprefix\url{http://link.aps.org/abstract/RMP/v74/p601}.

\bibitem[{\citenamefont{Qian and Vignale}(2002)}]{Qian2002}
\bibinfo{author}{\bibfnamefont{Z.}~\bibnamefont{Qian}} \bibnamefont{and}
  \bibinfo{author}{\bibfnamefont{G.}~\bibnamefont{Vignale}},
  \bibinfo{journal}{Physical Review Letters} \textbf{\bibinfo{volume}{88}},
  \bibinfo{eid}{056404} (pages~\bibinfo{numpages}{4}) (\bibinfo{year}{2002}),
  \urlprefix\url{http://link.aps.org/abstract/PRL/v88/e056404}.

\bibitem[{\citenamefont{Kurth and Eich}(2009)}]{Kurth2009}
\bibinfo{author}{\bibfnamefont{S.}~\bibnamefont{Kurth}} \bibnamefont{and}
  \bibinfo{author}{\bibfnamefont{F.~G.} \bibnamefont{Eich}},
  \bibinfo{journal}{Phys. Rev. B} \textbf{\bibinfo{volume}{80}},
  \bibinfo{pages}{125120} (\bibinfo{year}{2009}),
  \urlprefix\url{http://link.aps.org/doi/10.1103/PhysRevB.80.125120}.

\bibitem[{\citenamefont{Eich et~al.}(2010)\citenamefont{Eich, Kurth, Proetto,
  Sharma, and Gross}}]{Eich2010}
\bibinfo{author}{\bibfnamefont{F.~G.} \bibnamefont{Eich}},
  \bibinfo{author}{\bibfnamefont{S.}~\bibnamefont{Kurth}},
  \bibinfo{author}{\bibfnamefont{C.~R.} \bibnamefont{Proetto}},
  \bibinfo{author}{\bibfnamefont{S.}~\bibnamefont{Sharma}}, \bibnamefont{and}
  \bibinfo{author}{\bibfnamefont{E.~K.~U.} \bibnamefont{Gross}},
  \bibinfo{journal}{Phys. Rev. B} \textbf{\bibinfo{volume}{81}},
  \bibinfo{pages}{024430} (\bibinfo{year}{2010}),
  \urlprefix\url{http://link.aps.org/doi/10.1103/PhysRevB.81.024430}.

\bibitem[{\citenamefont{Callen and Welton}(1951)}]{Callen1951}
\bibinfo{author}{\bibfnamefont{H.~B.} \bibnamefont{Callen}} \bibnamefont{and}
  \bibinfo{author}{\bibfnamefont{T.~A.} \bibnamefont{Welton}},
  \bibinfo{journal}{Phys. Rev.} \textbf{\bibinfo{volume}{83}},
  \bibinfo{pages}{34} (\bibinfo{year}{1951}),
  \urlprefix\url{http://link.aps.org/abstract/PR/v83/p34}.

\bibitem[{\citenamefont{Nyquist}(1928)}]{Nyquist1928}
\bibinfo{author}{\bibfnamefont{H.}~\bibnamefont{Nyquist}},
  \bibinfo{journal}{Phys. Rev.} \textbf{\bibinfo{volume}{32}},
  \bibinfo{pages}{110} (\bibinfo{year}{1928}),
  \urlprefix\url{http://link.aps.org/abstract/PR/v32/p110}.

\bibitem[{\citenamefont{Kubo}(1966)}]{Kubo1966}
\bibinfo{author}{\bibfnamefont{R.}~\bibnamefont{Kubo}},
  \bibinfo{journal}{Reports on Progress in Physics}
  \textbf{\bibinfo{volume}{29}}, \bibinfo{pages}{255} (\bibinfo{year}{1966}),
  ISSN \bibinfo{issn}{0034-4885},
  \urlprefix\url{http://stacks.iop.org/0034-4885/29/i=1/a=306}.

\bibitem[{\citenamefont{van Hove}(1954)}]{VanHove1954}
\bibinfo{author}{\bibfnamefont{L.}~\bibnamefont{van Hove}},
  \bibinfo{journal}{Phys. Rev.} \textbf{\bibinfo{volume}{95}},
  \bibinfo{pages}{1374} (\bibinfo{year}{1954}),
  \urlprefix\url{http://link.aps.org/abstract/PR/v95/p1374}.

\bibitem[{\citenamefont{Katsnelson and Lichtenstein}(2004)}]{Katsnelson2004}
\bibinfo{author}{\bibfnamefont{M.~I.} \bibnamefont{Katsnelson}}
  \bibnamefont{and} \bibinfo{author}{\bibfnamefont{A.~I.}
  \bibnamefont{Lichtenstein}}, \bibinfo{journal}{Journal of Physics: Condensed
  Matter} \textbf{\bibinfo{volume}{16}}, \bibinfo{pages}{7439}
  (\bibinfo{year}{2004}),
  \urlprefix\url{http://stacks.iop.org/0953-8984/16/7439}.

\bibitem[{\citenamefont{Edwards and Rahman}(1978)}]{Edwards1978}
\bibinfo{author}{\bibfnamefont{D.~M.} \bibnamefont{Edwards}} \bibnamefont{and}
  \bibinfo{author}{\bibfnamefont{M.~A.} \bibnamefont{Rahman}},
  \bibinfo{journal}{Journal of Physics F: Metal Physics}
  \textbf{\bibinfo{volume}{8}}, \bibinfo{pages}{1501} (\bibinfo{year}{1978}),
  \urlprefix\url{http://stacks.iop.org/0305-4608/8/1501}.

\bibitem[{\citenamefont{Abrikosov et~al.}(1975)\citenamefont{Abrikosov, Gorkov,
  and Dzyaloshinski}}]{Abrikosov1975}
\bibinfo{author}{\bibfnamefont{A.}~\bibnamefont{Abrikosov}},
  \bibinfo{author}{\bibfnamefont{L.}~\bibnamefont{Gorkov}}, \bibnamefont{and}
  \bibinfo{author}{\bibfnamefont{I.}~\bibnamefont{Dzyaloshinski}},
  \emph{\bibinfo{title}{Methods of quantum field theory in statistical
  physics}}, Dover Books on Physics and Chemistry (\bibinfo{publisher}{Dover},
  \bibinfo{address}{New York}, \bibinfo{year}{1975}).

\bibitem[{\citenamefont{Schmalian et~al.}(1996)\citenamefont{Schmalian, Langer,
  Grabowski, and Bennemann}}]{Schmalian1996}
\bibinfo{author}{\bibfnamefont{J.}~\bibnamefont{Schmalian}},
  \bibinfo{author}{\bibfnamefont{M.}~\bibnamefont{Langer}},
  \bibinfo{author}{\bibfnamefont{S.}~\bibnamefont{Grabowski}},
  \bibnamefont{and} \bibinfo{author}{\bibfnamefont{K.~H.}
  \bibnamefont{Bennemann}}, \bibinfo{journal}{Computer Physics Communications}
  \textbf{\bibinfo{volume}{93}}, \bibinfo{pages}{141} (\bibinfo{year}{1996}),
  ISSN \bibinfo{issn}{0010-4655},
  \urlprefix\url{http://www.sciencedirect.com/science/article/B6TJ5-3VS8H3K-1/%
2/af18c96af2ced65be732656572c4530d}.

\bibitem[{\citenamefont{Savrasov}(1990)}]{Savrasov1990}
\bibinfo{author}{\bibfnamefont{S.~Y.} \bibnamefont{Savrasov}},
  \bibinfo{journal}{Solid State Communications} \textbf{\bibinfo{volume}{74}},
  \bibinfo{pages}{69} (\bibinfo{year}{1990}),
  \urlprefix\url{http://www.sciencedirect.com/science/article/B6TVW-46SPG4P-1B%
/2/e486637f6f834b7ead11104d423db890}.

\bibitem[{\citenamefont{Savrasov}(1992)}]{Savrasov1992}
\bibinfo{author}{\bibfnamefont{S.~Y.} \bibnamefont{Savrasov}},
  \bibinfo{journal}{Physical Review Letters} \textbf{\bibinfo{volume}{69}},
  \bibinfo{pages}{2819} (\bibinfo{year}{1992}),
  \urlprefix\url{http://link.aps.org/abstract/PRL/v69/p2819}.

\bibitem[{\citenamefont{Savrasov}(1996)}]{Savrasov1996}
\bibinfo{author}{\bibfnamefont{S.~Y.} \bibnamefont{Savrasov}},
  \bibinfo{journal}{Physical Review B (Condensed Matter)}
  \textbf{\bibinfo{volume}{54}}, \bibinfo{pages}{16470} (\bibinfo{year}{1996}),
  \urlprefix\url{http://link.aps.org/abstract/PRB/v54/p16470}.

\bibitem[{\citenamefont{Wildberger et~al.}(1997)\citenamefont{Wildberger,
  Zeller, and Dederichs}}]{Wildberger1997}
\bibinfo{author}{\bibfnamefont{K.}~\bibnamefont{Wildberger}},
  \bibinfo{author}{\bibfnamefont{R.}~\bibnamefont{Zeller}}, \bibnamefont{and}
  \bibinfo{author}{\bibfnamefont{P.~H.} \bibnamefont{Dederichs}},
  \bibinfo{journal}{Phys. Rev. B} \textbf{\bibinfo{volume}{55}},
  \bibinfo{pages}{10074} (\bibinfo{year}{1997}).

\bibitem[{\citenamefont{Zabloudil}(2005)}]{Zabloudil2005}
\bibinfo{editor}{\bibfnamefont{J.}~\bibnamefont{Zabloudil}}, ed.,
  \emph{\bibinfo{title}{Electron scattering in solid matter: a theoretical and
  computational treatise}}, vol. \bibinfo{volume}{147} of
  \emph{\bibinfo{series}{Springer series in solid-state sciences}}
  (\bibinfo{publisher}{Springer}, \bibinfo{address}{Berlin},
  \bibinfo{year}{2005}).

\bibitem[{\citenamefont{Graham et~al.}(2005)\citenamefont{Graham, Woodall, and
  Squyres}}]{Graham2005}
\bibinfo{author}{\bibfnamefont{R.~L.} \bibnamefont{Graham}},
  \bibinfo{author}{\bibfnamefont{T.~S.} \bibnamefont{Woodall}},
  \bibnamefont{and} \bibinfo{author}{\bibfnamefont{J.~M.}
  \bibnamefont{Squyres}}, in \emph{\bibinfo{booktitle}{Proceedings, 6th Annual
  International Conference on Parallel Processing and Applied Mathematics}}
  (\bibinfo{address}{Pozna\'n, Poland}, \bibinfo{year}{2005}).

\bibitem[{\citenamefont{Graham et~al.}(2006)\citenamefont{Graham, Shipman,
  Barrett, Castain, Bosilca, and Lumsdaine}}]{Graham2006}
\bibinfo{author}{\bibfnamefont{R.~L.} \bibnamefont{Graham}},
  \bibinfo{author}{\bibfnamefont{G.~M.} \bibnamefont{Shipman}},
  \bibinfo{author}{\bibfnamefont{B.~W.} \bibnamefont{Barrett}},
  \bibinfo{author}{\bibfnamefont{R.~H.} \bibnamefont{Castain}},
  \bibinfo{author}{\bibfnamefont{G.}~\bibnamefont{Bosilca}}, \bibnamefont{and}
  \bibinfo{author}{\bibfnamefont{A.}~\bibnamefont{Lumsdaine}}, in
  \emph{\bibinfo{booktitle}{Proceedings, Fifth International Workshop on
  Algorithms, Models and Tools for Parallel Computing on Heterogeneous
  Networks}} (\bibinfo{address}{Barcelona, Spain}, \bibinfo{year}{2006}).

\bibitem[{\citenamefont{Hursey et~al.}(2007)\citenamefont{Hursey, Mallove,
  Squyres, and Lumsdaine}}]{Hursey2007}
\bibinfo{author}{\bibfnamefont{J.}~\bibnamefont{Hursey}},
  \bibinfo{author}{\bibfnamefont{E.}~\bibnamefont{Mallove}},
  \bibinfo{author}{\bibfnamefont{J.~M.} \bibnamefont{Squyres}},
  \bibnamefont{and}
  \bibinfo{author}{\bibfnamefont{A.}~\bibnamefont{Lumsdaine}}, in
  \emph{\bibinfo{booktitle}{Proceedings, Euro PVM/MPI}}
  (\bibinfo{address}{Paris, France}, \bibinfo{year}{2007}).

\bibitem[{\citenamefont{Mook and Nicklow}(1973)}]{Mook1973}
\bibinfo{author}{\bibfnamefont{H.~A.} \bibnamefont{Mook}} \bibnamefont{and}
  \bibinfo{author}{\bibfnamefont{R.~M.} \bibnamefont{Nicklow}},
  \bibinfo{journal}{Phys. Rev. B} \textbf{\bibinfo{volume}{7}},
  \bibinfo{pages}{336} (\bibinfo{year}{1973}),
  \urlprefix\url{http://link.aps.org/abstract/PRB/v7/p336}.

\bibitem[{\citenamefont{Shirane et~al.}(1968)\citenamefont{Shirane, Minkiewicz,
  and Nathans}}]{Shirane1968}
\bibinfo{author}{\bibfnamefont{G.}~\bibnamefont{Shirane}},
  \bibinfo{author}{\bibfnamefont{V.~J.} \bibnamefont{Minkiewicz}},
  \bibnamefont{and} \bibinfo{author}{\bibfnamefont{R.}~\bibnamefont{Nathans}},
  \bibinfo{journal}{Journal of Applied Physics} \textbf{\bibinfo{volume}{39}},
  \bibinfo{pages}{383} (\bibinfo{year}{1968}),
  \urlprefix\url{http://link.aip.org/link/?JAP/39/383/1}.

\bibitem[{\citenamefont{Collins et~al.}(1969)\citenamefont{Collins, Minkiewicz,
  Nathans, Passel, and Shirane}}]{Collins1969}
\bibinfo{author}{\bibfnamefont{M.~F.} \bibnamefont{Collins}},
  \bibinfo{author}{\bibfnamefont{V.~J.} \bibnamefont{Minkiewicz}},
  \bibinfo{author}{\bibfnamefont{R.}~\bibnamefont{Nathans}},
  \bibinfo{author}{\bibfnamefont{L.}~\bibnamefont{Passel}}, \bibnamefont{and}
  \bibinfo{author}{\bibfnamefont{G.}~\bibnamefont{Shirane}},
  \bibinfo{journal}{Phys. Rev.} \textbf{\bibinfo{volume}{179}},
  \bibinfo{pages}{417} (\bibinfo{year}{1969}),
  \urlprefix\url{http://link.aps.org/abstract/PR/v179/p417}.

\bibitem[{\citenamefont{Loong et~al.}(1984)\citenamefont{Loong, Carpenter,
  Lynn, Robinson, and Mook}}]{Loong1984}
\bibinfo{author}{\bibfnamefont{C.-K.} \bibnamefont{Loong}},
  \bibinfo{author}{\bibfnamefont{J.~M.} \bibnamefont{Carpenter}},
  \bibinfo{author}{\bibfnamefont{J.~W.} \bibnamefont{Lynn}},
  \bibinfo{author}{\bibfnamefont{R.~A.} \bibnamefont{Robinson}},
  \bibnamefont{and} \bibinfo{author}{\bibfnamefont{H.~A.} \bibnamefont{Mook}},
  \bibinfo{journal}{J. Appl. Phys.} \textbf{\bibinfo{volume}{55}},
  \bibinfo{pages}{1895} (\bibinfo{year}{1984}),
  \urlprefix\url{http://link.aip.org/link/?JAP/55/1895/1}.

\bibitem[{\citenamefont{Buczek et~al.}(2010{\natexlab{b}})\citenamefont{Buczek,
  Ernst, Sandratskii, and Bruno}}]{Buczek2010}
\bibinfo{author}{\bibfnamefont{P.}~\bibnamefont{Buczek}},
  \bibinfo{author}{\bibfnamefont{A.}~\bibnamefont{Ernst}},
  \bibinfo{author}{\bibfnamefont{L.}~\bibnamefont{Sandratskii}},
  \bibnamefont{and} \bibinfo{author}{\bibfnamefont{P.}~\bibnamefont{Bruno}},
  \bibinfo{journal}{Journal of Magnetism and Magnetic Materials}
  \textbf{\bibinfo{volume}{322}}, \bibinfo{pages}{1396}
  (\bibinfo{year}{2010}{\natexlab{b}}), ISSN \bibinfo{issn}{0304-8853},
  \urlprefix\url{http://dx.doi.org/10.1016/j.jmmm.2009.03.010}.

\bibitem[{\citenamefont{Baldacchini et~al.}(2003)\citenamefont{Baldacchini,
  Chiodo, Allegretti, Mariani, Betti, Monachesi, and
  Del~Sole}}]{Baldacchini2003}
\bibinfo{author}{\bibfnamefont{C.}~\bibnamefont{Baldacchini}},
  \bibinfo{author}{\bibfnamefont{L.}~\bibnamefont{Chiodo}},
  \bibinfo{author}{\bibfnamefont{F.}~\bibnamefont{Allegretti}},
  \bibinfo{author}{\bibfnamefont{C.}~\bibnamefont{Mariani}},
  \bibinfo{author}{\bibfnamefont{M.~G.} \bibnamefont{Betti}},
  \bibinfo{author}{\bibfnamefont{P.}~\bibnamefont{Monachesi}},
  \bibnamefont{and} \bibinfo{author}{\bibfnamefont{R.}~\bibnamefont{Del~Sole}},
  \bibinfo{journal}{Phys. Rev. B} \textbf{\bibinfo{volume}{68}},
  \bibinfo{pages}{195109} (\bibinfo{year}{2003}),
  \urlprefix\url{http://link.aps.org/doi/10.1103/PhysRevB.68.195109}.

\bibitem[{\citenamefont{Achilli et~al.}(2007)\citenamefont{Achilli, Caravati,
  and Trioni}}]{Achilli2007}
\bibinfo{author}{\bibfnamefont{S.}~\bibnamefont{Achilli}},
  \bibinfo{author}{\bibfnamefont{S.}~\bibnamefont{Caravati}}, \bibnamefont{and}
  \bibinfo{author}{\bibfnamefont{M.~I.} \bibnamefont{Trioni}},
  \bibinfo{journal}{Journal of Physics: Condensed Matter}
  \textbf{\bibinfo{volume}{19}}, \bibinfo{pages}{305021}
  (\bibinfo{year}{2007}), ISSN \bibinfo{issn}{0953-8984},
  \urlprefix\url{http://stacks.iop.org/0953-8984/19/i=30/a=305021}.

\bibitem[{\citenamefont{Prokop et~al.}(2009)\citenamefont{Prokop, Tang, Zhang,
  Tudosa, Peixoto, Zakeri, and Kirschner}}]{Prokop2009}
\bibinfo{author}{\bibfnamefont{J.}~\bibnamefont{Prokop}},
  \bibinfo{author}{\bibfnamefont{W.~X.} \bibnamefont{Tang}},
  \bibinfo{author}{\bibfnamefont{Y.}~\bibnamefont{Zhang}},
  \bibinfo{author}{\bibfnamefont{I.}~\bibnamefont{Tudosa}},
  \bibinfo{author}{\bibfnamefont{T.~R.~F.} \bibnamefont{Peixoto}},
  \bibinfo{author}{\bibfnamefont{K.}~\bibnamefont{Zakeri}}, \bibnamefont{and}
  \bibinfo{author}{\bibfnamefont{J.}~\bibnamefont{Kirschner}},
  \bibinfo{journal}{Physical Review Letters} \textbf{\bibinfo{volume}{102}},
  \bibinfo{eid}{177206} (pages~\bibinfo{numpages}{4}) (\bibinfo{year}{2009}),
  \urlprefix\url{http://link.aps.org/abstract/PRL/v102/e177206}.

\bibitem[{\citenamefont{Tang et~al.}(2007)\citenamefont{Tang, Zhang, Tudosa,
  Prokop, Etzkorn, and Kirschner}}]{Tang2007}
\bibinfo{author}{\bibfnamefont{W.~X.} \bibnamefont{Tang}},
  \bibinfo{author}{\bibfnamefont{Y.}~\bibnamefont{Zhang}},
  \bibinfo{author}{\bibfnamefont{I.}~\bibnamefont{Tudosa}},
  \bibinfo{author}{\bibfnamefont{J.}~\bibnamefont{Prokop}},
  \bibinfo{author}{\bibfnamefont{M.}~\bibnamefont{Etzkorn}}, \bibnamefont{and}
  \bibinfo{author}{\bibfnamefont{J.}~\bibnamefont{Kirschner}},
  \bibinfo{journal}{Phys. Rev. Lett.} \textbf{\bibinfo{volume}{99}},
  \bibinfo{pages}{087202} (\bibinfo{year}{2007}),
  \urlprefix\url{http://link.aps.org/abstract/PRL/v99/e087202}.

\bibitem[{\citenamefont{Qian and H\"ubner}(1999)}]{Qian1999}
\bibinfo{author}{\bibfnamefont{X.}~\bibnamefont{Qian}} \bibnamefont{and}
  \bibinfo{author}{\bibfnamefont{W.}~\bibnamefont{H\"ubner}},
  \bibinfo{journal}{Phys. Rev. B} \textbf{\bibinfo{volume}{60}},
  \bibinfo{pages}{16192} (\bibinfo{year}{1999}),
  \urlprefix\url{http://link.aps.org/abstract/PRB/v60/p16192}.

\bibitem[{\citenamefont{Sander et~al.}(1999)\citenamefont{Sander, Enders, and
  Kirschner}}]{Sander1999a}
\bibinfo{author}{\bibfnamefont{D.}~\bibnamefont{Sander}},
  \bibinfo{author}{\bibfnamefont{A.}~\bibnamefont{Enders}}, \bibnamefont{and}
  \bibinfo{author}{\bibfnamefont{J.}~\bibnamefont{Kirschner}},
  \bibinfo{journal}{EPL (Europhysics Letters)} \textbf{\bibinfo{volume}{45}},
  \bibinfo{pages}{208} (\bibinfo{year}{1999}), ISSN \bibinfo{issn}{0295-5075}.

\bibitem[{\citenamefont{Meyerheim et~al.}(2001)\citenamefont{Meyerheim, Sander,
  Popescu, Kirschner, Steadman, and Ferrer}}]{Meyerheim2001}
\bibinfo{author}{\bibfnamefont{H.~L.} \bibnamefont{Meyerheim}},
  \bibinfo{author}{\bibfnamefont{D.}~\bibnamefont{Sander}},
  \bibinfo{author}{\bibfnamefont{R.}~\bibnamefont{Popescu}},
  \bibinfo{author}{\bibfnamefont{J.}~\bibnamefont{Kirschner}},
  \bibinfo{author}{\bibfnamefont{P.}~\bibnamefont{Steadman}}, \bibnamefont{and}
  \bibinfo{author}{\bibfnamefont{S.}~\bibnamefont{Ferrer}},
  \bibinfo{journal}{Phys. Rev. B} \textbf{\bibinfo{volume}{64}},
  \bibinfo{pages}{045414} (\bibinfo{year}{2001}),
  \urlprefix\url{http://link.aps.org/abstract/PRB/v64/e045414}.

\bibitem[{\citenamefont{Zhang et~al.}(2010)\citenamefont{Zhang, Buczek,
  Sandratskii, Tang, Prokop, Tudosa, Peixoto, Zakeri, and
  Kirschner}}]{Zhang2010}
\bibinfo{author}{\bibfnamefont{Y.}~\bibnamefont{Zhang}},
  \bibinfo{author}{\bibfnamefont{P.}~\bibnamefont{Buczek}},
  \bibinfo{author}{\bibfnamefont{L.}~\bibnamefont{Sandratskii}},
  \bibinfo{author}{\bibfnamefont{W.~X.} \bibnamefont{Tang}},
  \bibinfo{author}{\bibfnamefont{J.}~\bibnamefont{Prokop}},
  \bibinfo{author}{\bibfnamefont{I.}~\bibnamefont{Tudosa}},
  \bibinfo{author}{\bibfnamefont{T.~R.~F.} \bibnamefont{Peixoto}},
  \bibinfo{author}{\bibfnamefont{K.}~\bibnamefont{Zakeri}}, \bibnamefont{and}
  \bibinfo{author}{\bibfnamefont{J.}~\bibnamefont{Kirschner}},
  \bibinfo{journal}{Phys. Rev. B} \textbf{\bibinfo{volume}{81}},
  \bibinfo{pages}{094438} (\bibinfo{year}{2010}),
  \urlprefix\url{http://link.aps.org/doi/10.1103/PhysRevB.81.094438}.

\bibitem[{\citenamefont{Muniz et~al.}(2008)\citenamefont{Muniz, Costa, and
  Mills}}]{Muniz2008}
\bibinfo{author}{\bibfnamefont{R.}~\bibnamefont{Muniz}},
  \bibinfo{author}{\bibfnamefont{A.}~\bibnamefont{Costa}}, \bibnamefont{and}
  \bibinfo{author}{\bibfnamefont{D.}~\bibnamefont{Mills}},
  \bibinfo{journal}{Magnetics, IEEE Transactions on DOI -
  10.1109/TMAG.2008.924546} \textbf{\bibinfo{volume}{44}},
  \bibinfo{pages}{1974} (\bibinfo{year}{2008}), ISSN \bibinfo{issn}{0018-9464}.

\bibitem[{\citenamefont{Pajda et~al.}(2000)\citenamefont{Pajda, Kudrnovsk\'y,
  Turek, Drchal, and Bruno}}]{Pajda2000}
\bibinfo{author}{\bibfnamefont{M.}~\bibnamefont{Pajda}},
  \bibinfo{author}{\bibfnamefont{J.}~\bibnamefont{Kudrnovsk\'y}},
  \bibinfo{author}{\bibfnamefont{I.}~\bibnamefont{Turek}},
  \bibinfo{author}{\bibfnamefont{V.}~\bibnamefont{Drchal}}, \bibnamefont{and}
  \bibinfo{author}{\bibfnamefont{P.}~\bibnamefont{Bruno}},
  \bibinfo{journal}{Phys. Rev. Lett.} \textbf{\bibinfo{volume}{85}},
  \bibinfo{pages}{5424} (\bibinfo{year}{2000}),
  \urlprefix\url{http://link.aps.org/abstract/PRL/v85/p5424}.

\bibitem[{\citenamefont{Udvardi et~al.}(2003)\citenamefont{Udvardi, Szunyogh,
  Palotas, and Weinberger}}]{Udvardi2003}
\bibinfo{author}{\bibfnamefont{L.}~\bibnamefont{Udvardi}},
  \bibinfo{author}{\bibfnamefont{L.}~\bibnamefont{Szunyogh}},
  \bibinfo{author}{\bibfnamefont{K.}~\bibnamefont{Palotas}}, \bibnamefont{and}
  \bibinfo{author}{\bibfnamefont{P.}~\bibnamefont{Weinberger}},
  \bibinfo{journal}{Physical Review B (Condensed Matter and Materials Physics)}
  \textbf{\bibinfo{volume}{68}}, \bibinfo{eid}{104436}
  (pages~\bibinfo{numpages}{11}) (\bibinfo{year}{2003}),
  \urlprefix\url{http://link.aps.org/abstract/PRB/v68/e104436}.

\bibitem[{\citenamefont{Sandratskii et~al.}(2007)\citenamefont{Sandratskii,
  Singer, and \c{S}a\c{s}io\u{g}lu}}]{Sandratskii2007}
\bibinfo{author}{\bibfnamefont{L.~M.} \bibnamefont{Sandratskii}},
  \bibinfo{author}{\bibfnamefont{R.}~\bibnamefont{Singer}}, \bibnamefont{and}
  \bibinfo{author}{\bibfnamefont{E.}~\bibnamefont{\c{S}a\c{s}io\u{g}lu}},
  \bibinfo{journal}{Phys. Rev. B} \textbf{\bibinfo{volume}{76}},
  \bibinfo{pages}{184406} (\bibinfo{year}{2007}),
  \urlprefix\url{http://link.aps.org/abstract/PRB/v76/e184406}.

\bibitem[{\citenamefont{Yoo et~al.}(1998)\citenamefont{Yoo, Soderlind, and
  Cynn}}]{Yoo1998}
\bibinfo{author}{\bibfnamefont{C.-S.} \bibnamefont{Yoo}},
  \bibinfo{author}{\bibfnamefont{P.}~\bibnamefont{Soderlind}},
  \bibnamefont{and} \bibinfo{author}{\bibfnamefont{H.}~\bibnamefont{Cynn}},
  \bibinfo{journal}{Journal of Physics: Condensed Matter}
  \textbf{\bibinfo{volume}{10}}, \bibinfo{pages}{L311} (\bibinfo{year}{1998}),
  \urlprefix\url{http://stacks.iop.org/0953-8984/10/L311}.

\bibitem[{\citenamefont{Taylor and Floyd}(1950)}]{Taylor1950}
\bibinfo{author}{\bibfnamefont{A.}~\bibnamefont{Taylor}} \bibnamefont{and}
  \bibinfo{author}{\bibfnamefont{R.~W.} \bibnamefont{Floyd}},
  \bibinfo{journal}{Acta Crystallographica} \textbf{\bibinfo{volume}{3}},
  \bibinfo{pages}{285} (\bibinfo{year}{1950}),
  \urlprefix\url{http://dx.doi.org/10.1107/S0365110X50000732}.

\bibitem[{\citenamefont{Perring et~al.}(1995)\citenamefont{Perring, Taylor, and
  Squires}}]{Perring1995}
\bibinfo{author}{\bibfnamefont{T.}~\bibnamefont{Perring}},
  \bibinfo{author}{\bibfnamefont{A.}~\bibnamefont{Taylor}}, \bibnamefont{and}
  \bibinfo{author}{\bibfnamefont{G.}~\bibnamefont{Squires}},
  \bibinfo{journal}{Physica B: Condensed Matter}
  \textbf{\bibinfo{volume}{213-214}}, \bibinfo{pages}{348}
  (\bibinfo{year}{1995}),
  \urlprefix\url{http://www.sciencedirect.com/science/article/B6TVH-40087S6-BX%
/2/02be76d1df0c67afea117c118c335d32}.

\bibitem[{\citenamefont{Buczek et~al.}(2010{\natexlab{c}})\citenamefont{Buczek,
  Ernst, and Sandratskii}}]{Buczek2010a}
\bibinfo{author}{\bibfnamefont{P.}~\bibnamefont{Buczek}},
  \bibinfo{author}{\bibfnamefont{A.}~\bibnamefont{Ernst}}, \bibnamefont{and}
  \bibinfo{author}{\bibfnamefont{L.~M.} \bibnamefont{Sandratskii}},
  \bibinfo{journal}{Journal of Physics: Conference Series}
  \textbf{\bibinfo{volume}{200}}, \bibinfo{pages}{042006}
  (\bibinfo{year}{2010}{\natexlab{c}}),
  \urlprefix\url{http://iopscience.iop.org/1742-6596/200/4/042006/}.

\bibitem[{\citenamefont{Trohidou et~al.}(1991)\citenamefont{Trohidou, Blackman,
  and Cooke}}]{Trohidou1991}
\bibinfo{author}{\bibfnamefont{K.~N.} \bibnamefont{Trohidou}},
  \bibinfo{author}{\bibfnamefont{J.~A.} \bibnamefont{Blackman}},
  \bibnamefont{and} \bibinfo{author}{\bibfnamefont{J.~F.} \bibnamefont{Cooke}},
  \bibinfo{journal}{Phys. Rev. Lett.} \textbf{\bibinfo{volume}{67}},
  \bibinfo{pages}{2561} (\bibinfo{year}{1991}),
  \urlprefix\url{http://link.aps.org/abstract/PRL/v67/p2561}.

\bibitem[{\citenamefont{Bass et~al.}(1992)\citenamefont{Bass, Blackman, and
  Cooke}}]{Bass1992}
\bibinfo{author}{\bibfnamefont{J.~M.} \bibnamefont{Bass}},
  \bibinfo{author}{\bibfnamefont{J.~A.} \bibnamefont{Blackman}},
  \bibnamefont{and} \bibinfo{author}{\bibfnamefont{J.~F.} \bibnamefont{Cooke}},
  \bibinfo{journal}{Journal of Physics: Condensed Matter}
  \textbf{\bibinfo{volume}{4}}, \bibinfo{pages}{L275} (\bibinfo{year}{1992}),
  \urlprefix\url{http://stacks.iop.org/0953-8984/4/L275}.

\bibitem[{\citenamefont{Balashov}(2009)}]{Balashov2009a}
\bibinfo{author}{\bibfnamefont{T.}~\bibnamefont{Balashov}}, Ph.D. thesis,
  \bibinfo{school}{Physikalisches Institut, Universit\"at Karlsruhe (TH)},
  \bibinfo{address}{Karlsruhe} (\bibinfo{year}{2009}),
  \urlprefix\url{http://digbib.ubka.uni-karlsruhe.de/volltexte/1000011569}.

\bibitem[{\citenamefont{Sinclair and Brockhouse}(1960)}]{Sinclair1960}
\bibinfo{author}{\bibfnamefont{R.~N.} \bibnamefont{Sinclair}} \bibnamefont{and}
  \bibinfo{author}{\bibfnamefont{B.~N.} \bibnamefont{Brockhouse}},
  \bibinfo{journal}{Phys. Rev.} \textbf{\bibinfo{volume}{120}},
  \bibinfo{pages}{1638} (\bibinfo{year}{1960}),
  \urlprefix\url{http://link.aps.org/abstract/PR/v120/p1638}.

\bibitem[{\citenamefont{Pickart et~al.}(1967)\citenamefont{Pickart, Alperin,
  Minkiewicz, Nathans, Shirane, and Steinsvoll}}]{Pickart1967}
\bibinfo{author}{\bibfnamefont{S.~J.} \bibnamefont{Pickart}},
  \bibinfo{author}{\bibfnamefont{H.~A.} \bibnamefont{Alperin}},
  \bibinfo{author}{\bibfnamefont{V.~J.} \bibnamefont{Minkiewicz}},
  \bibinfo{author}{\bibfnamefont{R.}~\bibnamefont{Nathans}},
  \bibinfo{author}{\bibfnamefont{G.}~\bibnamefont{Shirane}}, \bibnamefont{and}
  \bibinfo{author}{\bibfnamefont{O.}~\bibnamefont{Steinsvoll}},
  \bibinfo{journal}{Phys. Rev.} \textbf{\bibinfo{volume}{156}},
  \bibinfo{pages}{623} (\bibinfo{year}{1967}),
  \urlprefix\url{http://link.aps.org/abstract/PR/v156/p623}.

\bibitem[{\citenamefont{Mook et~al.}(1969)\citenamefont{Mook, Nicklow,
  Thompson, and Wilkinson}}]{Mook1969}
\bibinfo{author}{\bibfnamefont{H.~A.} \bibnamefont{Mook}},
  \bibinfo{author}{\bibfnamefont{R.~M.} \bibnamefont{Nicklow}},
  \bibinfo{author}{\bibfnamefont{E.~D.} \bibnamefont{Thompson}},
  \bibnamefont{and} \bibinfo{author}{\bibfnamefont{M.~K.}
  \bibnamefont{Wilkinson}}, \bibinfo{journal}{J. Appl. Phys.}
  \textbf{\bibinfo{volume}{40}}, \bibinfo{pages}{1450} (\bibinfo{year}{1969}),
  \urlprefix\url{http://link.aip.org/link/?JAP/40/1450/1}.

\bibitem[{\citenamefont{Mook and Tocchetti}(1979)}]{Mook1979}
\bibinfo{author}{\bibfnamefont{H.~A.} \bibnamefont{Mook}} \bibnamefont{and}
  \bibinfo{author}{\bibfnamefont{D.}~\bibnamefont{Tocchetti}},
  \bibinfo{journal}{Phys. Rev. Lett.} \textbf{\bibinfo{volume}{43}},
  \bibinfo{pages}{2029} (\bibinfo{year}{1979}),
  \urlprefix\url{http://link.aps.org/abstract/PRL/v43/p2029}.

\bibitem[{\citenamefont{L\"owdin}(1950)}]{Lowdin1950}
\bibinfo{author}{\bibfnamefont{P.-O.} \bibnamefont{L\"owdin}},
  \bibinfo{journal}{The Journal of Chemical Physics}
  \textbf{\bibinfo{volume}{18}}, \bibinfo{pages}{365} (\bibinfo{year}{1950}),
  \urlprefix\url{http://dx.doi.org/10.1063/1.1747632}.

\bibitem[{\citenamefont{Mayer}(2002)}]{Mayer2002}
\bibinfo{author}{\bibfnamefont{I.}~\bibnamefont{Mayer}},
  \bibinfo{journal}{International Journal of Quantum Chemistry}
  \textbf{\bibinfo{volume}{90}}, \bibinfo{pages}{63} (\bibinfo{year}{2002}),
  \urlprefix\url{http://dx.doi.org/10.1002/qua.981}.

\bibitem[{\citenamefont{Denman and Beavers}(1976)}]{Denman1976}
\bibinfo{author}{\bibfnamefont{E.~D.} \bibnamefont{Denman}} \bibnamefont{and}
  \bibinfo{author}{\bibfnamefont{A.~N.} \bibnamefont{Beavers}},
  \bibinfo{journal}{Applied Mathematics and Computation}
  \textbf{\bibinfo{volume}{2}}, \bibinfo{pages}{63} (\bibinfo{year}{1976}),
  ISSN \bibinfo{issn}{0096-3003},
  \urlprefix\url{http://www.sciencedirect.com/science/article/B6TY8-45DHTDS-1P%
/2/72f0208a7d533e941c6f9037dde2ba3c}.

\bibitem[{\citenamefont{Wildberger et~al.}(1995)\citenamefont{Wildberger, Lang,
  Zeller, and Dederichs}}]{Wildberger1995}
\bibinfo{author}{\bibfnamefont{K.}~\bibnamefont{Wildberger}},
  \bibinfo{author}{\bibfnamefont{P.}~\bibnamefont{Lang}},
  \bibinfo{author}{\bibfnamefont{R.}~\bibnamefont{Zeller}}, \bibnamefont{and}
  \bibinfo{author}{\bibfnamefont{P.~H.} \bibnamefont{Dederichs}},
  \bibinfo{journal}{Physical Review B (Condensed Matter)}
  \textbf{\bibinfo{volume}{52}}, \bibinfo{pages}{11502} (\bibinfo{year}{1995}),
  \urlprefix\url{http://link.aps.org/abstract/PRB/v52/p11502}.

\bibitem[{\citenamefont{Beach et~al.}(2000)\citenamefont{Beach, Gooding, and
  Marsiglio}}]{Beach2000}
\bibinfo{author}{\bibfnamefont{K.~S.~D.} \bibnamefont{Beach}},
  \bibinfo{author}{\bibfnamefont{R.~J.} \bibnamefont{Gooding}},
  \bibnamefont{and}
  \bibinfo{author}{\bibfnamefont{F.}~\bibnamefont{Marsiglio}},
  \bibinfo{journal}{Phys. Rev. B} \textbf{\bibinfo{volume}{61}},
  \bibinfo{pages}{5147} (\bibinfo{year}{2000}).

\bibitem[{\citenamefont{Eschrig et~al.}(1986)\citenamefont{Eschrig, Richter,
  and Velick\'y}}]{Eschrig1986}
\bibinfo{author}{\bibfnamefont{H.}~\bibnamefont{Eschrig}},
  \bibinfo{author}{\bibfnamefont{R.}~\bibnamefont{Richter}}, \bibnamefont{and}
  \bibinfo{author}{\bibfnamefont{B.}~\bibnamefont{Velick\'y}},
  \bibinfo{journal}{Journal of Physics C: Solid State Physics}
  \textbf{\bibinfo{volume}{19}}, \bibinfo{pages}{7173} (\bibinfo{year}{1986}),
  \urlprefix\url{http://stacks.iop.org/0022-3719/19/7173}.

\bibitem[{\citenamefont{Lee and Chang}(1996)}]{Lee1996}
\bibinfo{author}{\bibfnamefont{K.-H.} \bibnamefont{Lee}} \bibnamefont{and}
  \bibinfo{author}{\bibfnamefont{K.~J.} \bibnamefont{Chang}},
  \bibinfo{journal}{Phys. Rev. B} \textbf{\bibinfo{volume}{54}},
  \bibinfo{pages}{R8285} (\bibinfo{year}{1996}).

\bibitem[{\citenamefont{Hass et~al.}(1984)\citenamefont{Hass, Velick\'y, and
  Ehrenreich}}]{Hass1984}
\bibinfo{author}{\bibfnamefont{K.~C.} \bibnamefont{Hass}},
  \bibinfo{author}{\bibfnamefont{B.}~\bibnamefont{Velick\'y}},
  \bibnamefont{and}
  \bibinfo{author}{\bibfnamefont{H.}~\bibnamefont{Ehrenreich}},
  \bibinfo{journal}{Phys. Rev. B} \textbf{\bibinfo{volume}{29}},
  \bibinfo{pages}{3697} (\bibinfo{year}{1984}),
  \urlprefix\url{http://link.aps.org/abstract/PRB/v29/p3697}.

\bibitem[{\citenamefont{Gray and Kaplan}(1986)}]{Gray1986}
\bibinfo{author}{\bibfnamefont{L.~J.} \bibnamefont{Gray}} \bibnamefont{and}
  \bibinfo{author}{\bibfnamefont{T.}~\bibnamefont{Kaplan}},
  \bibinfo{journal}{Journal of Physics A: Mathematical and General}
  \textbf{\bibinfo{volume}{19}}, \bibinfo{pages}{1555} (\bibinfo{year}{1986}),
  ISSN \bibinfo{issn}{0305-4470}.

\end{thebibliography}
